\documentclass[12pt,aps,prd,tightenlines,nofootinbib,floatfix]{revtex4-2}
\usepackage{latexsym,amsmath,amssymb,amsbsy,graphicx,tabularx}
\usepackage{multirow}
\usepackage{rotating}

\begin{document}

\title{Testing relativistic quark model predictions for semileptonic $D$ and $D_s$ decays in light of recent experimental data}

\author{V.\,O. \surname{Galkin$^1$}} 
\email{vgalkin@frccsc.ru}
\author{I.\,S. \surname{Sukhanov$^{1,2}$}}
\email{sukhanov.is17@physics.msu.ru}

\affiliation{$^1$
	Federal Research Center ``Computer Science and Control'' Russian Academy of Sciences, Vavilov Street 40, Moscow, 119333 Russia}
\affiliation{$^2$ %Department of Quantum Theory and High Energy Physics Physics,
 Faculty of Physics, M.V. Lomonosov Moscow State University, Leninskie Gory 1-2,
	Moscow 119991, Russia.}
\preprint{}	

\begin{abstract}
Weak decays of $D_{(s)}$ mesons are an important tool for precision tests of the Standard Model. Recent measurements by the BESIII Collaboration produced an extensive dataset on semileptonic $D$ and $D_s$ decays, providing increasingly precise constraints on theoretical predictions. In this paper, we compare predictions of the relativistic quark model for semileptonic $D$ and $D_s$ decays into pseudoscalar and vector mesons with recent experimental measurements. Using previously calculated form factors and updated input parameters, we evaluate the corresponding differential decay rates, angular observables, and branching fractions. We also extract the CKM matrix elements $|V_{cd}|$ and $|V_{cs}|$ from the available experimental measurements. The comparison reveals some channel-dependent discrepancies while supporting a unified description of a broad range of semileptonic charm-meson decays.
\end{abstract}


\maketitle

\section{INTRODUCTION} \label{sec:intro}

Semileptonic decays of charmed mesons provide an important environment for studies of weak interactions in the charm sector. In particular, the Cabibbo-favored and Cabibbo-suppressed transitions governed by the Cabibbo-Kobayashi-Maskawa (CKM) matrix elements $V_{cs}$ and $V_{cd}$, respectively, provide an important tool for determining these quantities and testing the consistency of the CKM framework. The extraction of CKM parameters from exclusive semileptonic decays requires both precise experimental measurements of decay rates and reliable knowledge of the corresponding hadronic matrix elements. Considerable experimental effort has therefore been devoted to measurements of $D$ and $D_s$ semileptonic decays into pseudoscalar and vector mesons. Measurements of the $D\to K\ell\nu$ and $D\to\pi\ell\nu$ channels, which are particularly relevant for determinations of $|V_{cs}|$ and $|V_{cd}|$, have been performed by Belle~\cite{Belle2006}, BaBar~\cite{BaBar2015Pi}, CLEO~\cite{CLEO2009PiK}, and, more recently, BESIII \cite{BESIII2026Pi,BESIII2026K}. The experimental program has also been extended to a broad range of final states containing $\eta^{(\prime)}$, $\rho$, $\omega$, $K^{(*)}$, and $\phi$ mesons, providing complementary information on the dynamics of charm semileptonic transitions \cite{BESIII2019Rho,BESIII2019DsKstar,BESIII2020OmegaMu,BESIII2023DsEta,BESIII2023DsPhi}.

Experimental studies of charm semileptonic decays have progressed substantially during the last decade, with BESIII providing measurements for an increasingly broad set of decay modes. Among the results obtained in recent years are $D_s^+\to\phi\ell^+\nu_\ell$, $D\to\rho\ell\nu_\ell$, $D^+\to\omega\mu^+\nu_\mu$, and $D_s^+\to K^{*0}e^+\nu_e$, together with measurements of their branching fractions and, in several cases, decay dynamics and form factors \cite{BESIII2018DsPhi,BESIII2019Rho,BESIII2019DsKstar,BESIII2020OmegaMu,BESIII2021RhoMu}. Particularly rapid progress has been achieved during the last four years. Precision measurements of $D_s^+\to\eta^{(\prime)}e^+\nu_e$ and studies of $D_s^+\to K^+K^-\mu^+\nu_\mu$ were reported in 2023 \cite{BESIII2023DsEta,BESIII2023DsPhi}; these were followed in 2024 by improved measurements of $D_s^+\to K^0e^+\nu_e$ and studies of $D_s^+\to\eta^{(\prime)}\mu^+\nu_\mu$, including tests of lepton-flavor universality and measurements of decay dynamics \cite{BESIII2024DsKe,BESIII2024DsEtaMu}. More recent results include improved analyses of $D^+\to\eta^{(\prime)}\ell^+\nu_\ell$ \cite{BESIII2025Eta,BESIII2025EtaPrime} and, in 2026, precise measurements of the $D\to\pi\ell\nu_\ell$, $D\to K\ell\nu_\ell$, and $D\to K^*(892)\ell\nu_\ell$ channels, as well as new measurements of Cabibbo-suppressed $D_s$ transitions \cite{BESIII2026Pi,BESIII2026K,BESIII2026Kstar,BESIII2025DsK,BESIII2026DsKstar}. 
Recent measurements increasingly provide detailed differential distributions and angular observables in addition to branching fractions.

The theoretical description of exclusive semileptonic decays is encoded in transition form factors, which contain the nonperturbative hadronic dynamics of the initial and final mesons. From the experimental point of view, measurements of differential decay distributions provide access not only to form-factor normalizations but also to their momentum-transfer dependence and, for decays into vector mesons, to ratios and combinations of several independent form factors. On the theory side, reliable calculations of these quantities are necessary for extracting CKM matrix elements from measured decay rates and for making predictions for channels with limited experimental information. At the same time, experimentally determined form factors constitute useful benchmarks for testing different approaches to nonperturbative QCD, in particular lattice QCD (LQCD; ETMC, HPQCD, and CLQCD) \cite{HPQCD2014,Lubicz2017,Fan2025}, QCD sum rules (QCDSR) \cite{Du2004}, light-cone sum rules (LCSR) \cite{Khodjamirian2009,Fu2020,Zhong2024,Lin2025}, LCSR within heavy-quark effective field theory (LCSR/HQEFT) \cite{Wu2006}, the covariant light-front quark model (CLFQM) \cite{Cheng2017,Yang2026}, and the covariant confining quark model (CCQM) \cite{Ivanov2019}, as well as continuum Schwinger function methods (CSM) \cite{Yao2022}, including the symmetry-preserving contact-interaction (SCI) framework \cite{Xing2022}. Comparison with experimentally extracted form factors over the accessible $q^2$ region thus provides a more detailed test of theoretical predictions for their normalization and momentum-transfer dependence, beyond comparisons of integrated branching fractions.

In this paper, we reassess our predictions of Ref.~\cite{Faustov2020} in light of recent experimental measurements. The form factors are taken from that calculation, with modifications restricted to corrections of typographical errors; the present analysis focuses on their comparison with measured $q^2$ distributions and decay observables and on the effect of $\eta$--$\eta'$ mixing, including a possible gluonic component.

The paper is organized as follows. In Sec.~\ref{sec:FF}, we introduce the hadronic matrix elements and the corresponding form factors for the $D_{(s)}\to P$ and $D_{(s)}\to V$ transitions, and describe their $q^2$ dependence and the helicity formalism for decay observables. The resulting form factors and decay observables are compared with the available experimental measurements in Sec.~\ref{sec:comparison}, with particular attention paid to the quality of the agreement over different kinematic regions. Finally, the main results and conclusions are summarized in Sec.~\ref{sec:conclusion}.

\section{Form factors}\label{sec:FF}
The nonperturbative dynamics of the weak
$D_{(s)}\to P$ and $D_{(s)}\to V$ transitions is conventionally
described in terms of invariant form factors. Throughout this section, $q=p_{D_{(s)}}-p_F$ is the four-momentum transferred to the lepton pair, with $F=P$ or $V$. We employ the conventions of Ref.~\cite{Faustov2020}.

For a transition of an initial $D_{(s)}$ meson with four-momentum
$p_{D_{(s)}}$ to a pseudoscalar meson $P$, the matrix element of the vector current is parameterized by two
independent form factors
\begin{align}
\left\langle
P(p_P)
\left|
\bar q\gamma^\mu c
\right|
D_{(s)}(p_{D_{(s)}})
\right\rangle
={}&
f_+(q^2)
\left[
p_{D_{(s)}}^\mu+p_P^\mu
-
\frac{M_{D_{(s)}}^2-M_P^2}{q^2}
q^\mu
\right]
\nonumber\\
&+
f_0(q^2)
\frac{M_{D_{(s)}}^2-M_P^2}{q^2}
q^\mu .
\label{eq:FF_P_matrix_element}
\end{align}
The absence of a kinematic singularity at $q^2=0$ requires $f_+(0)=f_0(0)$. The axial-vector-current matrix element vanishes for transitions between two pseudoscalar states by parity.

For transitions to a vector meson $V$ with four-momentum $p_V$,
mass $M_V$, and polarization vector $\epsilon^\mu$, the vector- and axial-vector-current matrix elements are parameterized by four independent form factors,

\begin{equation}
\left\langle
V(p_V,\epsilon)
\left|
\bar q\gamma^\mu c
\right|
D_{(s)}(p_{D_{(s)}})
\right\rangle
=
\frac{2iV(q^2)}
{M_{D_{(s)}}+M_V}
\varepsilon^{\mu\nu\rho\sigma}
\epsilon_\nu^*
p_{D_{(s)}\rho}
p_{V\sigma},
\label{eq:FF_V_vector}
\end{equation}

\begin{align}
\left\langle
V(p_V,\epsilon)
\left|
\bar q\gamma^\mu\gamma_5 c
\right|
D_{(s)}(p_{D_{(s)}})
\right\rangle
={}&
2M_V A_0(q^2)
\frac{\epsilon^*\cdot q}{q^2}
q^\mu
\nonumber\\
&+
\left(M_{D_{(s)}}+M_V\right)
A_1(q^2)
\left[
\epsilon^{*\mu}
-
\frac{\epsilon^*\cdot q}{q^2}
q^\mu
\right]
\nonumber\\
&-
A_2(q^2)
\frac{\epsilon^*\cdot q}
{M_{D_{(s)}}+M_V}
\left[
p_{D_{(s)}}^\mu+p_V^\mu
-
\frac{M_{D_{(s)}}^2-M_V^2}{q^2}
q^\mu
\right].
\label{eq:FF_V_axial}
\end{align}
At maximum recoil, the axial form factors satisfy
\begin{equation}
A_0(0)
=
\frac{M_{D_{(s)}}+M_V}{2M_V}A_1(0)
-
\frac{M_{D_{(s)}}-M_V}{2M_V}A_2(0),
\label{eq:FF_A_constraint}
\end{equation}
which removes the apparent kinematic singularity at $q^2=0$.

For weak $D_{(s)}$ transitions, the hadronic matrix element of the local current $J_\mu^{W}=\bar f\gamma_\mu(1-\gamma_5)c$, where $f=d,s$, is evaluated in the quasipotential approach as \cite{Faustov2020,Ebert2003b,Ebert2003c}
\begin{equation}
\langle F(p_F)|J_\mu^{W}|D_{(s)}(p_{D_{(s)}})\rangle
=
\int \frac{d^3p\,d^3q}{(2\pi)^6}
\,\overline{\Psi}_{F,\mathbf{p}_F}(\mathbf{p})\Gamma_\mu(\mathbf{p},\mathbf{q})
\Psi_{D_{(s)},\mathbf{p}_{D_{(s)}}}(\mathbf{q}).
\label{eq:RQM_matrix_element}
\end{equation}
The two-particle vertex function is decomposed as
\begin{equation}
\Gamma_\mu
=
\Gamma_\mu^{(1)}+\Gamma_\mu^{(2)}.
\label{eq:RQM_vertex_sum}
\end{equation}

The leading contribution $\Gamma_\mu^{(1)}$ corresponds to the impulse approximation, in which the weak current acts on the active charm quark while the antiquark remains a spectator.
The contribution $\Gamma_\mu^{(2)}$ accounts for the interaction of the active quark with the spectator antiquark and contains the negative-energy part of the intermediate active-quark propagator. It arises from the projection onto positive-energy states and depends explicitly on the quark--antiquark interaction kernel. Explicit expressions can be found in Refs.~\cite{Ebert2003b,Ebert2003c}.

Another essential relativistic effect is associated with the recoil of the final meson. The decay calculation is performed in the rest frame of the initial $D_{(s)}$ meson, where the final meson carries the recoil momentum $\boldsymbol{\Delta}=\mathbf{p}_F$. Its wave function must therefore be boosted from its rest frame to the moving frame. The corresponding transformation is given in Refs.~\cite{Faustov2020,Ebert2003b}.

The wave function of a meson of mass $M$ appearing in Eq.~(\ref{eq:RQM_matrix_element}) is determined from the Schr\"odinger-type quasipotential equation~\cite{Faustov2020,Ebert2003a}

\begin{equation}
\left(
\frac{b^2(M)}{2\mu_R}
-
\frac{\mathbf{p}^2}{2\mu_R}
\right)
\Psi_M(\mathbf{p})
=
\int \frac{d^3q}{(2\pi)^3}
V(\mathbf{p},\mathbf{q};M)
\Psi_M(\mathbf{q}),
\label{eq:RQM_quasipotential}
\end{equation}
where $\mathbf{p}$ is the relative quark momentum. The relative momentum squared on the mass shell in the center-of-mass frame is
\begin{equation}
b^2(M)
=
\frac{
\left[M^2-(m_1+m_2)^2\right]
\left[M^2-(m_1-m_2)^2\right]
}{4M^2},
\label{eq:RQM_b2}
\end{equation}
and the relativistic reduced mass is defined as
\begin{equation}
\mu_R
=
\frac{M^4-(m_1^2-m_2^2)^2}{4M^3},
\label{eq:RQM_muR}
\end{equation}
where $m_1$ and $m_2$ are the constituent-quark masses.

The kernel $V(\mathbf{p},\mathbf{q};M)$ in Eq.~\eqref{eq:RQM_quasipotential} is a QCD-motivated quark--antiquark quasipotential constructed from the off-shell scattering amplitude projected onto positive-energy states. It contains a one-gluon-exchange contribution, which dominates at short distances, and a mixture of Lorentz-vector and Lorentz-scalar linear confining interactions, which dominate at large distances \cite{Ebert2003a}. 
All parameters of the model were fixed in previous studies of hadron spectroscopy and decays and are not refitted to the semileptonic $D$-meson data \cite{Ebert2003a,Ebert2009,Ebert2010}. 

Within the RQM, the form factors entering
Eqs.~\eqref{eq:FF_P_matrix_element}--\eqref{eq:FF_V_axial}
are calculated as overlap integrals of the initial and final
meson wave functions. The explicit expressions, including the relevant relativistic contributions, are given in Refs.~\cite{Ebert2003b,Ebert2003c}. 

For the form factors $f_+(q^2)$, $V(q^2)$, and $A_0(q^2)$,
the following modified-pole parameterization is used:
\begin{equation}
F(q^2)
=
\frac{F(0)}
{
\displaystyle
\left(1-\frac{q^2}{M^2}\right)
\left(
1-\sigma_1\frac{q^2}{M_{D_{(s)}^*}^2}
+\sigma_2\frac{q^4}{M_{D_{(s)}^*}^4}
\right)
}.
\label{eq:FF_Faustov_pole}
\end{equation}
For $f_0(q^2)$, $A_1(q^2)$, and $A_2(q^2)$, the explicit
pole factor is absent,
\begin{equation}
F(q^2)
=
\frac{F(0)}
{
\displaystyle
1-\sigma_1\frac{q^2}{M_{D_{(s)}^*}^2}
+\sigma_2\frac{q^4}{M_{D_{(s)}^*}^4}
}.
\end{equation}

The mass entering the explicit pole in Eq.~\eqref{eq:FF_Faustov_pole} is chosen according to the quantum numbers of the corresponding intermediate state. Following Ref.~\cite{Faustov2020}, the $c\to s$ transitions use $M=2.112~\mathrm{GeV}$ for $f_+$ and $V$, and $M=1.968~\mathrm{GeV}$ for $A_0$; the corresponding $c\to d$ values are $2.010~\mathrm{GeV}$ and $1.870~\mathrm{GeV}$, respectively.

The numerical values of $F(0)$, $\sigma_1$, and $\sigma_2$
for individual $D$ and $D_s$ transitions are given in
Ref.~\cite{Faustov2020}.

The hadron masses, meson lifetimes, and CKM inputs used to evaluate decay observables are taken from the latest PDG compilation \cite{PDG2026}, while the mixing-angle inputs are taken from LHCb \cite{LHCb2025EtaMix}. An independent relative uncertainty of $5\%$ is assigned to each RQM form factor, and CKM and relevant mixing-angle uncertainties contribute to the predicted observables.

Once the transition form factors are determined, the differential and total decay rates of the semileptonic $D_{(s)}$ decays can be evaluated using the helicity formalism. The decay amplitude for the process
$D_{(s)}\to F\ell^{+}\nu_{\ell}$, where $F=P$ or $V$, is written as
\begin{equation}
\mathcal{M}
\left(
D_{(s)}\to F\ell^{+}\nu_{\ell}
\right)
=
\frac{G_F}{\sqrt{2}}
V_{cq}
H^{\mu}L_{\mu},
\label{eq:decay_amplitude}
\end{equation}
where $G_F$ is the Fermi constant, $V_{cq}$ is the corresponding CKM matrix element with $q=d,s$, and
\begin{equation}
L_{\mu}
=
\bar{\nu}_{\ell}\gamma_{\mu}(1-\gamma_5)\ell
\label{eq:leptonic_current}
\end{equation}
is the leptonic current. The hadronic current
\begin{equation}
H^{\mu}
=
\left\langle
F
\left|
\bar q\gamma^{\mu}(1-\gamma_5)c
\right|
D_{(s)}
\right\rangle
\label{eq:hadronic_current_decay}
\end{equation}
is expressed in terms of the invariant form factors introduced in
Eqs.~\eqref{eq:FF_P_matrix_element}--\eqref{eq:FF_V_axial}.

For convenience, we define the K\"all\'en function as
\begin{equation}
\lambda
\equiv
\lambda
\left(
M_{D_{(s)}}^2,M_F^2,q^2
\right)
=
M_{D_{(s)}}^4+M_F^4+q^4
-2
\left(
M_{D_{(s)}}^2M_F^2
+M_F^2q^2
+M_{D_{(s)}}^2q^2
\right),
\label{eq:decay_lambda}
\end{equation}
so that the magnitude of the final-meson three-momentum in the
$D_{(s)}$ rest frame is
\begin{equation}
|\mathbf{p}_F|
=
\frac{\sqrt{\lambda}}
{2M_{D_{(s)}}}.
\label{eq:final_momentum}
\end{equation}

The hadronic dynamics can be described in terms of the
helicity amplitudes $H_{\pm}$, $H_0$, and $H_t$, corresponding to
the transverse, longitudinal, and timelike polarizations of the
virtual $W$ boson.

For transitions to a pseudoscalar meson,
$D_{(s)}\to P$, the transverse amplitudes vanish 
\begin{equation}
H_{\pm}(q^2)=0,
\label{eq:HP_Hpm}
\end{equation}
and the remaining
helicity amplitudes are given by \cite{Faustov2020}
\begin{equation}
H_0(q^2)
=
\frac{\sqrt{\lambda}}{\sqrt{q^2}}
f_+(q^2),
\label{eq:HP_H0}
\end{equation}
%
%and
%
\begin{equation}
H_t(q^2)
=
\frac{M_{D_{(s)}}^2-M_P^2}
{\sqrt{q^2}}
f_0(q^2).
\label{eq:HP_Ht}
\end{equation}

For transitions to a vector meson,
$D_{(s)}\to V$, all four helicity amplitudes contribute. In the
conventions of Ref.~\cite{Faustov2020}, they have the form
\begin{equation}
H_{\pm}(q^2)
=
\frac{\sqrt{\lambda}}
{M_{D_{(s)}}+M_V}
\left[
V(q^2)
\mp
\frac{
\left(M_{D_{(s)}}+M_V\right)^2
}
{\sqrt{\lambda}}
A_1(q^2)
\right],
\label{eq:HV_Hpm}
\end{equation}
\begin{equation}
H_0(q^2)
=
\frac{1}
{2M_V\sqrt{q^2}}
\left[
\left(M_{D_{(s)}}+M_V\right)
\left(
M_{D_{(s)}}^2-M_V^2-q^2
\right)
A_1(q^2)
-
\frac{\lambda}
{M_{D_{(s)}}+M_V}
A_2(q^2)
\right],
\label{eq:HV_H0}
\end{equation}
and
\begin{equation}
H_t(q^2)
=
\frac{\sqrt{\lambda}}
{\sqrt{q^2}}
A_0(q^2).
\label{eq:HV_Ht}
\end{equation}

Following the helicity representation used in
Ref.~\cite{Faustov2020}, it is convenient to introduce the
bilinear combinations
\begin{align}
\mathcal{H}_U
&=
|H_+|^2+|H_-|^2,
&
\mathcal{H}_L
&=
|H_0|^2,
\nonumber\\
\mathcal{H}_P
&=
|H_+|^2-|H_-|^2,
&
\mathcal{H}_S
&=
|H_t|^2,
\nonumber\\
\mathcal{H}_{SL}
&=
\operatorname{Re}
\left(
H_0H_t^{\dagger}
\right).
\label{eq:helicity_structure_functions}
\end{align}
Here $\mathcal{H}_U$ and $\mathcal{H}_L$ describe the transverse
and longitudinal contributions, respectively, while
$\mathcal{H}_S$ represents the scalar contribution associated
with the timelike helicity amplitude. The functions
$\mathcal{H}_P$ and $\mathcal{H}_{SL}$ enter the angular and
polarization observables.

The differential decay rate can be expressed as
\cite{Faustov2020}

\begin{equation}
\frac{
d\Gamma
\left(
D_{(s)}\to F\ell^+\nu_{\ell}
\right)
}
{dq^2}
=
\frac{G_F^2}{(2\pi)^3}
|V_{cq}|^2
\frac{
\sqrt{\lambda}
\left(q^2-m_{\ell}^2\right)^2
}
{
24M_{D_{(s)}}^3q^2
}
\mathcal{H}_{\mathrm{total}},
\label{eq:differential_width}
\end{equation}
where
\begin{equation}
\mathcal{H}_{\mathrm{total}}
=
\left(
\mathcal{H}_U+\mathcal{H}_L
\right)
\left(
1+\frac{m_{\ell}^2}{2q^2}
\right)
+
\frac{3m_{\ell}^2}{2q^2}
\mathcal{H}_S.
\label{eq:Htotal}
\end{equation}

Additional information on the weak transition can be extracted from angular observables. We define $\theta$ as the angle between the charged-lepton momentum and the direction opposite to the final-meson momentum, both considered in the dilepton rest frame. With this convention, the normalized angular distribution can be written as
\cite{Faustov2020}
\begin{align}
\frac{1}{d\Gamma/dq^2}
\frac{d^2\Gamma}
{dq^2\,d(\cos\theta)}
={}&
\frac{1}{2}
\left[
1-\frac{1}{3}C_F^{\ell}(q^2)
\right]
+
A_{FB}(q^2)\cos\theta
\nonumber\\
&+
\frac{1}{2}
C_F^{\ell}(q^2)
\cos^2\theta .
\label{eq:normalized_angular_distribution}
\end{align}

The forward--backward asymmetry of the charged lepton is defined as
\begin{equation}
A_{FB}(q^2)
=
\frac{
\displaystyle
\int_0^1d(\cos\theta)
\frac{d^2\Gamma}{dq^2d(\cos\theta)}
-
\int_{-1}^0d(\cos\theta)
\frac{d^2\Gamma}{dq^2d(\cos\theta)}
}
{
\displaystyle
\int_0^1d(\cos\theta)
\frac{d^2\Gamma}{dq^2d(\cos\theta)}
+
\int_{-1}^0d(\cos\theta)
\frac{d^2\Gamma}{dq^2d(\cos\theta)}
},
\label{eq:AFB_definition}
\end{equation}
which in terms of the helicity amplitudes becomes
\begin{equation}
A_{FB}(q^2)
=
\frac{3}{4}
\frac{
\displaystyle
\mathcal{H}_P
-
2\frac{m_{\ell}^2}{q^2}\mathcal{H}_{SL}
}
{
\mathcal{H}_{\mathrm{total}}
}.
\label{eq:AFB}
\end{equation}

The total decay width is obtained by integration over the physical region, $m_\ell^2\leq q^2\leq(M_{D_{(s)}}-M_F)^2$,
\begin{equation}
\Gamma(D_{(s)}\to F\ell^+\nu_\ell)=
\int_{m_\ell^2}^{(M_{D_{(s)}}-M_F)^2}
\frac{d\Gamma}{dq^2}\,dq^2.
\label{eq:total_partial_width}
\end{equation}
The branching fraction is then obtained from this total decay width and the lifetime of the initial meson,
\begin{equation}
\mathcal{B}
\left(
D_{(s)}\to F\ell^+\nu_{\ell}
\right)
=
\tau_{D_{(s)}}
\,
\Gamma
\left(
D_{(s)}\to F\ell^+\nu_{\ell}
\right).
\label{eq:branching_fraction}
\end{equation}

To test electron--muon lepton-flavor universality, Ref.~\cite{Faustov2020} considers the integrated ratio
\begin{equation}
R_F
=
\frac{
\Gamma
\left(
D_{(s)}\to F\mu^+\nu_{\mu}
\right)
}
{
\Gamma
\left(
D_{(s)}\to Fe^+\nu_e
\right)
}.
\label{eq:RF}
\end{equation}
Since the same initial meson is involved in the numerator and
denominator, the lifetime cancels, and the same ratio can be
constructed directly from the corresponding branching fractions,
\begin{equation}
R_F
=
\frac{
\mathcal{B}
\left(
D_{(s)}\to F\mu^+\nu_{\mu}
\right)
}
{
\mathcal{B}
\left(
D_{(s)}\to Fe^+\nu_e
\right)
}.
\label{eq:RF_branching}
\end{equation}
The differential lepton-flavor ratio shown in the figures is defined as
\begin{equation}
R_{\mathrm{LFU}}(q^2)=
\frac{d\Gamma(D_{(s)}\to F\mu^+\nu_\mu)/dq^2}{d\Gamma(D_{(s)}\to Fe^+\nu_e)/dq^2},
\qquad q^2\geq m_\mu^2.
\label{eq:RF_differential}
\end{equation}

\section{Comparison with experimental data and other theoretical predictions}
\label{sec:comparison}

In the following figures, the thick solid dark-blue curve denotes the central RQM prediction, while the outer blue curves delimit its $1\sigma$ theoretical uncertainty. Experimental data are shown as red circles (BESIII), magenta squares (BaBar), and green triangles (CLEO), where applicable. The dashed curves in the $\eta'$ channels show the central predictions after including the glueball mixing factor; we do not give separate uncertainty bands for this case. The $\eta'$ form factors and branching fractions in the tables are quoted without the glueball contribution. 

\subsection{$D\to\pi$}
\label{subsec:comparison_pi}

\begin{figure}[htbp]
    \centering

    \begin{minipage}{0.48\linewidth}
        \centering
        \includegraphics[width=\linewidth]{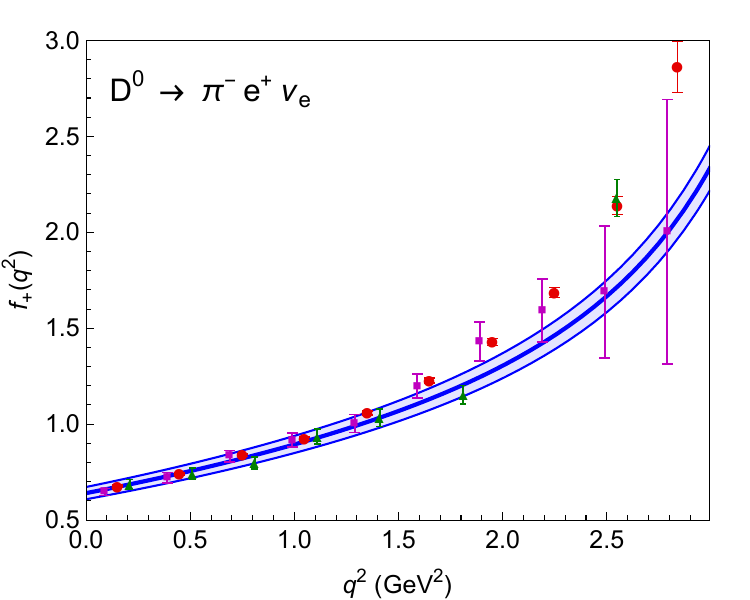}

        (a)
    \end{minipage}
    \hfill
    \begin{minipage}{0.48\linewidth}
        \centering
        \includegraphics[width=\linewidth]{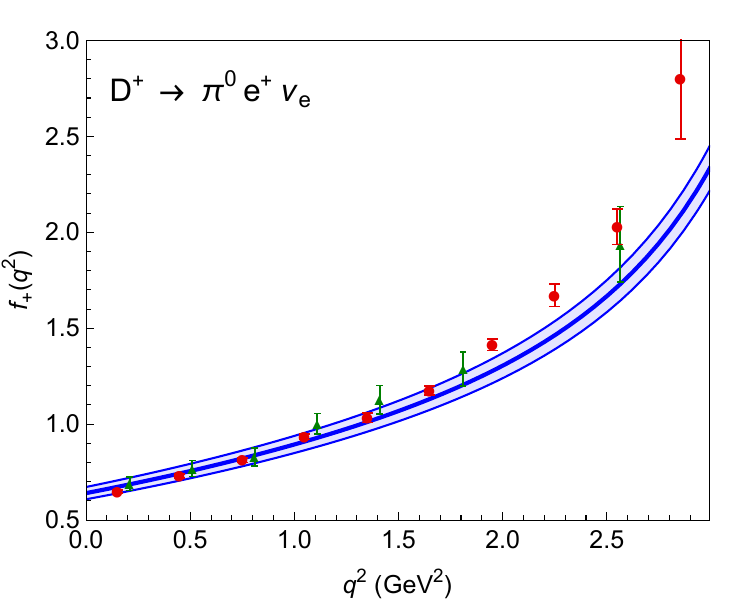}

        (b)
    \end{minipage}
    
	\begin{minipage}{0.48\linewidth}
        \centering
        \includegraphics[width=\linewidth]{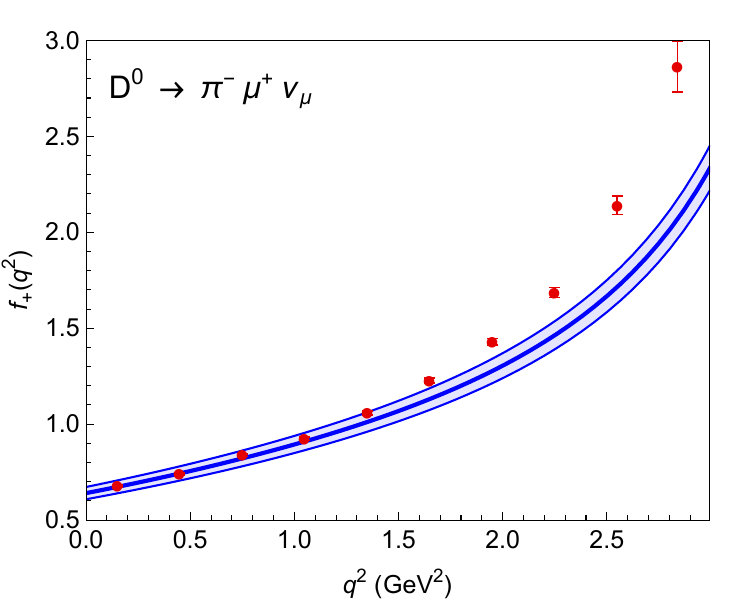}

        (c)
    \end{minipage}
    \hfill
    \begin{minipage}{0.48\linewidth}
        \centering
        \includegraphics[width=\linewidth]{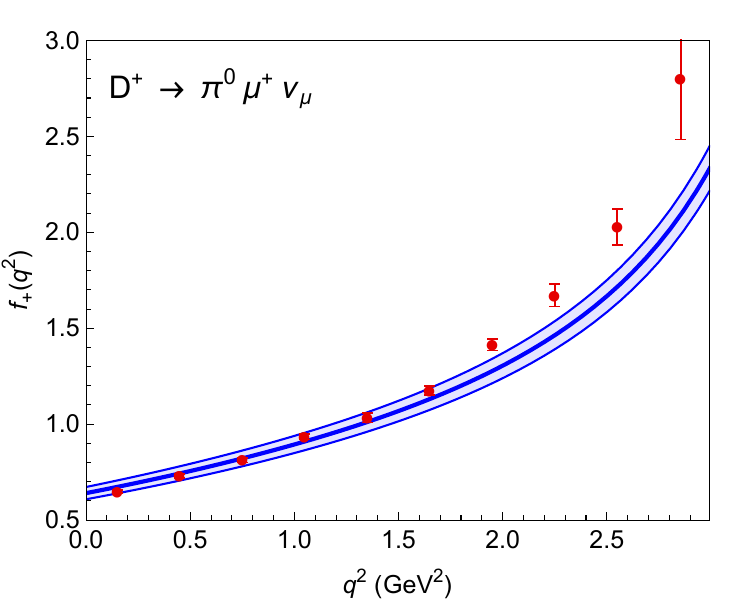}

        (d)
    \end{minipage}
    \caption{
        Form factors of $D\to\pi$ transitions.
    }
    \label{fig:FFDPi}
    
\end{figure}

\begin{figure}[htbp]
    \centering

    \begin{minipage}{0.48\linewidth}
        \centering
        \includegraphics[width=\linewidth]{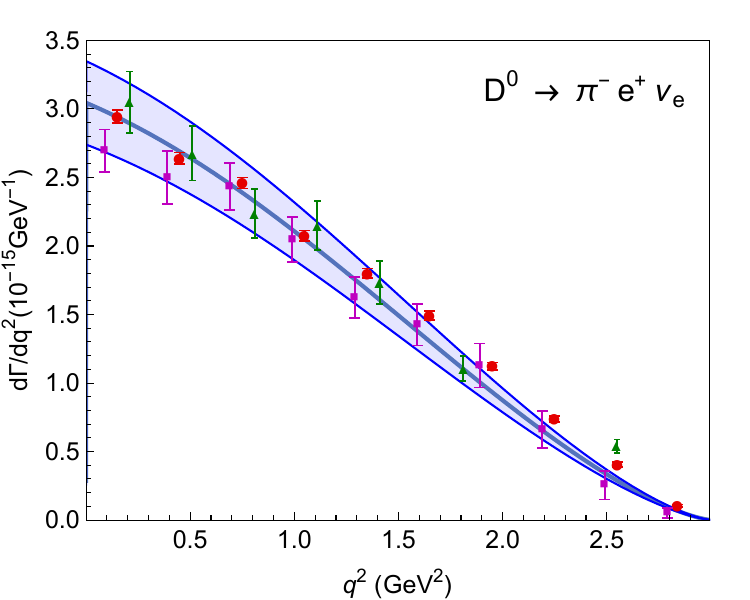}

        (a)
    \end{minipage}
    \hfill
    \begin{minipage}{0.48\linewidth}
        \centering
        \includegraphics[width=\linewidth]{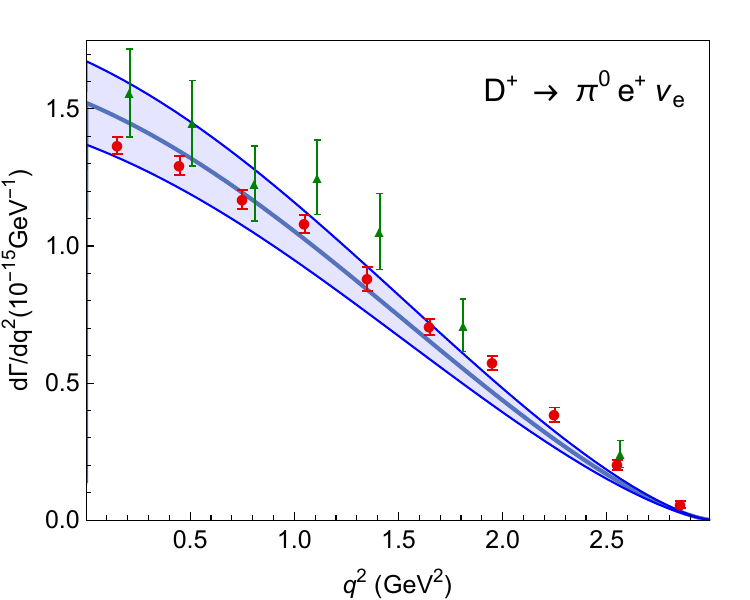}

        (b)
    \end{minipage}
    
	\begin{minipage}{0.48\linewidth}
        \centering
        \includegraphics[width=\linewidth]{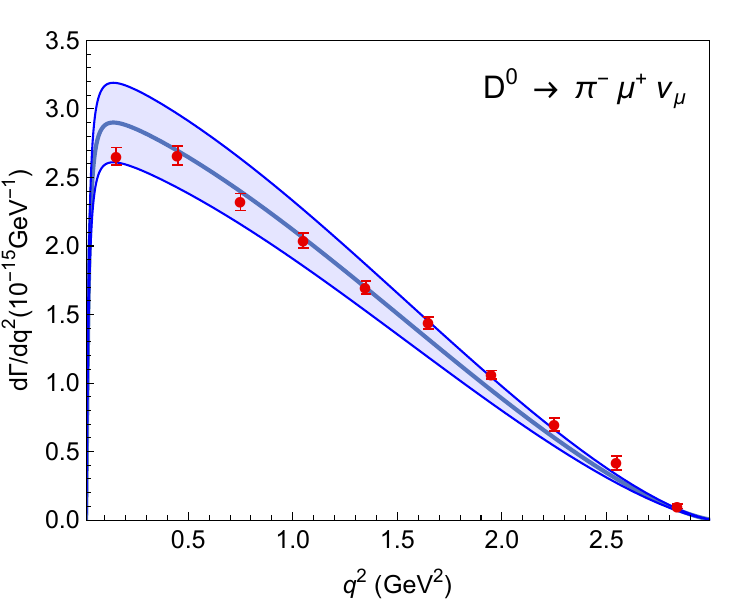}

        (c)
    \end{minipage}
    \hfill
    \begin{minipage}{0.48\linewidth}
        \centering
        \includegraphics[width=\linewidth]{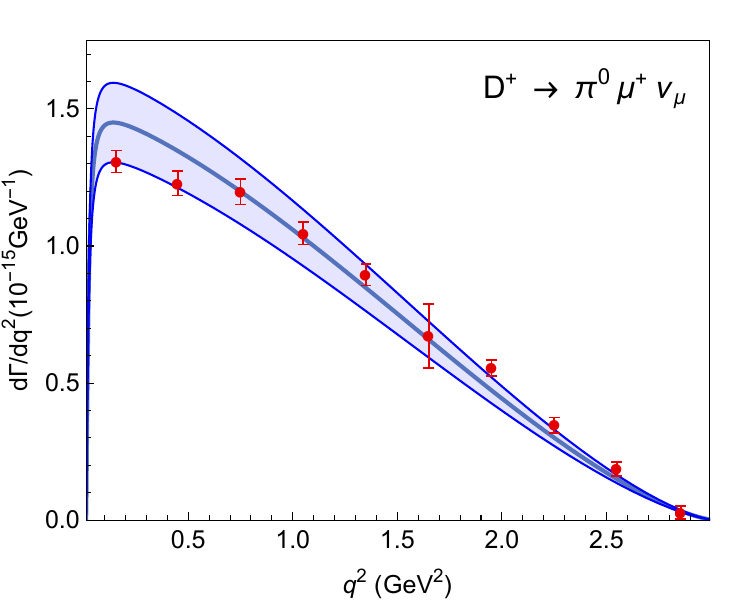}

        (d)
    \end{minipage}
    \caption{
        Differential decay rates for $D\to\pi$ transitions.
    }
    \label{fig:GammaDPi}
    
\end{figure}

\begin{figure}[htbp]
    \centering

    \begin{minipage}{0.48\linewidth}
        \centering
        \includegraphics[width=\linewidth]{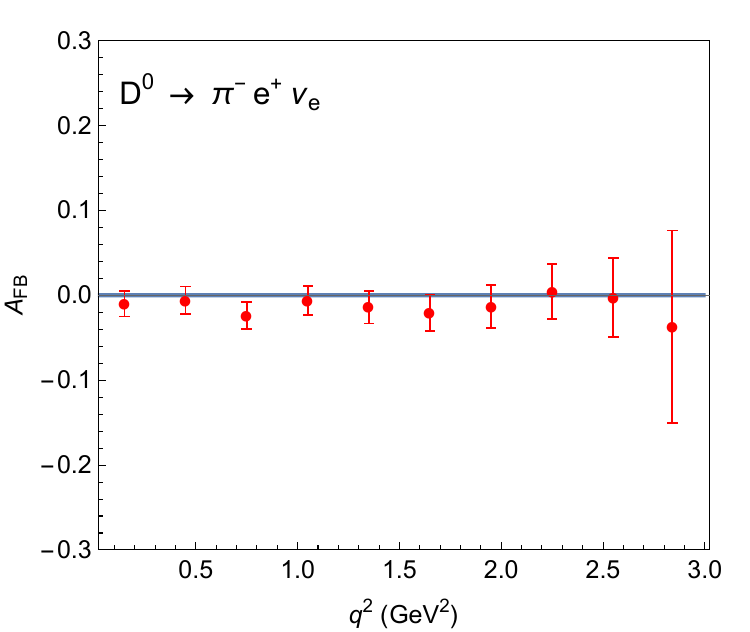}

        (a)
    \end{minipage}
    \hfill
    \begin{minipage}{0.48\linewidth}
        \centering
        \includegraphics[width=\linewidth]{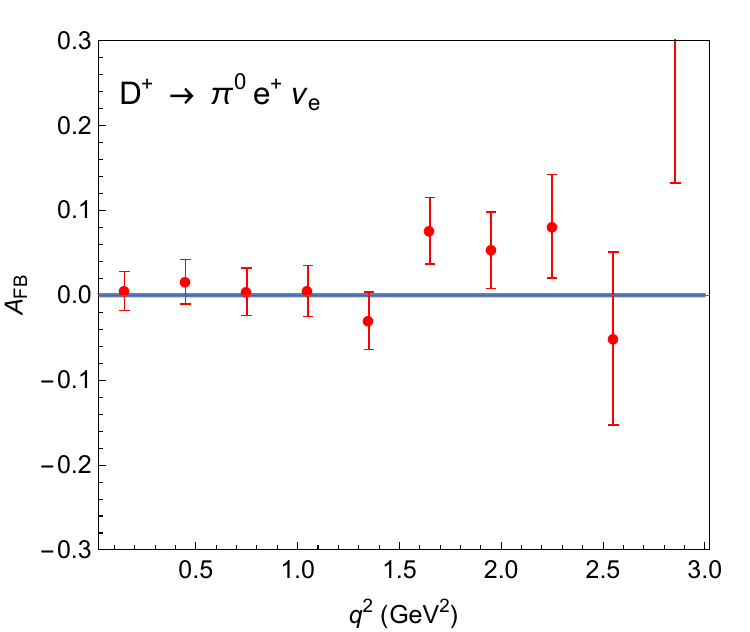}

        (b)
    \end{minipage}
 		
 	\begin{minipage}{0.48\linewidth}
        \centering
        \includegraphics[width=\linewidth]{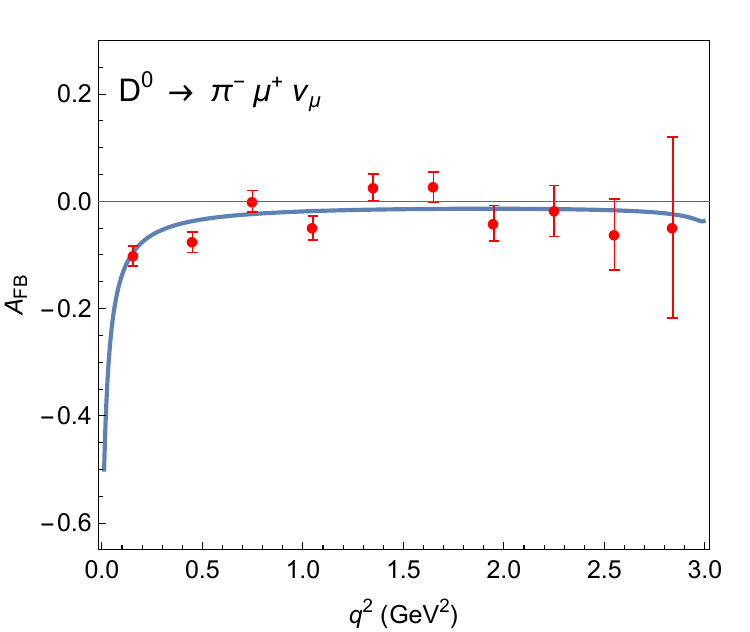}

        (c)
    \end{minipage}
    \hfill
    \begin{minipage}{0.48\linewidth}
        \centering
        \includegraphics[width=\linewidth]{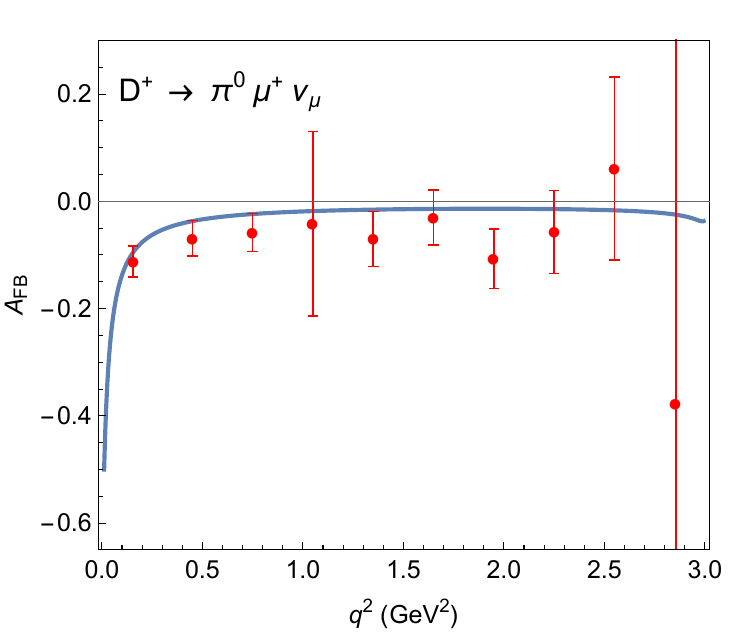}

        (d)
    \end{minipage}	   	
    	
	\begin{minipage}{0.48\linewidth}
        \centering
        \includegraphics[width=\linewidth]{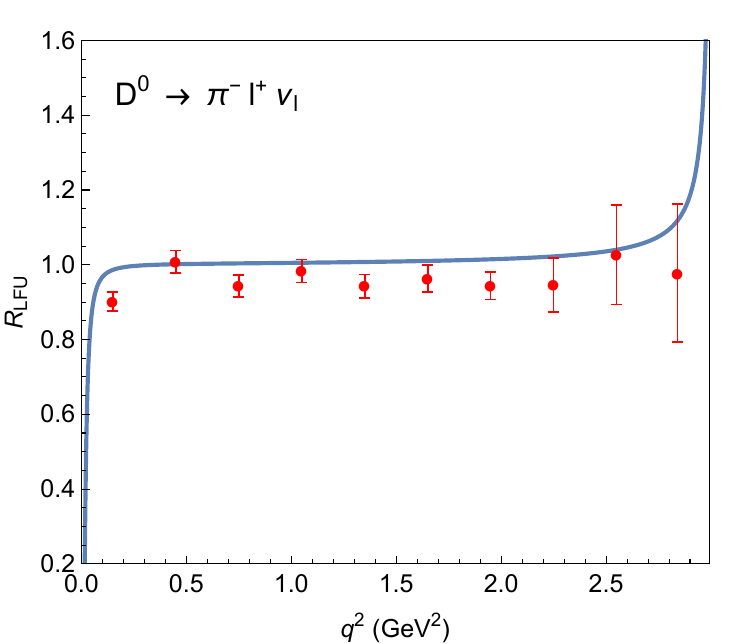}

        (e)
    \end{minipage}
    \hfill
    \begin{minipage}{0.48\linewidth}
        \centering
        \includegraphics[width=\linewidth]{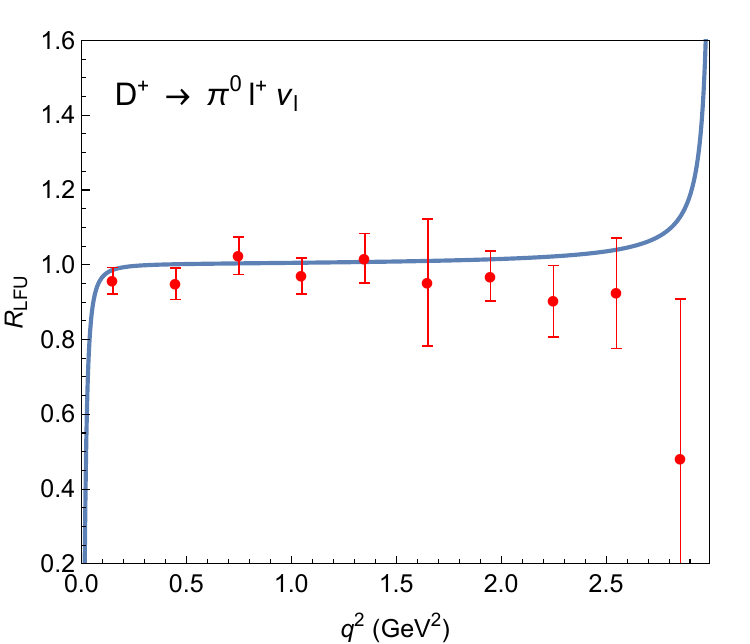}

        (f)
    \end{minipage}
    \caption{
        Forward-backward asymmetries and lepton-flavor-universality ratios for $D\to\pi$ transitions.
    }
    \label{fig:ObsDPi}
    
\end{figure}

In Figs.~\ref{fig:FFDPi}--\ref{fig:ObsDPi}, we compare the RQM results with the experimental form factors, differential decay rates, and the angular and lepton-flavor observables for $D\to\pi\ell\nu_\ell$. The form-factor points from BaBar and CLEO generally follow the predicted $q^2$ dependence. The more precise BESIII data are compatible with the calculation over most of the kinematic range, although deviations become more pronounced in the highest-$q^2$ bins. The differential decay rates follow the main trend of the data, and the measured forward--backward asymmetries and lepton-flavor ratios are consistent with the expected behavior within their uncertainties.

At $q^2=0$, the RQM value $f_+^{D\to\pi}(0)=0.640\pm0.032$ is compatible with the BESIII determination $0.6339\pm0.0028$ and with the other experimental results collected in Table~\ref{tab:FFDP}. Moreover, the four RQM branching fractions in Table~\ref{tab:BRDPi} are in good agreement with the corresponding BESIII values. 


\subsection{$D_{(s)}\to K$}
\label{subsec:comparison_K}

\begin{figure}[htbp]
    \centering

    \begin{minipage}{0.48\linewidth}
        \centering
        \includegraphics[width=\linewidth]{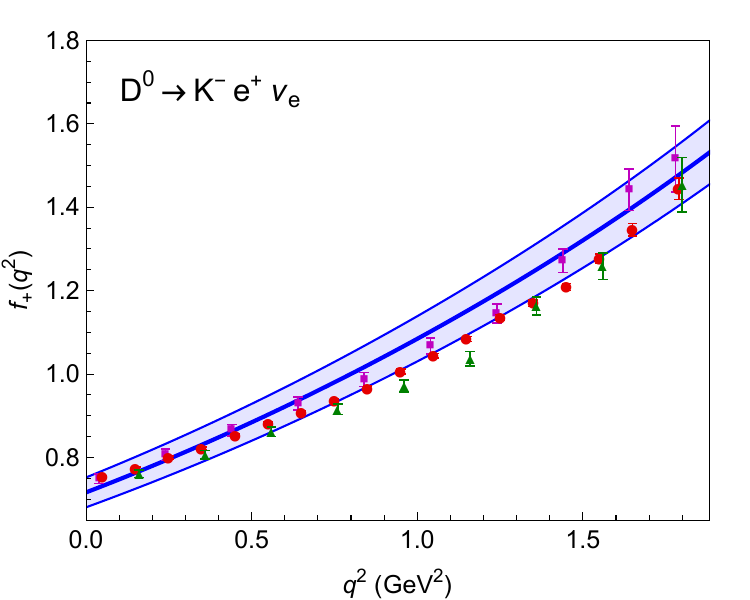}

        (a)
    \end{minipage}
    \hfill
    \begin{minipage}{0.48\linewidth}
        \centering
        \includegraphics[width=\linewidth]{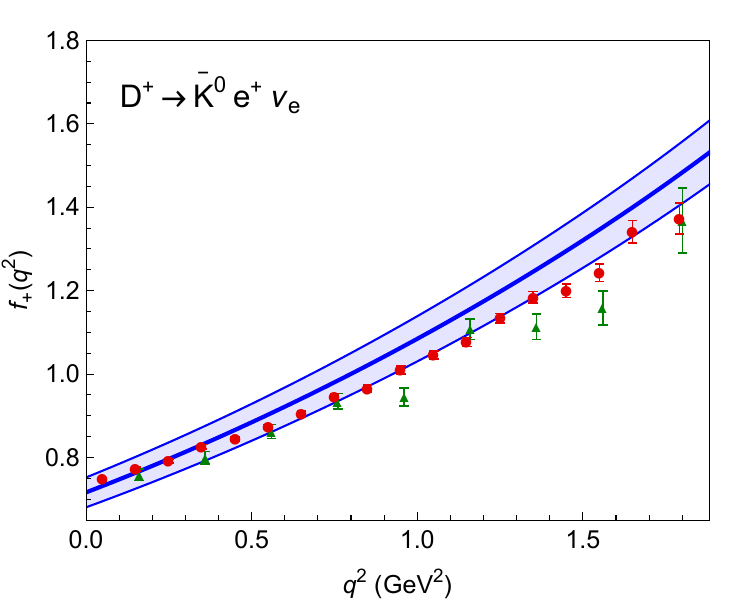}

        (b)
    \end{minipage}
    
	\begin{minipage}{0.48\linewidth}
        \centering
        \includegraphics[width=\linewidth]{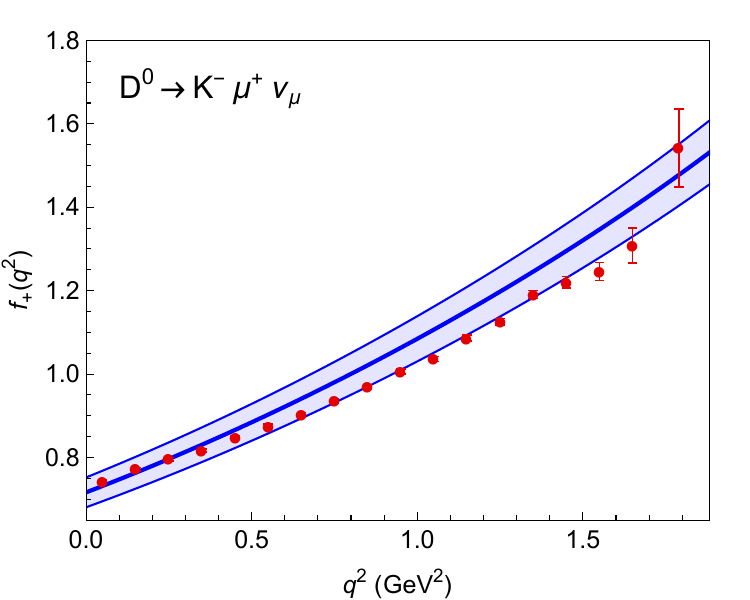}

        (c)
    \end{minipage}
    \hfill
    \begin{minipage}{0.48\linewidth}
        \centering
        \includegraphics[width=\linewidth]{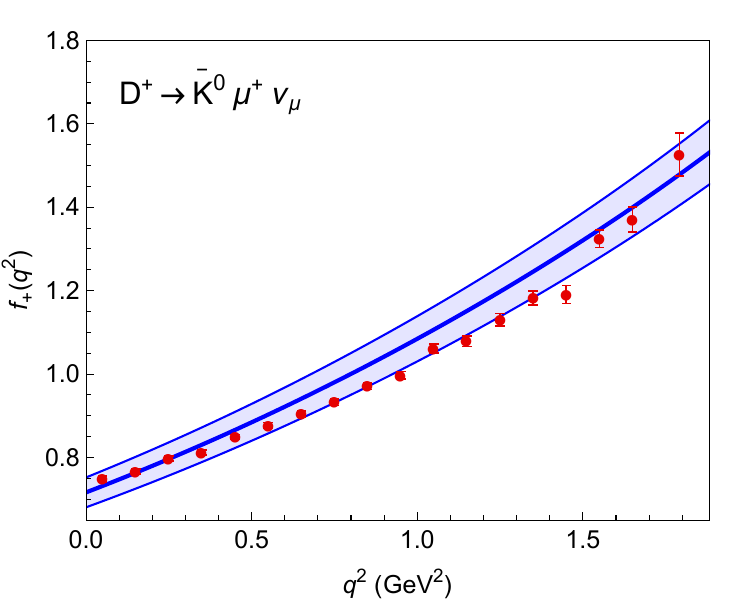}

        (d)
    \end{minipage}
    
	\begin{minipage}{0.48\linewidth}
        \centering
        \includegraphics[width=\linewidth]{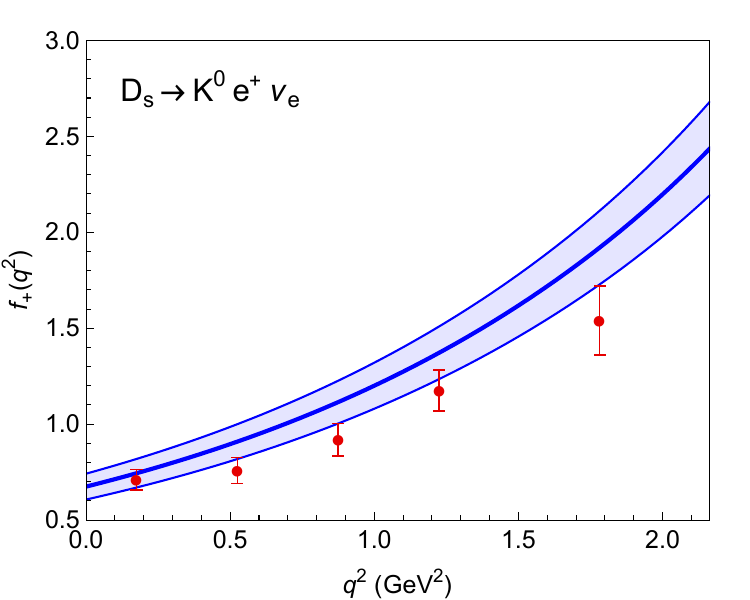}

        (e)
    \end{minipage}
    \hfill
    \begin{minipage}{0.48\linewidth}
        \centering
        \includegraphics[width=\linewidth]{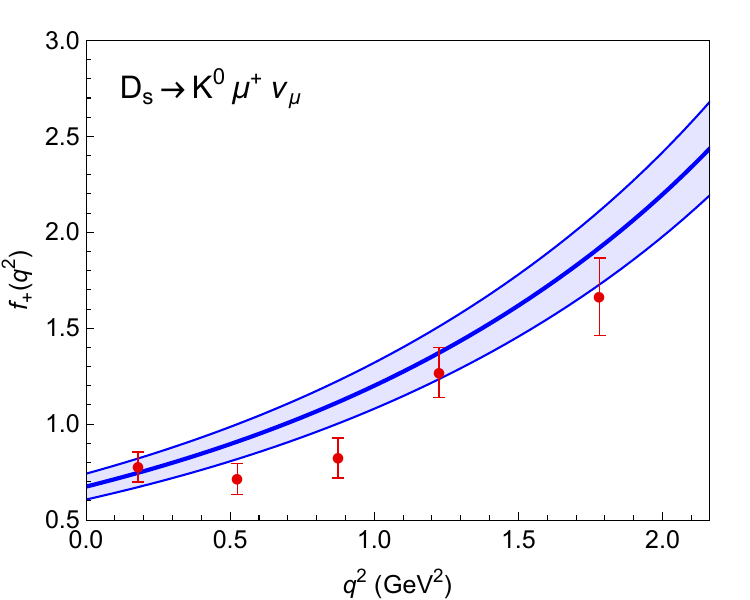}

        (f)
    \end{minipage}    
    
    \caption{
        Form factors of $D_{(s)}\to K$ transitions.
    }
    \label{fig:FFDK}
    
\end{figure}

\begin{figure}[htbp]
    \centering

    \begin{minipage}{0.48\linewidth}
        \centering
        \includegraphics[width=\linewidth]{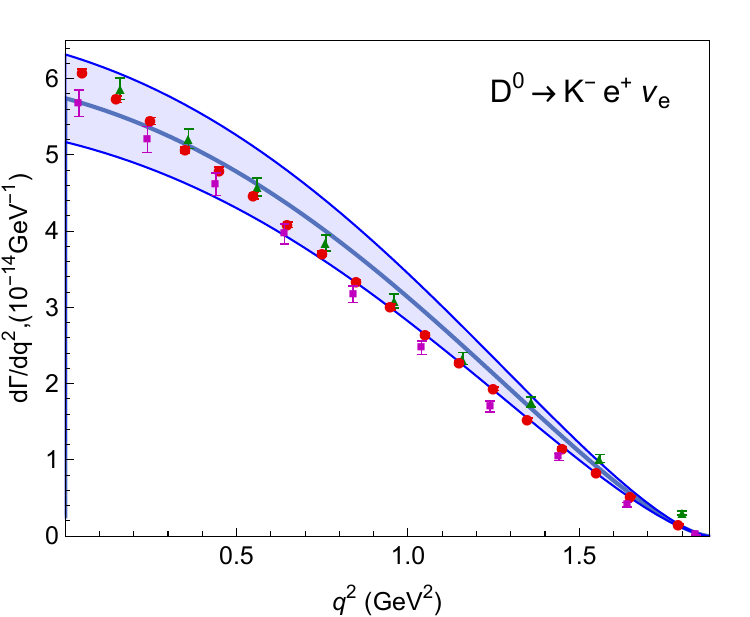}

        (a)
    \end{minipage}
    \hfill
    \begin{minipage}{0.48\linewidth}
        \centering
        \includegraphics[width=\linewidth]{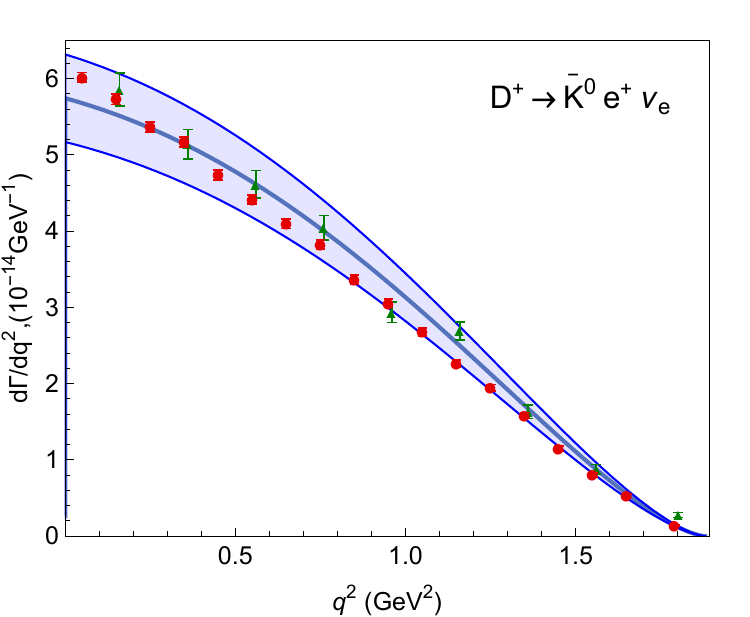}

        (b)
    \end{minipage}
    
	\begin{minipage}{0.48\linewidth}
        \centering
        \includegraphics[width=\linewidth]{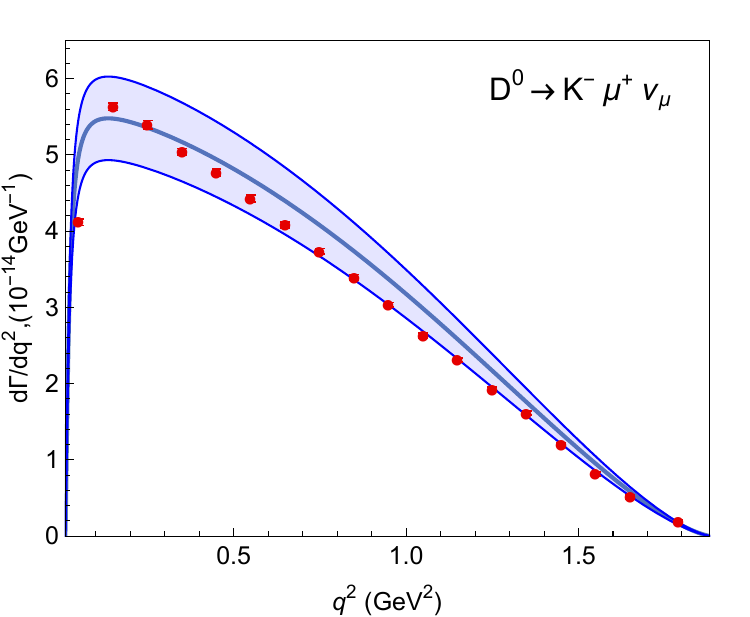}

        (c)
    \end{minipage}
    \hfill
    \begin{minipage}{0.48\linewidth}
        \centering
        \includegraphics[width=\linewidth]{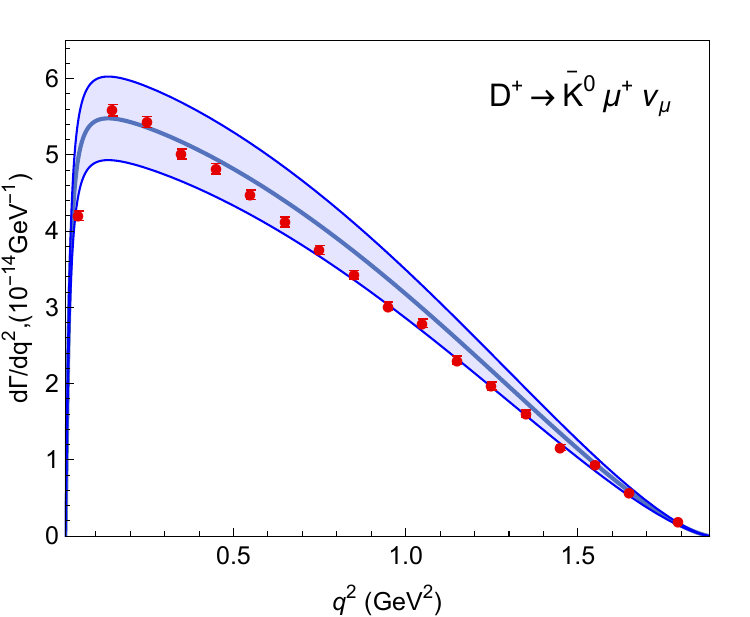}

        (d)
    \end{minipage}
    
    \begin{minipage}{0.48\linewidth}
        \centering
        \includegraphics[width=\linewidth]{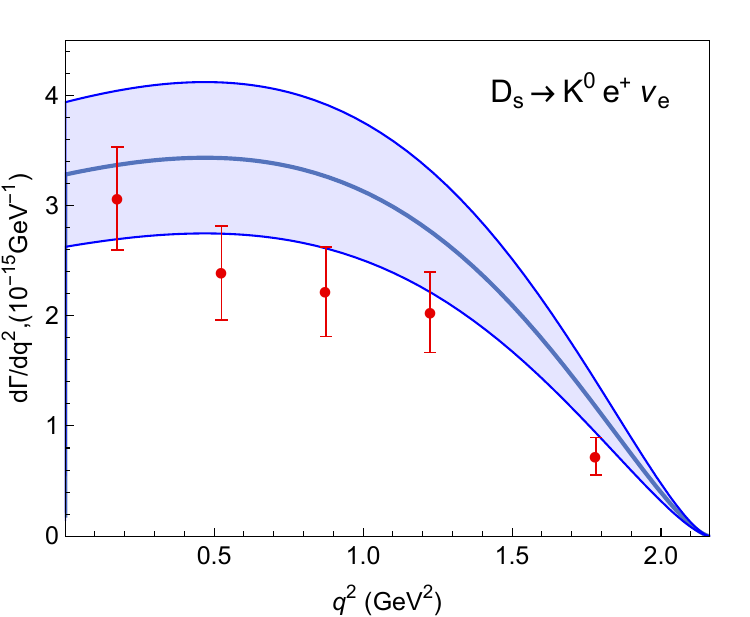}

        (e)
    \end{minipage}
    \hfill
    \begin{minipage}{0.48\linewidth}
        \centering
        \includegraphics[width=\linewidth]{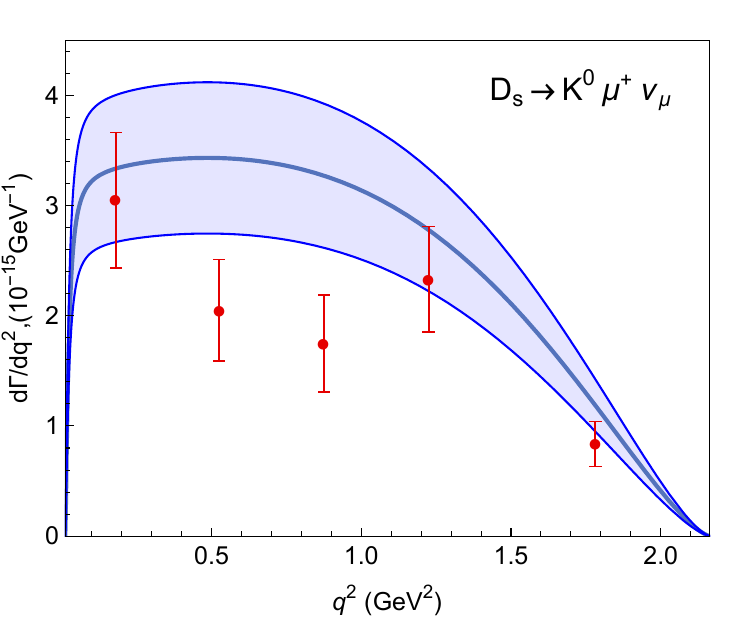}

        (f)
    \end{minipage}
    
    \caption{
        Differential decay rates of $D_{(s)}\to K$ transitions.
    }
    \label{fig:GammaDK}
    
\end{figure}

\begin{figure}[htbp]
    \centering

    \begin{minipage}{0.48\linewidth}
        \centering
        \includegraphics[width=\linewidth]{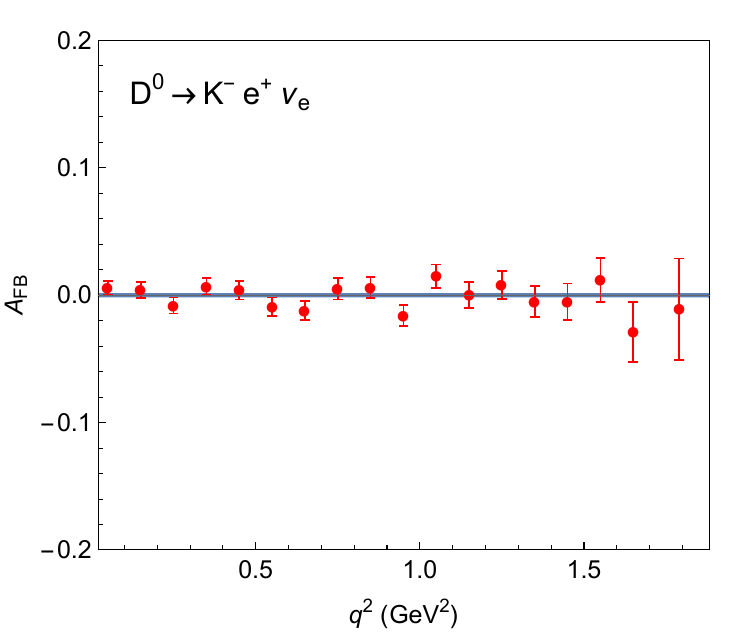}

        (a)
    \end{minipage}
    \hfill
    \begin{minipage}{0.48\linewidth}
        \centering
        \includegraphics[width=\linewidth]{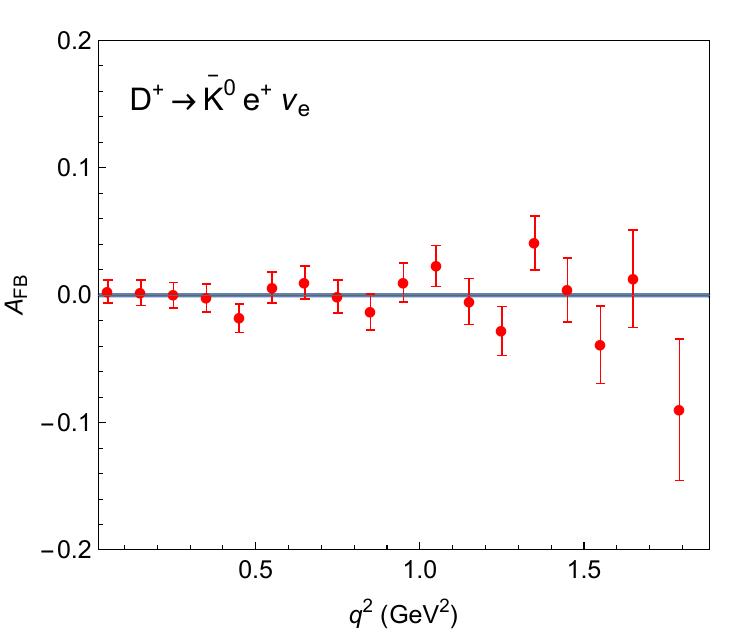}

        (b)
    \end{minipage}
 		
 	\begin{minipage}{0.48\linewidth}
        \centering
        \includegraphics[width=\linewidth]{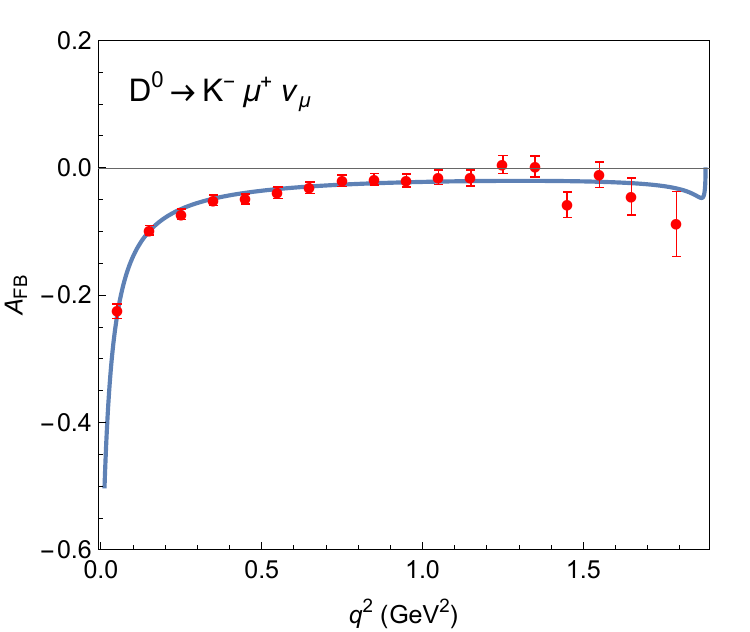}

        (c)
    \end{minipage}
    \hfill
    \begin{minipage}{0.48\linewidth}
        \centering
        \includegraphics[width=\linewidth]{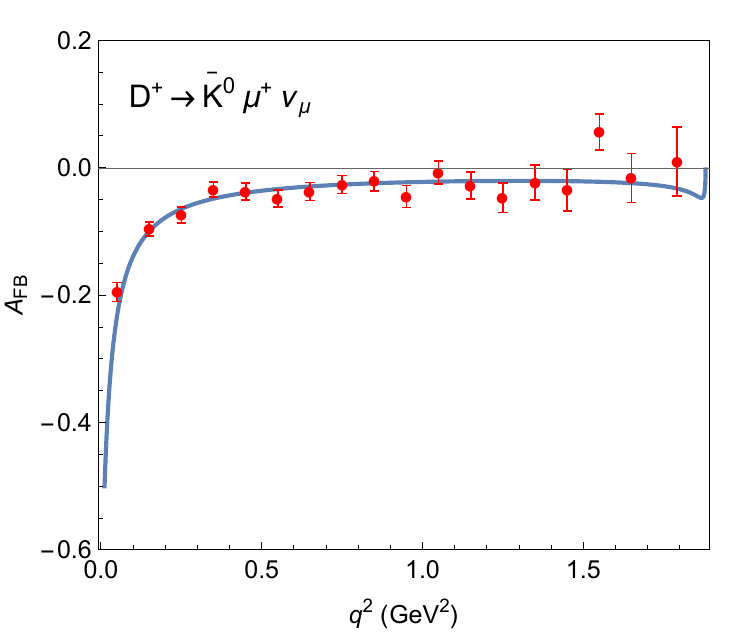}

        (d)
    \end{minipage}	   	
    	
	\begin{minipage}{0.48\linewidth}
        \centering
        \includegraphics[width=\linewidth]{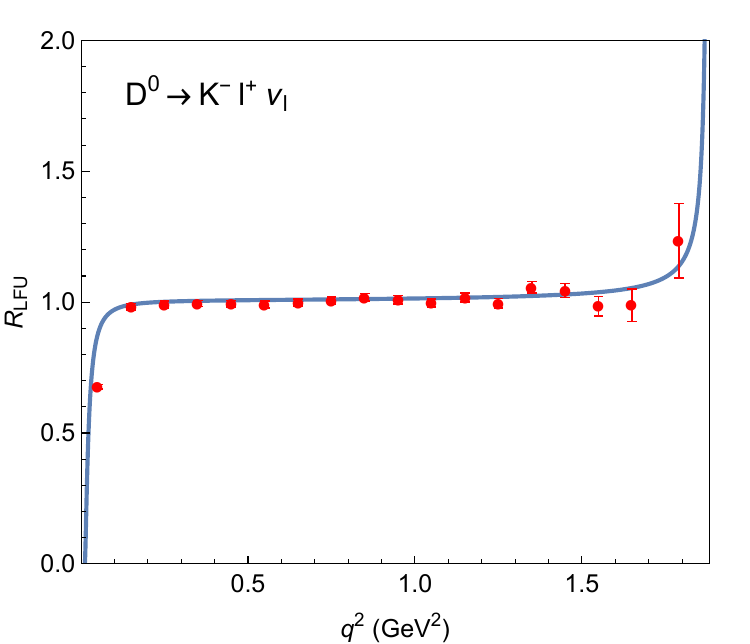}

        (e)
    \end{minipage}
    \hfill
    \begin{minipage}{0.48\linewidth}
        \centering
        \includegraphics[width=\linewidth]{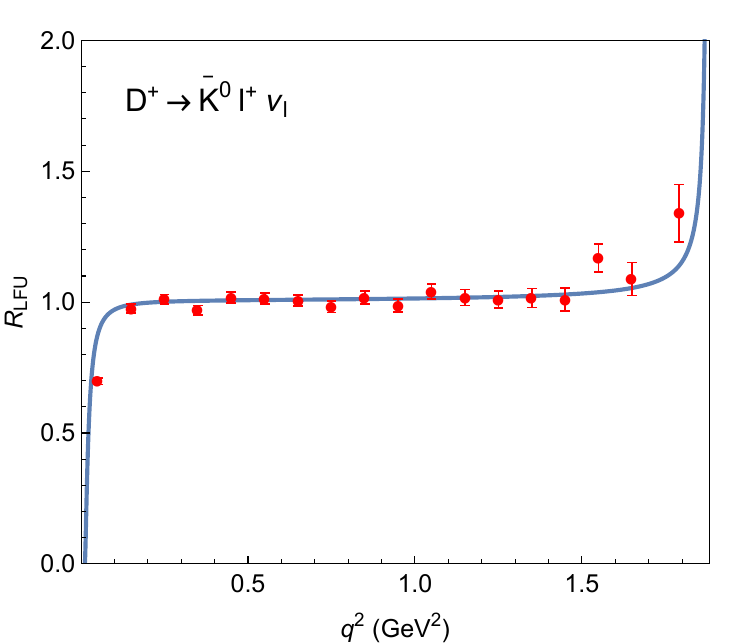}

        (f)
    \end{minipage}
    \caption{
        Forward-backward asymmetries and lepton-flavor-universality ratios for $D\to K$ transitions.
    }
    \label{fig:ObsDK}
    
\end{figure}

In Figs.~\ref{fig:FFDK}--\ref{fig:ObsDK}, we show the corresponding results for $D\to K$ and $D_s\to K$ transitions. For $D\to K$, the shape and normalization of $f_+(q^2)$ are in general compatible with the measurements by BESIII, BaBar, and CLEO, although small local deviations are visible. The RQM normalization $f_+^{D\to K}(0)=0.716\pm0.036$ agrees with the BESIII result $0.7383\pm0.0016$ within the combined uncertainty (Table~\ref{tab:FFDP}). The differential-rate distributions and the measured angular and lepton-flavor observables also agree with data. The integrated electron and muon branching fractions in Table~\ref{tab:BRDK} agree with BESIII within the uncertainties.

For $D_s\to K$, the situation is less uniform. The form-factor value $f_+(0)=0.674\pm0.034$ remains compatible with the BESIII result $0.623\pm0.037$ (Table~\ref{tab:FFDsP}), but the experimental points in the upper part of the $q^2$ range tend to lie below the RQM curves. The RQM branching fractions for the electron and muon channels, $(4.08\pm0.41)\times10^{-3}$ and $(4.00\pm0.40)\times10^{-3}$, exceed the BESIII measurements by slightly more than $2\sigma$ in both lepton channels (Table~\ref{tab:BRDsP}). 


\subsection{$D_{(s)}\to\eta^{(\prime)}$}
\label{subsec:comparison_eta}
The physical $\eta$ and $\eta'$ mesons are mixtures of the quark-flavor states,
$|\eta_q\rangle=(|u\bar u\rangle+|d\bar d\rangle)/\sqrt{2}$
and $|\eta_s\rangle=|s\bar s\rangle$.
A possible gluonic admixture can be included through
the extended mixing scheme~\cite{Escribano2007}
\begin{align}
|\eta\rangle
&=
\cos\phi_P|\eta_q\rangle
-\sin\phi_P|\eta_s\rangle,
\nonumber\\
|\eta'\rangle
&=
\cos\phi_G
\left(
\sin\phi_P|\eta_q\rangle
+\cos\phi_P|\eta_s\rangle
\right)
+\sin\phi_G|G\rangle,
\label{eq:eta_glue_mixing}
\end{align}
where $\phi_P$ is the quark-flavor mixing angle, $\phi_G$ describes the gluonic admixture, and $|G\rangle$ denotes a glueball state. We use the LHCb results $\phi_P=(41.6^{+1.0}_{-1.2})^\circ$ and $\phi_G=(28.1^{+3.9}_{-4.0})^\circ$ \cite{LHCb2025EtaMix}. The extracted gluonic angle is model dependent and may also reflect gluon-mediated contributions to the decays used in its determination.

The inclusion of the glueball admixture suppresses the quark-induced
$\eta'$ transition form factors by a factor $\cos\phi_G$,
leading to a suppression of the corresponding decay widths
and branching fractions by $\cos^2\phi_G$.
The $\eta$ transitions remain unchanged within this mixing scheme. All tabulated $\eta'$ form factors and branching fractions omit the additional gluonic suppression. The dashed curves illustrate the change in the central prediction for $\phi_G=28.1^\circ$; no separate uncertainty band for the gluonic scenario is shown. 
\begin{figure}[htbp]
    \centering

    \begin{minipage}{0.48\linewidth}
        \centering
        \includegraphics[width=\linewidth]{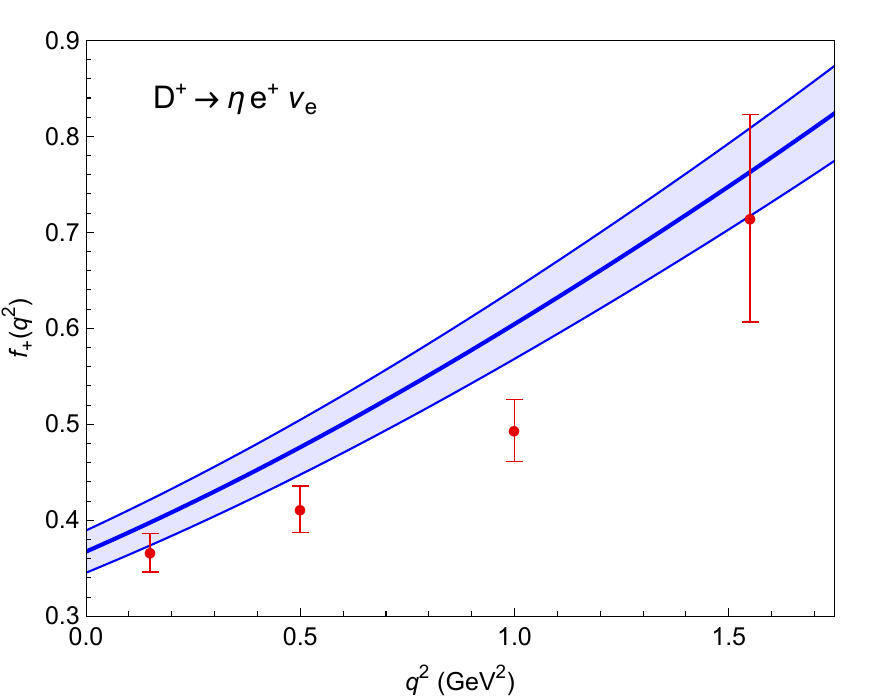}

        (a)
    \end{minipage}
    \hfill
    \begin{minipage}{0.48\linewidth}
        \centering
        \includegraphics[width=\linewidth]{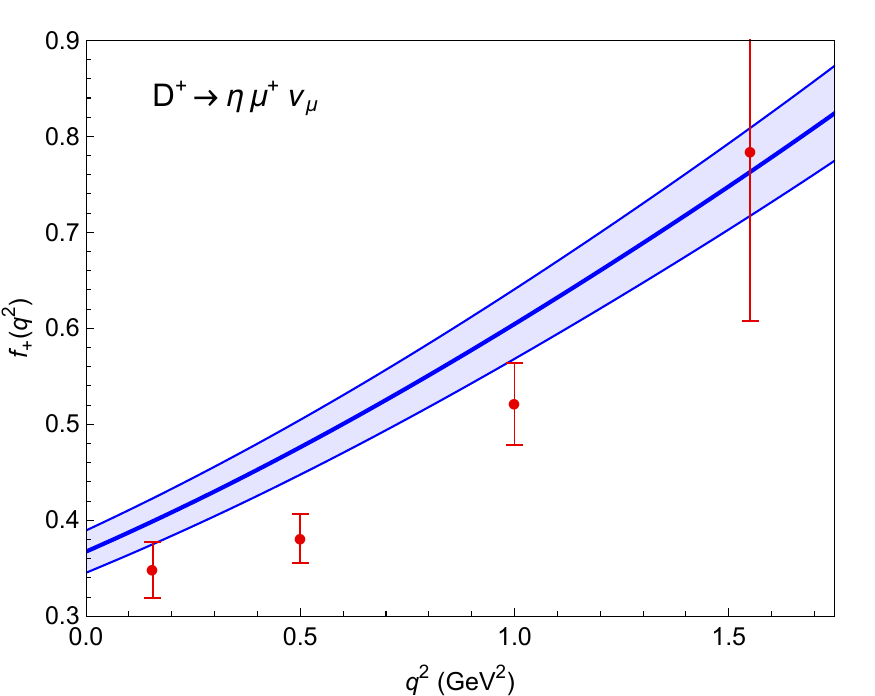}

        (b)
    \end{minipage}
    
	\begin{minipage}{0.48\linewidth}
        \centering
        \includegraphics[width=\linewidth]{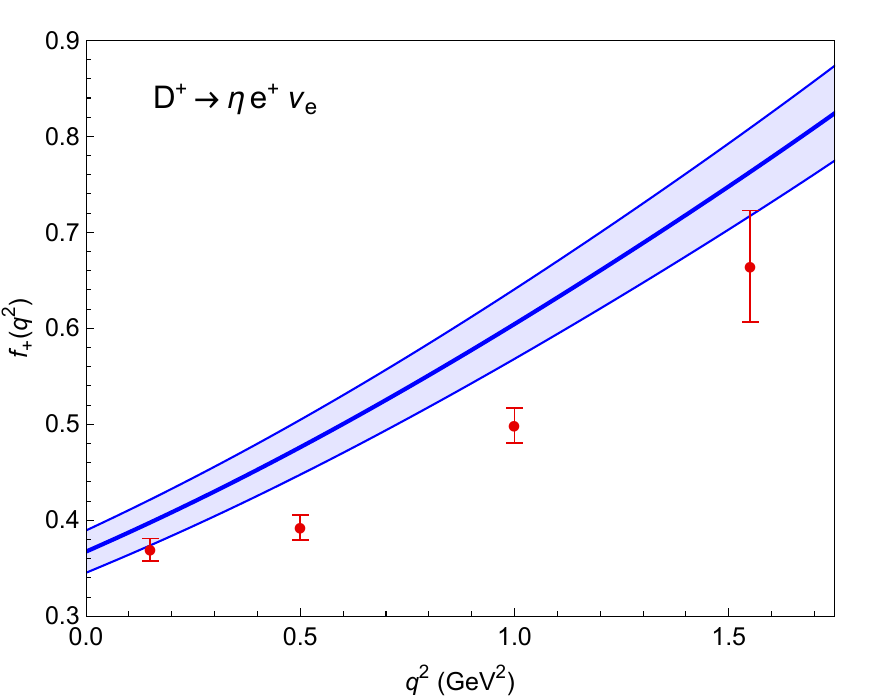}

        (c)
    \end{minipage}
    \hfill
    \begin{minipage}{0.48\linewidth}
        \centering
        \includegraphics[width=\linewidth]{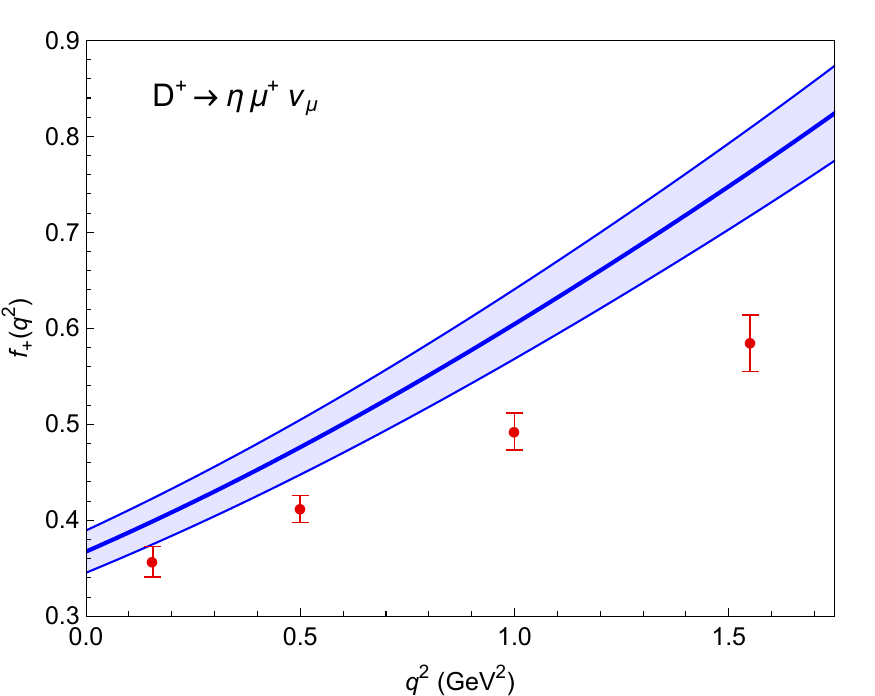}

        (d)
    \end{minipage}
    
    \begin{minipage}{0.48\linewidth}
        \centering
        \includegraphics[width=\linewidth]{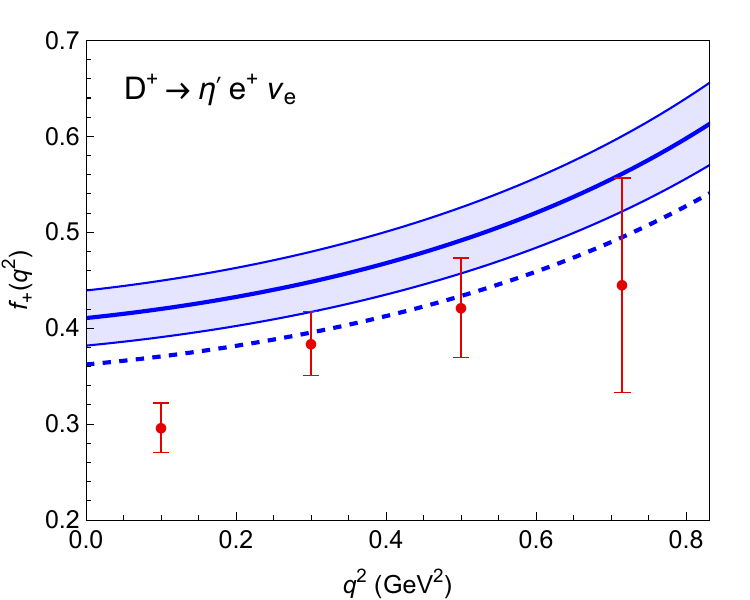}

        (e)
    \end{minipage}
    \hfill
    \begin{minipage}{0.48\linewidth}
        \centering
        \includegraphics[width=\linewidth]{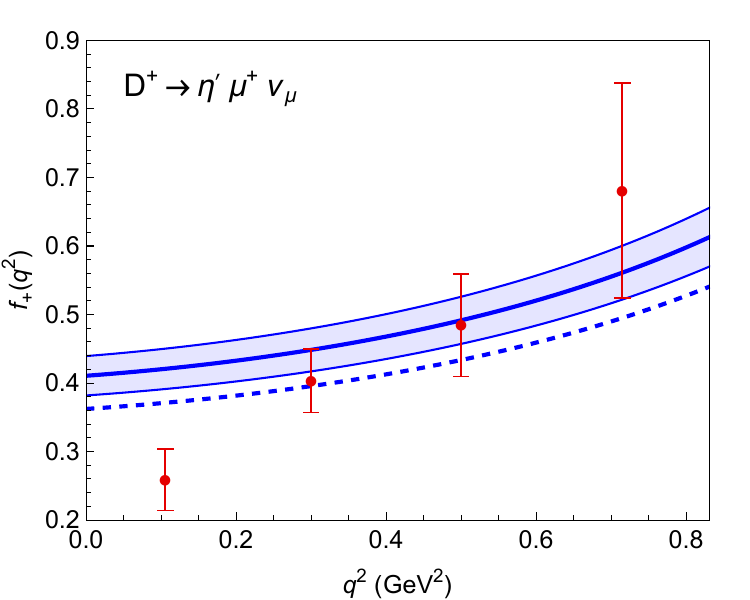}

        (f)
    \end{minipage}
    
    \caption{
        Form factors for $D\to\eta^{(\prime)}$ transitions.
        The experimental $\eta$ mesons are reconstructed through
        $\eta\to\pi^+\pi^-\pi^0$ in panels (a,b)
        and $\eta\to\gamma\gamma$ in panels (c,d).
        Panels (e,f) show combined experimental results from
        $\eta'\to\eta\pi^+\pi^-$ and $\eta'\to\rho^0\gamma$.
    }
    \label{fig:FFDeta}
    
\end{figure}

\begin{figure}[htbp]
    \centering

    \begin{minipage}{0.48\linewidth}
        \centering
        \includegraphics[width=\linewidth]{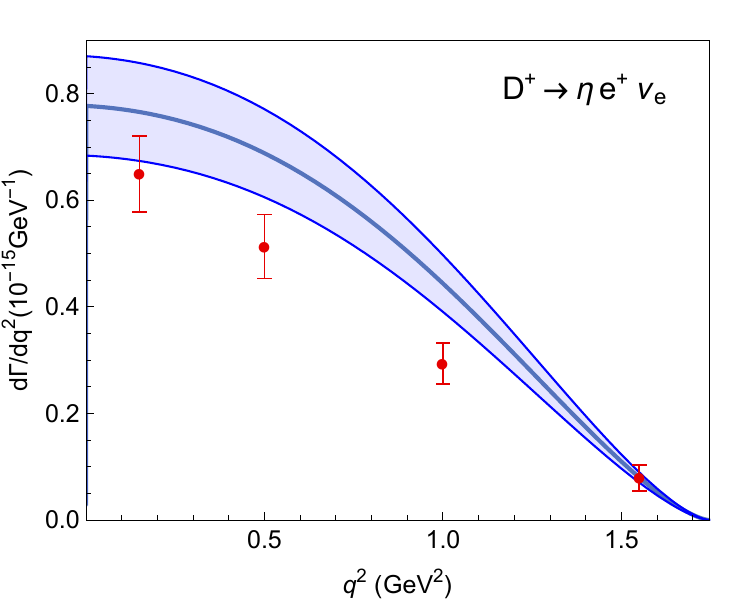}

        (a)
    \end{minipage}
    \hfill
    \begin{minipage}{0.48\linewidth}
        \centering
        \includegraphics[width=\linewidth]{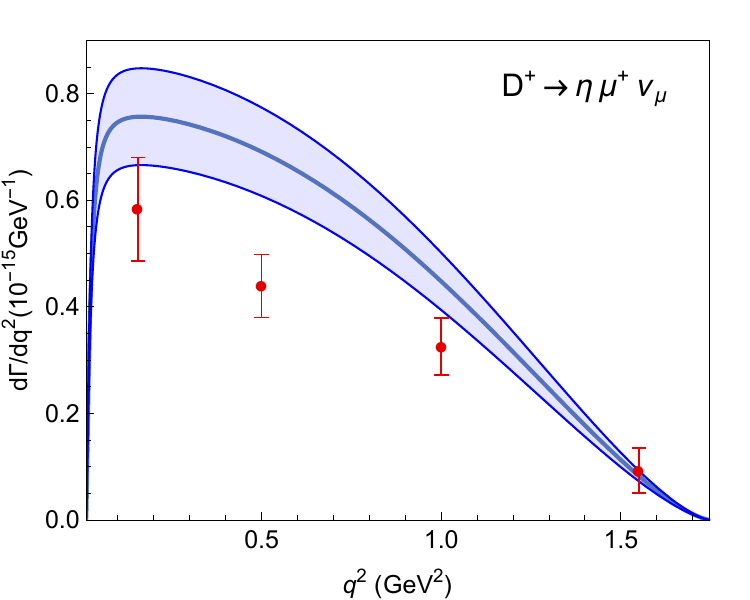}

        (b)
    \end{minipage}
    
	\begin{minipage}{0.48\linewidth}
        \centering
        \includegraphics[width=\linewidth]{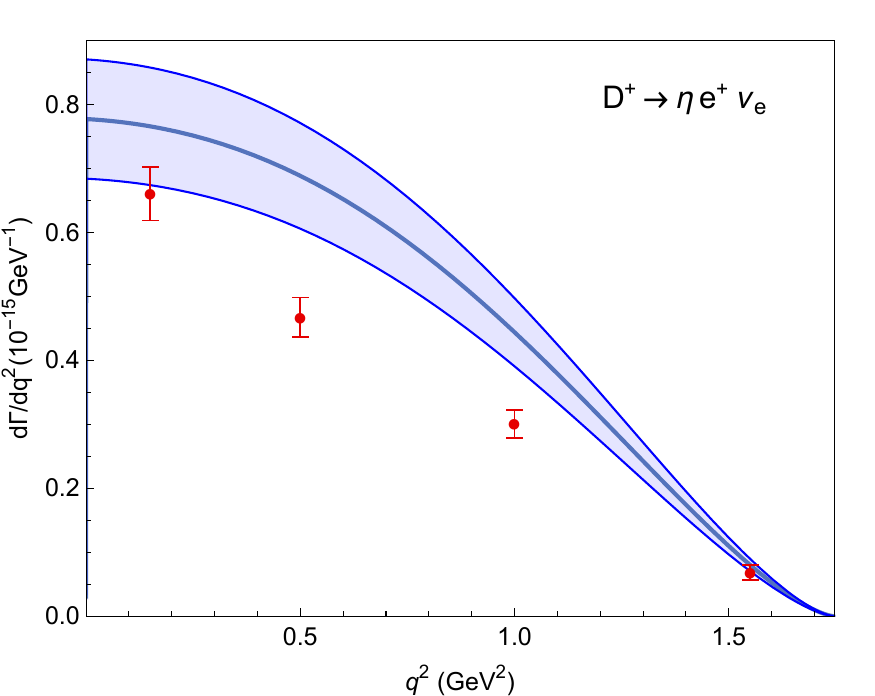}

        (c)
    \end{minipage}
    \hfill
    \begin{minipage}{0.48\linewidth}
        \centering
        \includegraphics[width=\linewidth]{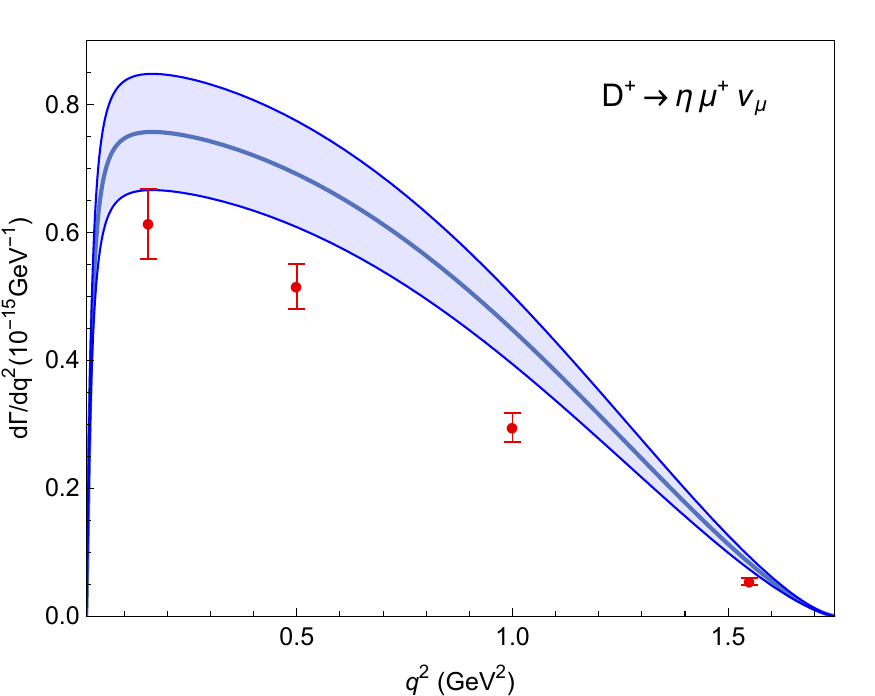}

        (d)
    \end{minipage}
    \begin{minipage}{0.48\linewidth}
        \centering
        \includegraphics[width=\linewidth]{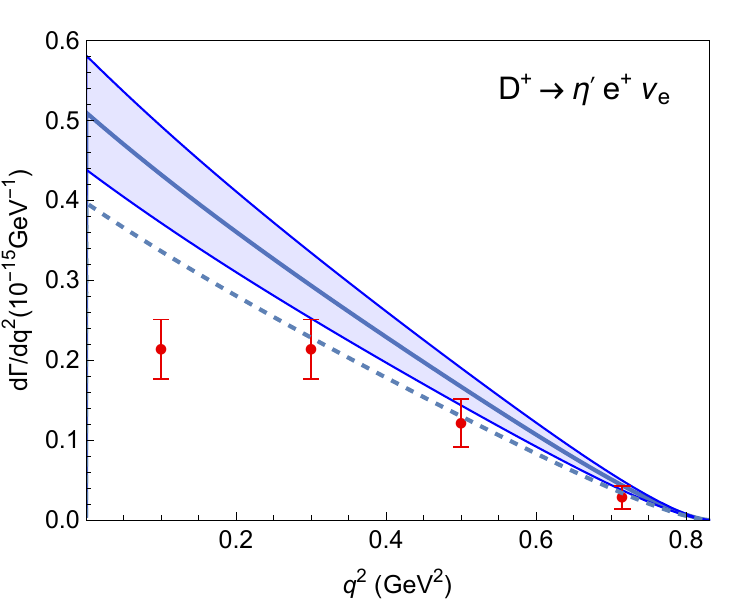}

        (e)
    \end{minipage}
    \hfill
    \begin{minipage}{0.48\linewidth}
        \centering
        \includegraphics[width=\linewidth]{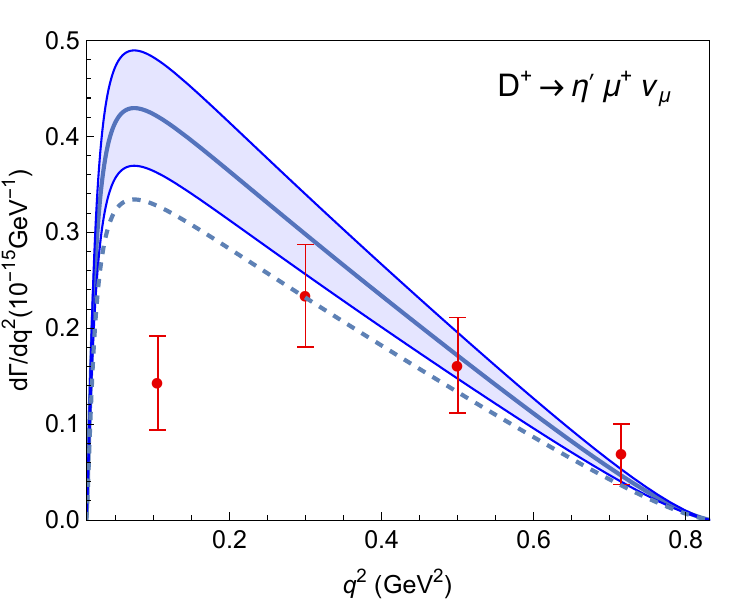}

        (f)
    \end{minipage}
    \caption{
        Differential decay rates for $D\to\eta^{(\prime)}$ transitions.
        The experimental $\eta$ mesons are reconstructed through
        $\eta\to\pi^+\pi^-\pi^0$ in panels (a,b)
        and $\eta\to\gamma\gamma$ in panels (c,d).
        Panels (e,f) show combined experimental results from
        $\eta'\to\eta\pi^+\pi^-$ and $\eta'\to\rho^0\gamma$.
    }
    \label{fig:GammaDeta}
    
\end{figure}

\begin{figure}[htbp]
    \centering

    \begin{minipage}{0.48\linewidth}
        \centering
        \includegraphics[width=\linewidth]{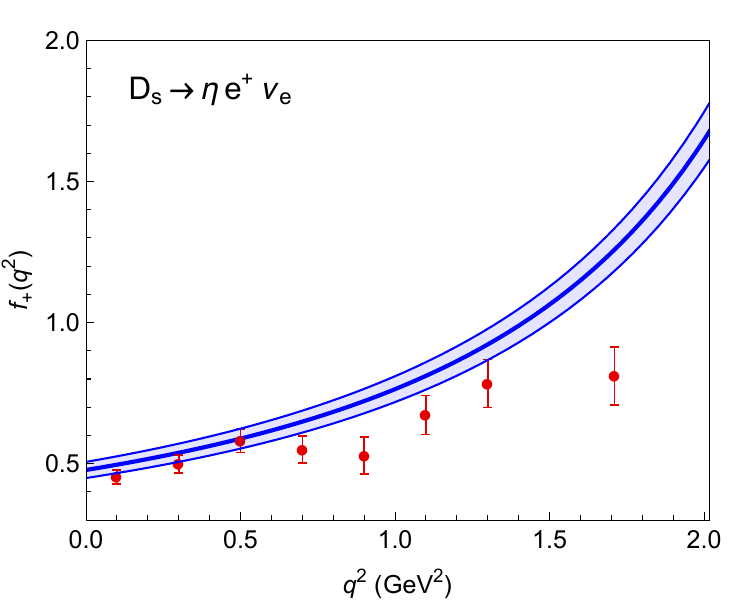}

        (a)
    \end{minipage}
    \hfill
    \begin{minipage}{0.48\linewidth}
        \centering
        \includegraphics[width=\linewidth]{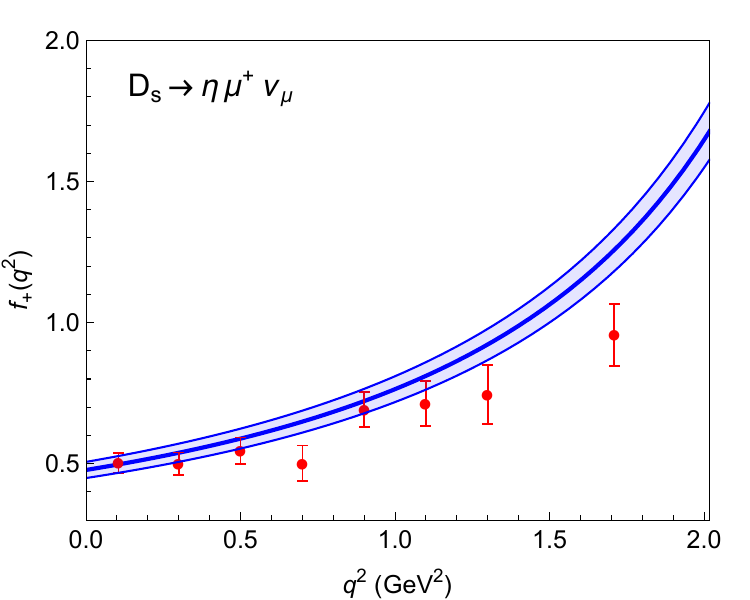}

        (b)
    \end{minipage}
    
	\begin{minipage}{0.48\linewidth}
        \centering
        \includegraphics[width=\linewidth]{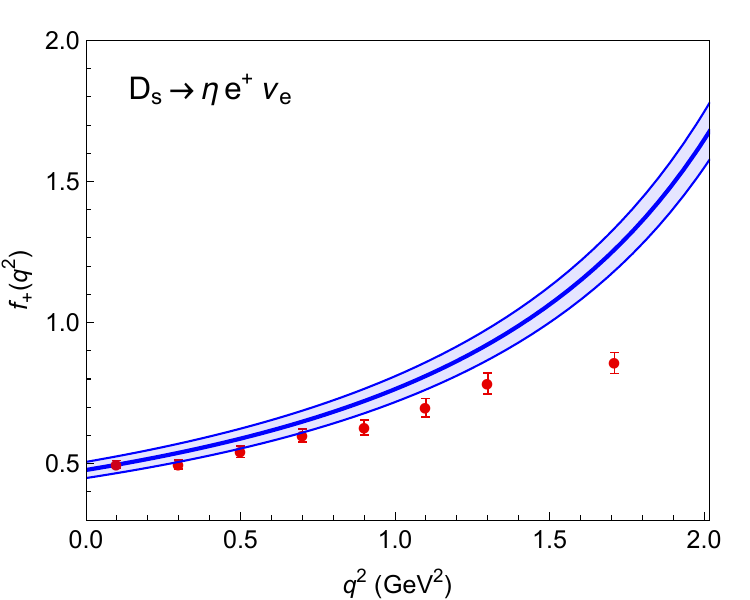}

        (c)
    \end{minipage}
    \hfill
    \begin{minipage}{0.48\linewidth}
        \centering
        \includegraphics[width=\linewidth]{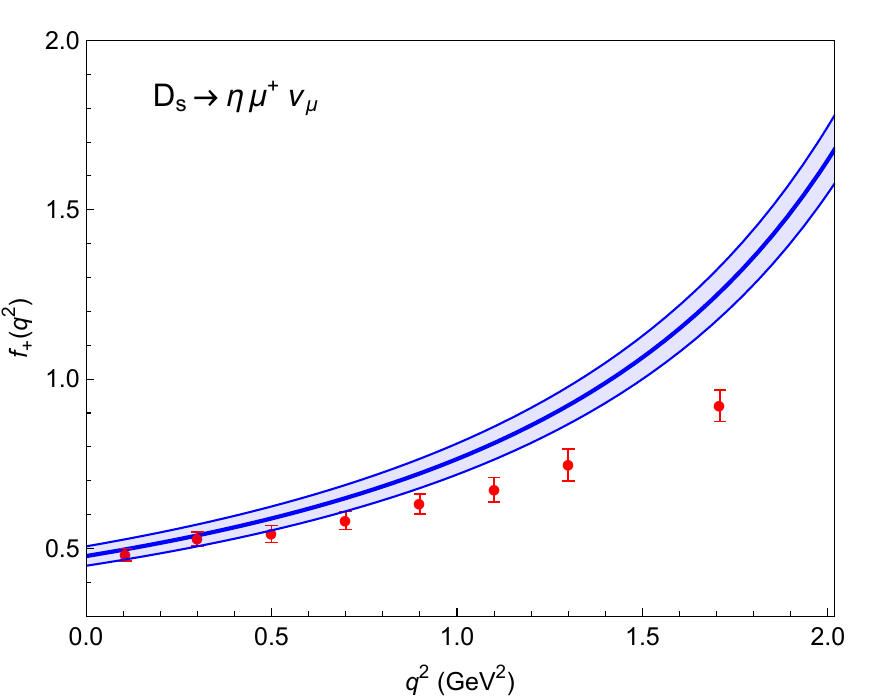}

        (d)
    \end{minipage}

    \caption{
        Form factors for $D_s\to\eta$ transitions.
        The experimental $\eta$ mesons are reconstructed through
        $\eta\to\pi^+\pi^-\pi^0$ in panels (a,b)
        and $\eta\to\gamma\gamma$ in panels (c,d).
    }
    \label{fig:FFDseta}
    
\end{figure}

\begin{figure}[htbp]
    \centering

    \begin{minipage}{0.48\linewidth}
        \centering
        \includegraphics[width=\linewidth]{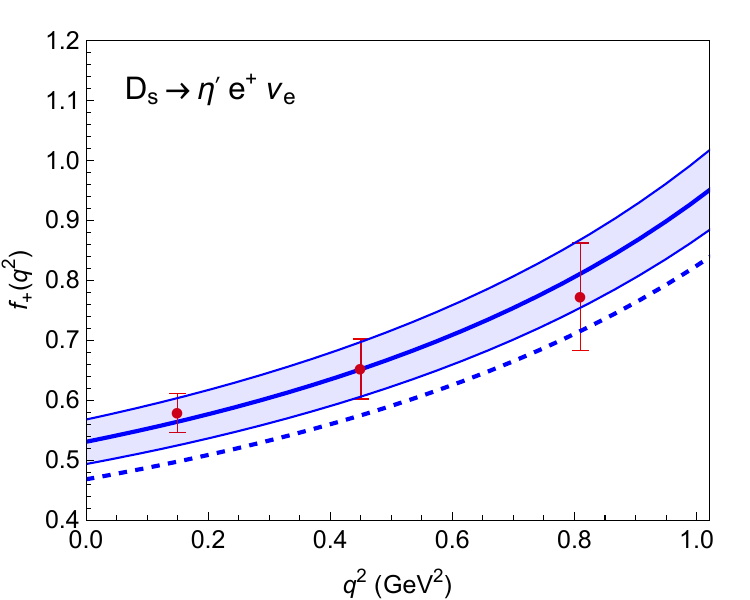}

        (a)
    \end{minipage}
    \hfill
    \begin{minipage}{0.48\linewidth}
        \centering
        \includegraphics[width=\linewidth]{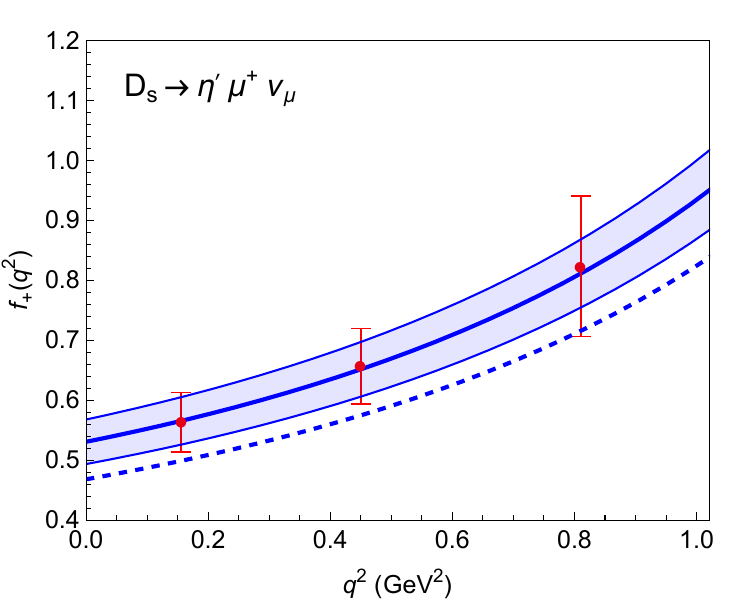}

        (b)
    \end{minipage}
    
	\begin{minipage}{0.48\linewidth}
        \centering
        \includegraphics[width=\linewidth]{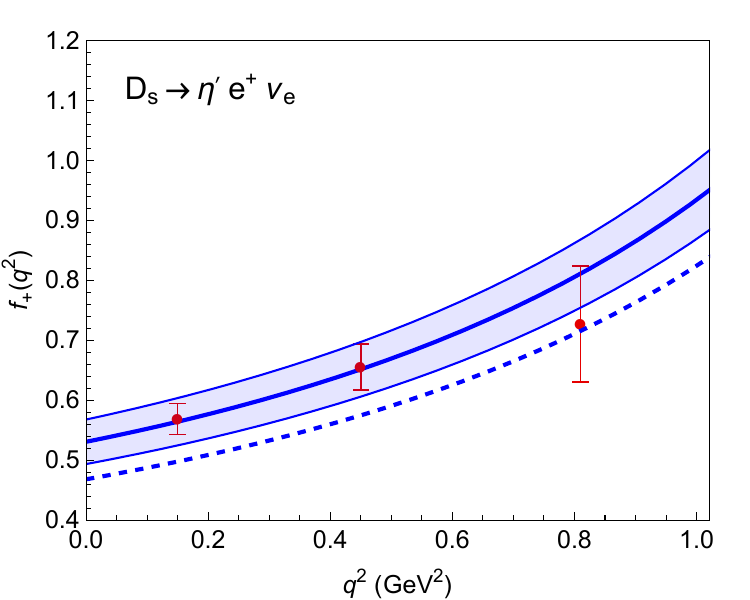}

        (c)
    \end{minipage}
    \hfill
    \begin{minipage}{0.48\linewidth}
        \centering
        \includegraphics[width=\linewidth]{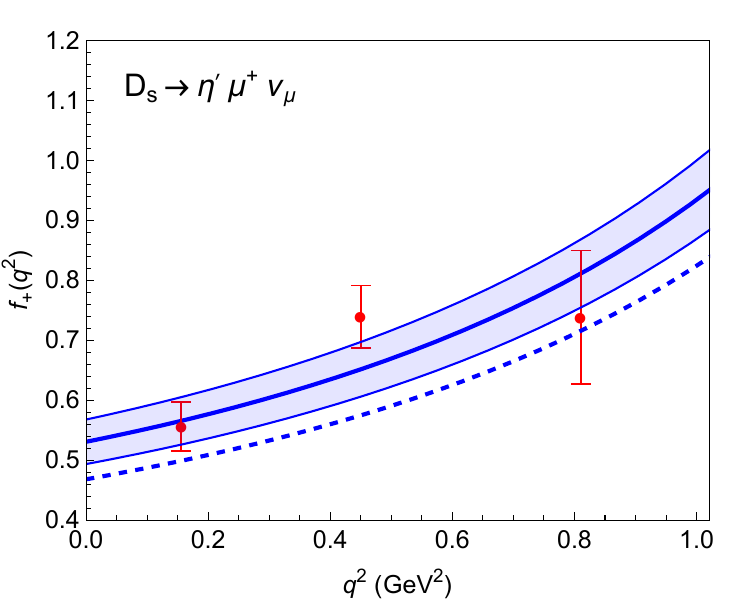}

        (d)
    \end{minipage}

    \caption{
        Form factors for $D_s\to\eta'$ transitions.
        The experimental $\eta'$ mesons are reconstructed through
        $\eta'\to\eta\pi^+\pi^-$ in panels (a,b)
        and $\eta'\to\rho^0\gamma$ in panels (c,d).
    }
    \label{fig:FFDsetaprime}
    
\end{figure}

\begin{figure}[htbp]
    \centering

    \begin{minipage}{0.48\linewidth}
        \centering
        \includegraphics[width=\linewidth]{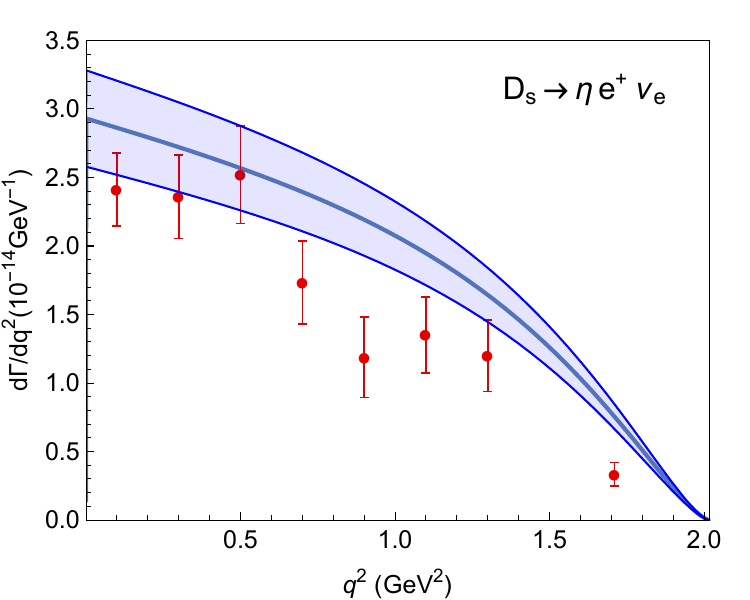}

        (a)
    \end{minipage}
    \hfill
    \begin{minipage}{0.48\linewidth}
        \centering
        \includegraphics[width=\linewidth]{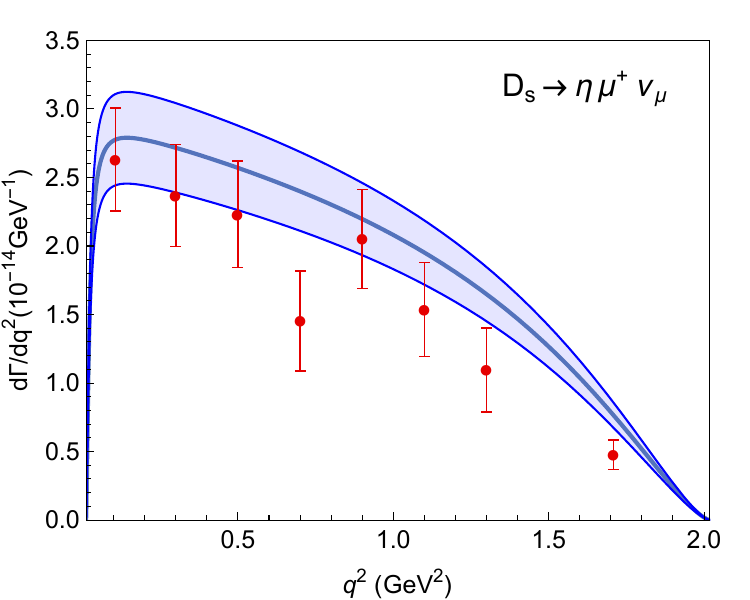}

        (b)
    \end{minipage}
    
	\begin{minipage}{0.48\linewidth}
        \centering
        \includegraphics[width=\linewidth]{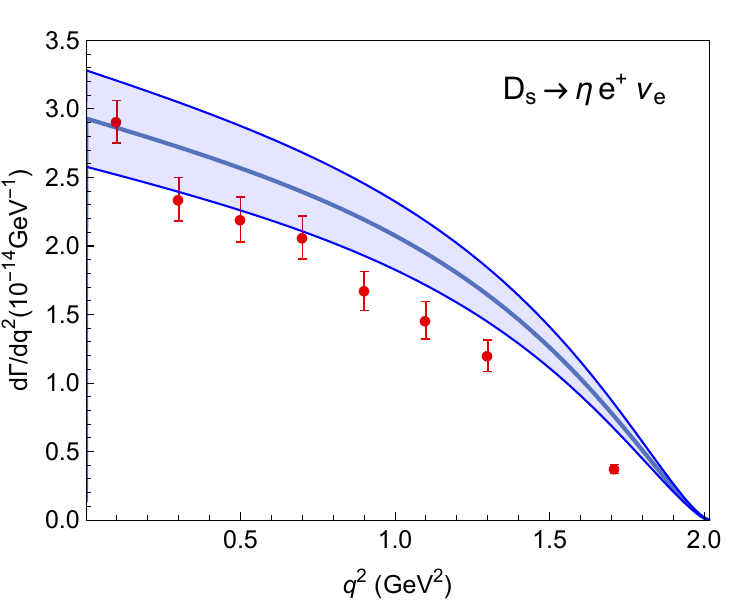}

        (c)
    \end{minipage}
    \hfill
    \begin{minipage}{0.48\linewidth}
        \centering
        \includegraphics[width=\linewidth]{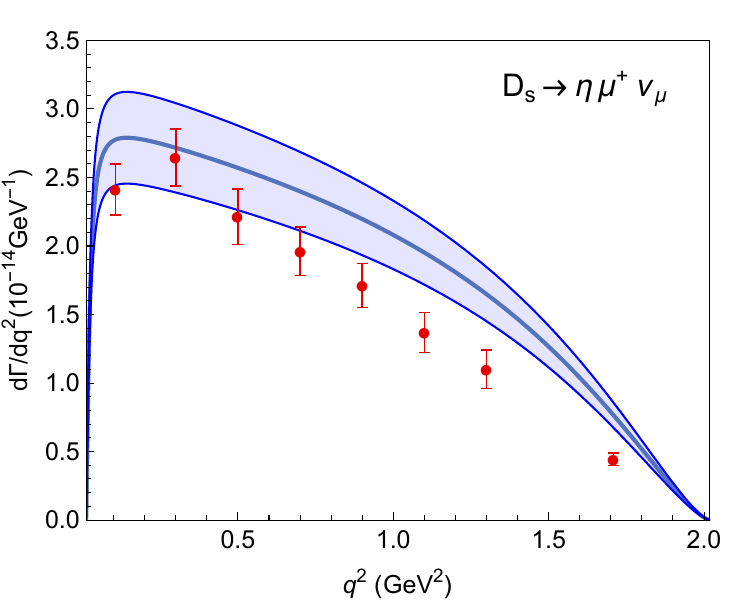}

        (d)
    \end{minipage}
    
    \caption{
        Differential decay rates for $D_s\to\eta$ transitions.
        The experimental $\eta$ mesons are reconstructed through
        $\eta\to\pi^+\pi^-\pi^0$ in panels (a,b)
        and $\eta\to\gamma\gamma$ in panels (c,d).
    }
    \label{fig:GammaDseta}
    
\end{figure}

\begin{figure}[htbp]
    \centering

    \begin{minipage}{0.48\linewidth}
        \centering
        \includegraphics[width=\linewidth]{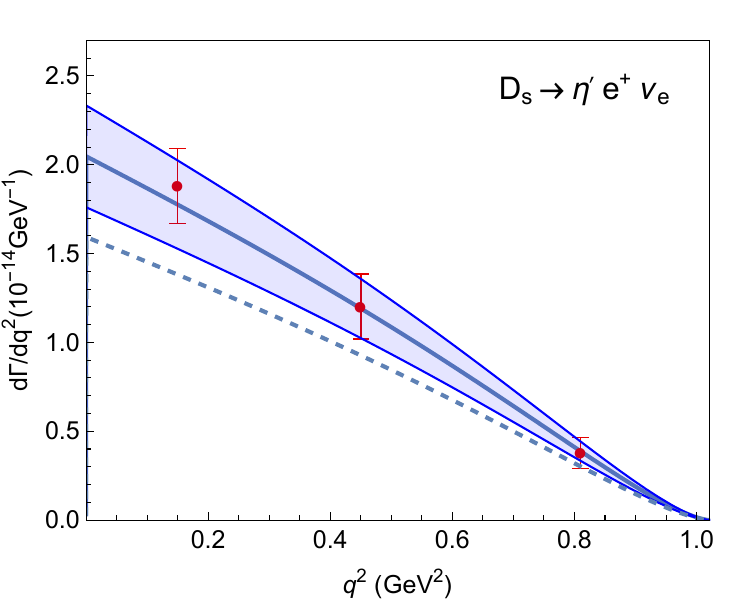}

        (a)
    \end{minipage}
    \hfill
    \begin{minipage}{0.48\linewidth}
        \centering
        \includegraphics[width=\linewidth]{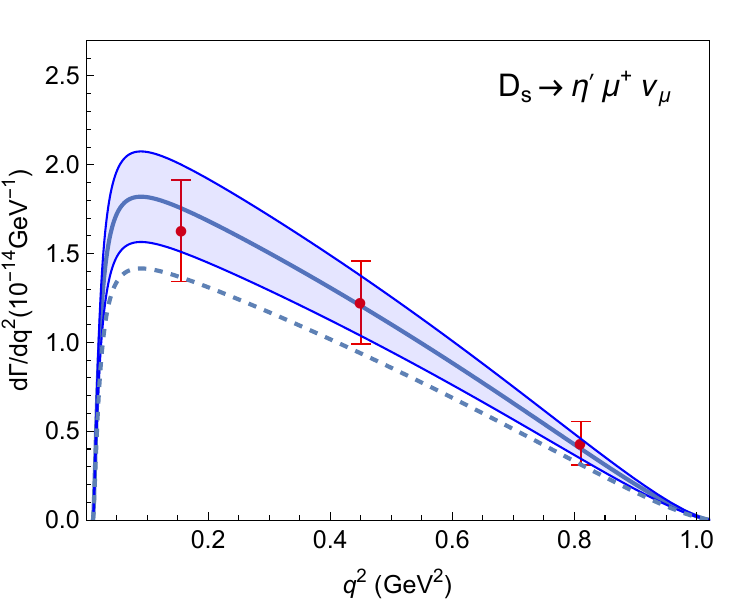}

        (b)
    \end{minipage}
    
	\begin{minipage}{0.48\linewidth}
        \centering
        \includegraphics[width=\linewidth]{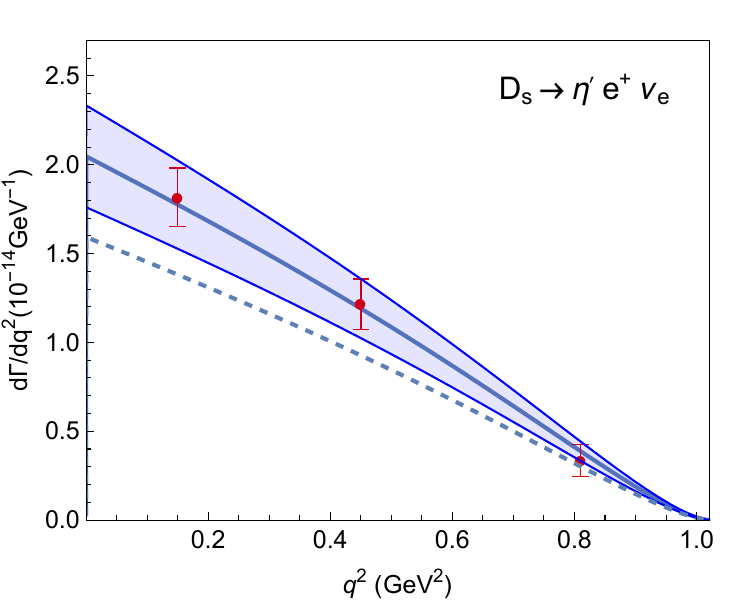}

        (c)
    \end{minipage}
    \hfill
    \begin{minipage}{0.48\linewidth}
        \centering
        \includegraphics[width=\linewidth]{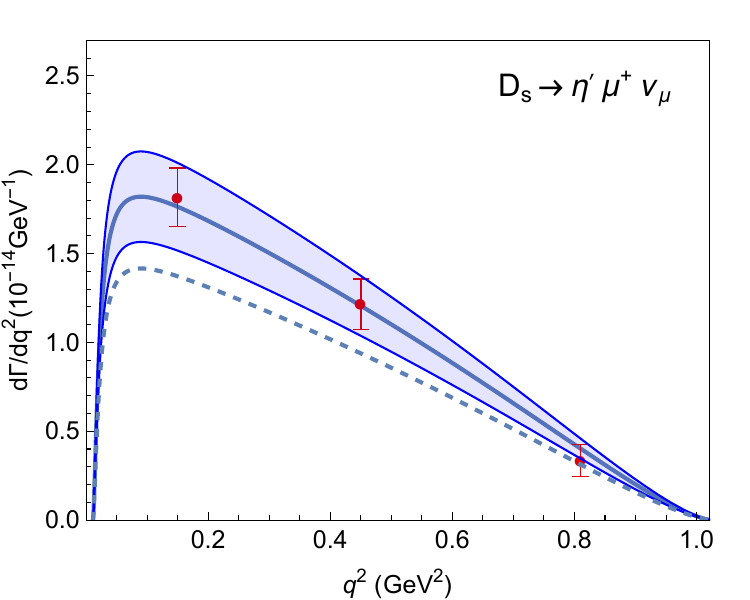}

        (d)
    \end{minipage}
    
    \caption{
        Differential decay rates for $D_s\to\eta'$ transitions.
        The experimental $\eta'$ mesons are reconstructed through
        $\eta'\to\eta\pi^+\pi^-$ in panels (a,b)
        and $\eta'\to\rho^0\gamma$ in panels (c,d).
    }
    \label{fig:GammaDsetaprime}
    
\end{figure}

In Figs.~\ref{fig:FFDeta}--\ref{fig:GammaDsetaprime}, we compare the RQM predictions with the measured $q^2$ distributions for $D_{(s)}\to\eta^{(\prime)}$ transitions. For $D\to\eta$, the form-factor $f_+(0)$ is compatible with BESIII and CLEO (Table~\ref{tab:FFDP}); however, several points, particularly at larger $q^2$, lie below the central theoretical curves. The integrated branching fractions agree with BESIII data: the deviations are close to $2\sigma$. In $D_s\to\eta$, the value at $q^2=0$ agrees with BESIII, but the data tend to fall below the theoretical form-factor curve as $q^2$ increases. The corresponding RQM branching fractions exceed the BESIII measurements, although the differences remain below the $2\sigma$ level (Table~\ref{tab:BRDsP}).

For $D\to\eta'$, the unmodified RQM form-factor $f_+^{D\to\eta'}(0)=0.411\pm0.029$ 
exceeds the BESIII value $0.264\pm0.026$ by more than $3\sigma$ (Table~\ref{tab:FFDP}). Using $\phi_G=28.1^\circ$ and assuming no direct weak-current coupling to the gluonic component, the quark-induced form factor is multiplied by $\cos\phi_G$ and the partial width by $\cos^2\phi_G$. The resulting form factor, $0.363\pm0.026$, reduces the discrepancy, although the remaining difference still exceeds the $2\sigma$ level and approaches $3\sigma$. Branching fractions without the gluonic admixture for $D^+\to\eta'e^+\nu_e$ and $D^+\to\eta'\mu^+\nu_\mu$ are $(2.97\pm0.42)\times10^{-4}$ and $(2.79\pm0.39)\times10^{-4}$ (Table~\ref{tab:BRDpEta}). The gluonic suppression changes these values to approximately $(2.31\pm0.33)\times10^{-4}$ and $(2.17\pm0.30)\times10^{-4}$, improving the agreement with BESIII. The dashed curves illustrate the corresponding reduction in the form factors and differential decay rates; their separation from the solid curves should not be interpreted as an additional uncertainty interval.

In contrast, for $D_s\to\eta'$ the unmodified RQM form-factor $0.531\pm0.037$ is close to the BESIII determination $0.542\pm0.027$ (Table~\ref{tab:FFDsP}). Including the same gluonic factor yields $0.468\pm0.033$, increasing the difference to $2\sigma$ level. The electron and muon branching fractions, $(0.80\pm0.11)\times10^{-2}$ and $(0.77\pm0.11)\times10^{-2}$, are initially in good agreement with BESIII (Table~\ref{tab:BRDsP}). After the gluonic correction they become approximately $(0.623\pm0.086)\times10^{-2}$ and $(0.599\pm0.086)\times10^{-2}$, with differences approaching the $2\sigma$ level from below. Hence the additional gluonic suppression worsens the agreement for $D_s\to\eta'$ while improving it for $D\to\eta'$. 

\begin{table}
\caption{$D\to P$ form factors at $q^2=0$. The theoretical $D\to\eta^{(\prime)}$ values include the $1/\sqrt{2}$ light-quark flavor factor.}
\label{tab:FFDP}
\scriptsize
\begin{ruledtabular}
\begin{tabular}{ccccc}
Model &
$f_+^{D\to\pi}(0)$ &
$f_+^{D\to K}(0)$ &
$f_+^{D\to\eta}(0)$ &
$f_+^{D\to\eta'}(0)$ \\
\hline
RQM~\cite{Faustov2020} &
$0.640\pm0.032$ &
$0.716\pm0.036$ &
$0.367\pm0.022$ &
$0.411\pm0.029$ \\
LCSR/HQEFT~\cite{Wu2006} &
$0.635^{+0.060}_{-0.057}$ &
$0.661^{+0.067}_{-0.066}$ &
$0.393^{+0.040}_{-0.037}$ &
 \\
LCSR~\cite{Khodjamirian2009,Offen2013} &
$0.67^{+0.10}_{-0.07}$ &
$0.75^{+0.11}_{-0.08}$ &
$0.390\pm0.036$ &
$0.324\pm0.074$ \\
LQCD (ETMC)~\cite{Lubicz2017} &
$0.612(35)$ &
$0.765(31)$ &
 &
 \\
CCQM~\cite{Ivanov2019} &
$0.63$ &
$0.77$ &
$0.36$ &
$0.36$ \\
CLFQM~\cite{Yang2026} &
$0.679^{+0.003}_{-0.004}$ &
$0.789^{+0.002}_{-0.001}$ &
$0.395^{+0.013}_{-0.018}$ &
$0.322^{+0.011}_{-0.015}$ \\
\hline
BESIII~\cite{BESIII2026Pi,BESIII2026K,BESIII2025Eta,BESIII2025EtaPrime} &
$0.6339\pm0.0028$ &
$0.7383\pm0.0016$ &
$0.347\pm0.011$ &
$0.264\pm0.026$ \\
Belle~\cite{Belle2006} &
$0.624\pm0.036$ &
$0.695\pm0.023$ &
 &
 \\
BaBar~\cite{BaBar2015Pi,BaBar2007K} &
$0.610\pm0.021$ &
$0.727\pm0.011$ &
 &
 \\
CLEO~\cite{CLEO2009PiK,CLEO2011Eta} &
$0.666\pm0.020$ &
$0.739\pm0.009$ &
$0.381\pm0.028$ &
 \\
PDG 2026~\cite{PDG2026} &
$0.637\pm0.009$ &
$0.736\pm0.004$ &
$0.374\pm0.018$ &
$0.264\pm0.026$ \\
\end{tabular}
\end{ruledtabular}
\end{table}

\begin{table}
\caption{$D_s\to P$ form factors at $q^2=0$.}
\label{tab:FFDsP}
\scriptsize
\begin{ruledtabular}
\begin{tabular}{cccc}
Model &
$f_+^{D_s\to K}(0)$ &
$f_+^{D_s\to\eta}(0)$ &
$f_+^{D_s\to\eta'}(0)$ \\
\hline
RQM~\cite{Faustov2020} &
$0.674\pm0.034$ &
$0.478\pm0.029$ &
$0.531\pm0.037$ \\
LCSR/HQEFT~\cite{Wu2006} &
$0.820^{+0.080}_{-0.071}$ &
$0.611^{+0.062}_{-0.054}$ &
 \\
LCSR~\cite{Tian2024,Duplancic2015,Hu2022} &
$0.692^{+0.027}_{-0.026}$ &
$0.495^{+0.030}_{-0.029}$ &
$0.544^{+0.046}_{-0.042}$ \\
LQCD~\cite{Bali2015} &
 &
$0.564(11)$ &
$0.437(18)$ \\
CCQM~\cite{Ivanov2019} &
$0.60$ &
$0.49$ &
$0.59$ \\
CLFQM~\cite{Yang2026} &
$0.714^{+0.004}_{-0.002}$ &
$0.490^{+0.014}_{-0.017}$ &
$0.599^{+0.017}_{-0.021}$ \\
\hline
BESIII~\cite{BESIII2025DsK,BESIII2023DsEta} &
$0.623\pm0.037$ &
$0.466\pm0.010$ &
$0.542\pm0.027$ \\
PDG 2026~\cite{PDG2026} &
$0.645\pm0.045$ &
$0.463\pm0.010$ &
$0.544\pm0.025$ \\
\end{tabular}
\end{ruledtabular}
\end{table}

\begin{table}
\caption{$D^0\to K^{*-}$ form factors at $q^2=0$.}
\label{tab:FFD0Kstar}
\scriptsize
\begin{ruledtabular}
\begin{tabular}{ccccccc}
Model& $V(0)$ & $A_0(0)$ & $A_1(0)$ & $A_2(0)$ & $r_V$ & $r_2$ \\
\hline
RQM~\cite{Faustov2020} &
$0.927\pm0.046$ & $0.655\pm0.033$ & $0.608\pm0.030$ & $0.520\pm0.026$ & $1.525\pm0.108$ & $0.856\pm0.061$ \\
CLFQM~\cite{Cheng2004} &
$0.94$ & $0.69$ & $0.65$ & $0.57$ & $1.446$ & $0.877$ \\
LCSR/HQEFT~\cite{Wu2006} &
$0.791^{+0.024}_{-0.026}$ & $0.694$ &
$0.571^{+0.020}_{-0.022}$ & $0.345^{+0.034}_{-0.037}$ & $1.385$ & $0.604$ \\
CLFQM~\cite{Verma2012} &
$0.98$ & $0.78$ & $0.72$ & $0.60$ & $1.361$ & $0.833$ \\
CCQM~\cite{Ivanov2019} &
$0.90$ & $0.77$ & $0.74$ & $0.68$ & $1.216$ & $0.919$ \\
CLFQM~\cite{Yang2026} &
$0.947^{+0.010}_{-0.011}$ & $0.635^{+0.002}_{-0.003}$ &
$0.656\pm0.004$ & $0.574^{+0.000}_{-0.001}$ & $1.444$ & $0.875$ \\
\hline
BESIII~\cite{BESIII2026Kstar} &
 &  & $0.618\pm0.0045$ &  & $1.444\pm0.028$ & $0.752\pm0.020$ \\
PDG 2026~\cite{PDG2026} &
 & & & & $1.45\pm0.04$ & $0.720\pm0.033$ \\
\end{tabular}
\end{ruledtabular}
\end{table}

\begin{table}
\caption{$D^0\to\rho^-$ form factors at $q^2=0$.}
\label{tab:FFD0Rho}
\scriptsize
\begin{ruledtabular}
\begin{tabular}{ccccccc}
Model& $V(0)$ & $A_0(0)$ & $A_1(0)$ & $A_2(0)$ & $r_V$ & $r_2$ \\
\hline
RQM~\cite{Faustov2020} &
$0.979\pm0.049$ & $0.712\pm0.036$ & $0.682\pm0.034$ & $0.640\pm0.032$ & $1.436\pm0.101$ & $0.939\pm0.066$ \\
CLFQM~\cite{Cheng2004} &
$0.86$ & $0.64$ & $0.58$ & $0.48$ & $1.483$ & $0.828$ \\
LCSR/HQEFT~\cite{Wu2006} &
$0.801^{+0.044}_{-0.036}$ & $0.759$ &
$0.599^{+0.035}_{-0.030}$ & $0.372^{+0.026}_{-0.031}$ & $1.337$ & $0.621$ \\
CLFQM~\cite{Verma2012} &
$0.88$ & $0.69$ & $0.60$ & $0.47$ & $1.467$ & $0.783$ \\
CCQM~\cite{Ivanov2019} &
$0.76$ & $0.64$ & $0.61$ & $0.57$ & $1.246$ & $0.934$ \\
LCSR~\cite{Zhong2024} &
$0.800^{+0.015}_{-0.014}$ & $0.525$ &
$0.498^{+0.014}_{-0.012}$ & $0.460^{+0.055}_{-0.047}$ & $1.607^{+0.071}_{-0.071}$ & $0.923^{+0.133}_{-0.119}$ \\
CLFQM~\cite{Yang2026} &
$0.845\pm0.009$ & $0.546\pm0.002$ &
$0.571\pm0.003$ & $0.485\pm0.001$ & $1.480$ & $0.849$ \\
\hline
CLEO~\cite{CLEO2013RhoOmega} &
$0.84^{+0.10}_{-0.11}$ &  &
$0.56^{+0.022}_{-0.032}$ & $0.47\pm0.072$ & $1.48\pm0.16$ & $0.83\pm0.12$ \\
PDG 2026~\cite{PDG2026} &
 & & & & $1.53\pm0.08$ & $0.82\pm0.05$ \\
\end{tabular}
\end{ruledtabular}
\end{table}

\begin{table}
\caption{$D^+\to\omega$ form factors at $q^2=0$.}
\label{tab:FFDpOmega}
\scriptsize
\begin{ruledtabular}
\begin{tabular}{ccccccc}
Model& $V(0)$ & $A_0(0)$ & $A_1(0)$ & $A_2(0)$ & $r_V$ & $r_2$ \\
\hline
RQM~\cite{Faustov2020} &
$0.871\pm0.043$ & $0.647\pm0.032$ & $0.674\pm0.034$ & $0.713\pm0.036$ & $1.292\pm0.091$ & $1.057\pm0.075$ \\
LCSR/HQEFT~\cite{Wu2006} &
$0.742^{+0.041}_{-0.034}$ & $0.711$ &
$0.556^{+0.033}_{-0.028}$ & $0.333^{+0.026}_{-0.030}$ & $1.335$ & $0.599$ \\
CLFQM~\cite{Verma2012} &
$0.85$ & $0.64$ & $0.58$ & $0.49$ & $1.466$ & $0.845$ \\
CCQM~\cite{Ivanov2019} &
$0.72$ & $0.60$ & $0.58$ & $0.55$ & $1.241$ & $0.948$ \\
CLFQM~\cite{Yang2026} &
$0.832\pm0.009$ & $0.539\pm0.002$ &
$0.559\pm0.003$ & $0.491^{+0.000}_{-0.001}$ & $1.488$ & $0.878$ \\
\hline
PDG 2026~\cite{PDG2026} &
 & & & & $1.24\pm0.11$ & $1.06\pm0.16$ \\
\end{tabular}
\end{ruledtabular}
\end{table}

\begin{table}
\caption{$D_s^+\to K^{*0}$ form factors at $q^2=0$.}
\label{tab:FFDsKstar}
\scriptsize
\begin{ruledtabular}
\begin{tabular}{ccccccc}
Model& $V(0)$ & $A_0(0)$ & $A_1(0)$ & $A_2(0)$ & $r_V$ & $r_2$ \\
\hline
RQM~\cite{Faustov2020} &
$0.959\pm0.048$ & $0.629\pm0.031$ & $0.596\pm0.030$ & $0.540\pm0.027$ & $1.609\pm0.113$ & $0.906\pm0.064$ \\
LCSR/HQEFT~\cite{Wu2006} &
$0.771\pm0.049$ & $0.754$ &
$0.589^{+0.040}_{-0.042}$ & $0.315^{+0.024}_{-0.018}$ & $1.309$ & $0.535$ \\
CLFQM~\cite{Verma2012} &
$0.87$ & $0.61$ & $0.56$ & $0.46$ & $1.554$ & $0.821$ \\
CCQM~\cite{Ivanov2019} &
$0.80$ & $0.58$ & $0.58$ & $0.57$ & $1.379$ & $0.983$ \\
CLFQM~\cite{Yang2026} &
$0.860\pm0.010$ & $0.531\pm0.002$ &
$0.549\pm0.003$ & $0.485^{+0.001}_{-0.002}$ & $1.566$ & $0.883$ \\
\hline
BESIII~\cite{BESIII2026DsKstar} &
 &  & $0.56\pm0.022$ &  & $1.63\pm0.16$ & $0.60\pm0.14$ \\
PDG 2026~\cite{PDG2026} &
 & & & & $1.67\pm0.38$ & $0.77\pm0.29$ \\
\end{tabular}
\end{ruledtabular}
\end{table}

\begin{table}
\caption{$D_s^+\to\phi$ form factors at $q^2=0$.}
\label{tab:FFDsPhi}
\scriptsize
\begin{ruledtabular}
\begin{tabular}{ccccccc}
Model& $V(0)$ & $A_0(0)$ & $A_1(0)$ & $A_2(0)$ & $r_V$ & $r_2$ \\
\hline
RQM~\cite{Faustov2020} &
$1.000\pm0.050$ & $0.713\pm0.036$ & $0.643\pm0.032$ & $0.492\pm0.025$ & $1.556\pm0.110$ & $0.766\pm0.054$ \\
QCDSR~\cite{Du2004} &
$1.21\pm0.33$ & $0.42\pm0.12$ & $0.55\pm0.15$ & $0.59\pm0.17$ & $2.20\pm0.85$ & $1.07\pm0.43$ \\
LCSR~\cite{Aliev2004} &
$0.90\pm0.20$ & $0.56\pm0.10$ & $0.65\pm0.15$ & $0.85\pm0.20$ & $1.385$ & $1.308$ \\
LCSR/HQEFT~\cite{Wu2006} &
$0.778^{+0.057}_{-0.062}$ & $0.692$ &
$0.569^{+0.046}_{-0.049}$ & $0.304^{+0.021}_{-0.017}$ & $1.367$ & $0.534$ \\
CLFQM~\cite{Verma2012} &
$0.98$ & $0.72$ & $0.69$ & $0.59$ & $1.420$ & $0.855$ \\
LQCD (HPQCD)~\cite{HPQCD2014} &
$1.059(124)$ & $0.706(37)$ & $0.615(24)$ & $0.457(78)$ & $1.720(210)$ & $0.740(120)$ \\
CCQM~\cite{Ivanov2019} &
$0.91$ & $0.68$ & $0.68$ & $0.67$ & $1.338$ & $0.985$ \\
LQCD (CLQCD)~\cite{Fan2025} &
$1.002(9)$ & $0.692(4)$ & $0.621(5)$ & $0.460(19)$ & $1.614(19)$ & $0.741(31)$ \\
CLFQM~\cite{Yang2026} &
$1.009\pm0.011$ & $0.634\pm0.002$ &
$0.656^{+0.003}_{-0.004}$ & $0.564\pm0.002$ & $1.538$ & $0.860$ \\
\hline
BaBar~\cite{BaBar2008DsPhi} &
 &  & $0.607\pm0.028$ &  & $1.849\pm0.11$ & $0.763\pm0.096$ \\
PDG 2026~\cite{PDG2026} &
 & & & & $1.76\pm0.07$ & $0.83\pm0.08$ \\
\end{tabular}
\end{ruledtabular}
\end{table}

\begin{table}
\caption{Branching fractions for $D\to\pi\ell\nu_\ell$ decays, in units of $10^{-3}$.}
\label{tab:BRDPi}
\scriptsize
\begin{ruledtabular}
\begin{tabular}{ccccc}
Model&
$\mathcal B(D^{0}\to\pi^{-}e^{+}\nu_{e})$ &
$\mathcal B(D^{0}\to\pi^{-}\mu^{+}\nu_{\mu})$ &
$\mathcal B(D^{+}\to\pi^{0}e^{+}\nu_{e})$ &
$\mathcal B(D^{+}\to\pi^{0}\mu^{+}\nu_{\mu})$ \\
\hline
RQM~\cite{Faustov2020}
& $2.78\pm0.28$ & $2.74\pm0.27$ & $3.50\pm0.35$ & $3.45\pm0.34$ \\
LCSR/HQEFT~\cite{Wu2006}
& $2.78$ & $2.75$ & $3.52$ & $3.49$ \\
CCQM~\cite{Ivanov2019}
& $2.2$ & $2.2$ & $2.9$ & $2.8$ \\
CSM~\cite{Yao2022}
& $2.73\pm0.22$ &  &  &  \\
CLFQM~\cite{Cheng2017}
&  &  & $4.1\pm0.3$ & $4.1\pm0.3$ \\
CLFQM~\cite{Yang2026}
& $2.69^{+0.10}_{-0.11}$
& $2.57^{+0.095}_{-0.099}$
& $3.46^{+0.13}_{-0.14}$
& $3.30^{+0.13}_{-0.14}$ \\
\hline
BESIII~\cite{BESIII2026Pi}
& $2.950\pm0.024$
& $2.817\pm0.042$
& $3.622\pm0.038$
& $3.507\pm0.050$ \\
PDG 2026~\cite{PDG2026}
& $2.91\pm0.04$
& $2.67\pm0.12$
& $3.72\pm0.17$
& $3.50\pm0.15$ \\
\end{tabular}
\end{ruledtabular}
\end{table}

\begin{table}
\caption{Branching fractions for $D\to K\ell\nu_\ell$ decays, in units of $10^{-2}$.}
\label{tab:BRDK}
\scriptsize
\begin{ruledtabular}
\begin{tabular}{ccccc}
Model&
$\mathcal B(D^{0}\to K^{-}e^{+}\nu_{e})$ &
$\mathcal B(D^{0}\to K^{-}\mu^{+}\nu_{\mu})$ &
$\mathcal B(D^{+}\to\bar{K}^{0}e^{+}\nu_{e})$ &
$\mathcal B(D^{+}\to\bar{K}^{0}\mu^{+}\nu_{\mu})$ \\
\hline
RQM~\cite{Faustov2020}
& $3.68\pm0.37$ & $3.61\pm0.36$ & $9.26\pm0.93$ & $9.08\pm0.91$ \\
LCSR/HQEFT~\cite{Wu2006}
& $3.20$ & $3.15$ & $8.12$ & $7.98$ \\
CCQM~\cite{Ivanov2019}
& $3.63$ & $3.53$ & $9.28$ & $9.02$ \\
CSM~\cite{Yao2022}
& $3.83\pm0.28$ &  &  &  \\
CLFQM~\cite{Cheng2017}
&  &  & $10.32\pm0.93$ & $10.07\pm0.91$ \\
CLFQM~\cite{Yang2026}
& $3.70^{+0.052}_{-0.055}$
& $3.45^{+0.042}_{-0.046}$
& $9.39^{+0.13}_{-0.14}$
& $8.75^{+0.12}_{-0.12}$ \\
\hline
BESIII~\cite{BESIII2026K}
& $3.548\pm0.018$
& $3.445\pm0.018$
& $8.928\pm0.056$
& $8.770\pm0.060$ \\
PDG 2026~\cite{PDG2026}
& $3.537\pm0.017$
& $3.418\pm0.019$
& $8.81\pm0.07$
& $8.68\pm0.10$ \\
\end{tabular}
\end{ruledtabular}
\end{table}

\begin{table}
\caption{Branching fractions for $D^{+}\to\eta^{(\prime)}\ell^{+}\nu_\ell$ decays, in units of $10^{-3}$ for $\eta$ and $10^{-4}$ for $\eta'$.}
\label{tab:BRDpEta}
\scriptsize
\begin{ruledtabular}
\begin{tabular}{ccccc}
Model&
$\mathcal B(D^{+}\to\eta e^{+}\nu_{e})$ &
$\mathcal B(D^{+}\to\eta\mu^{+}\nu_{\mu})$ &
$\mathcal B(D^{+}\to\eta' e^{+}\nu_{e})$ &
$\mathcal B(D^{+}\to\eta' \mu^{+}\nu_{\mu})$ \\
\hline
RQM~\cite{Faustov2020}
& $1.27\pm0.15$ & $1.24\pm0.15$ & $2.97\pm0.42$ & $2.79\pm0.39$ \\
LCSR/HQEFT~\cite{Wu2006}
& $0.86$ & $0.84$ &  &  \\
CCQM~\cite{Ivanov2019}
& $0.94$ & $0.91$ & $2.00$ & $1.90$ \\
CLFQM~\cite{Cheng2017}
& $1.20\pm0.10$ & $1.20\pm0.10$ & $1.80\pm0.20$ & $1.70\pm0.20$ \\
CLFQM~\cite{Yang2026}
& $1.06^{+0.081}_{-0.099}$
& $0.98^{+0.081}_{-0.098}$
& $1.69^{+0.13}_{-0.16}$
& $1.48^{+0.11}_{-0.14}$ \\
\hline
BESIII~\cite{BESIII2025Eta,BESIII2025EtaPrime}
& $0.975\pm0.040$
& $0.908\pm0.042$
& $1.79\pm0.20$
& $1.92\pm0.29$ \\
PDG 2026~\cite{PDG2026}
& $1.11\pm0.07$
& $1.04\pm0.11$
& $1.84\pm0.19$
& $1.92\pm0.29$ \\
\end{tabular}
\end{ruledtabular}
\end{table}

\begin{table}
\caption{Branching fractions for $D_s^{+}\to K^{0},\eta,\eta'$ semileptonic decays, in units of $10^{-3}$ for $K^0$ and $10^{-2}$ for $\eta^{(\prime)}$.}
\label{tab:BRDsP}
\tiny
\begin{ruledtabular}
\begin{tabular}{ccccccc}
Model&
$\mathcal B(D_s^{+}\to K^{0}e^{+}\nu_{e})$ &
$\mathcal B(D_s^{+}\to K^{0}\mu^{+}\nu_{\mu})$ &
$\mathcal B(D_s^{+}\to\eta e^{+}\nu_{e})$ &
$\mathcal B(D_s^{+}\to\eta\mu^{+}\nu_{\mu})$ &
$\mathcal B(D_s^{+}\to\eta' e^{+}\nu_{e})$ &
$\mathcal B(D_s^{+}\to\eta' \mu^{+}\nu_{\mu})$ \\
\hline
RQM~\cite{Faustov2020}
& $4.08\pm0.41$ & $4.01\pm0.40$ & $2.83\pm0.34$ & $2.77\pm0.33$ & $0.80\pm0.11$ & $0.77\pm0.11$ \\
LCSR/HQEFT~\cite{Wu2006}
& $3.90$ & $3.83$ & $1.27$ & $1.25$ &  &  \\
CCQM~\cite{Ivanov2019}
& $2.0$ & $2.0$ & $2.24$ & $2.18$ & $0.83$ & $0.79$ \\
CSM~\cite{Yao2022}
& $3.31\pm0.33$ &  &  &  &  &  \\
LCSR~\cite{Azizi2011}
&  &  & $3.15\pm0.97$ &  & $0.97\pm0.38\;(39.7^\circ)$ &  \\
CLFQM~\cite{Cheng2017}
& $2.7\pm0.2$ & $2.6\pm0.2$ & $2.26\pm0.21$ & $2.22\pm0.20$ & $0.89\pm0.09$ & $0.85\pm0.08$ \\
CLFQM~\cite{Yang2026}
& $2.65^{+0.10}_{-0.10}$
& $2.49^{+0.097}_{-0.093}$
& $2.14^{+0.13}_{-0.14}$
& $2.00^{+0.12}_{-0.14}$
& $0.88^{+0.051}_{-0.061}$
& $0.78^{+0.051}_{-0.051}$ \\
\hline
BESIII~\cite{BESIII2024DsKe,BESIII2025DsK,BESIII2023DsEta,BESIII2024DsEtaMu}
& $2.98\pm0.26$
& $2.89\pm0.30$
& $2.255\pm0.064$
& $2.235\pm0.073$
& $0.810\pm0.045$
& $0.801\pm0.062$ \\
CLEO~\cite{CLEO2015Ds}
& $3.9\pm0.85$ &  & $2.28\pm0.24$ &  & $0.68\pm0.16$ &  \\
PDG 2026~\cite{PDG2026}
& $2.88\pm0.26$ &  & $2.27\pm0.06$ & $2.24\pm0.07$ & $0.81\pm0.04$ & $0.80\pm0.06$ \\
\end{tabular}
\end{ruledtabular}
\end{table}

\begin{table}[htbp]
\caption{Branching fractions for $D\to\rho,\omega$ semileptonic decays, in units of $10^{-3}$.}
\label{tab:BRDRhoOmega}
\tiny
\begin{ruledtabular}
\resizebox{\textwidth}{!}{%
\begin{tabular}{ccccccc}
Model&
$\mathcal B(D^0\to\rho^-e^+\nu_e)$ &
$\mathcal B(D^0\to\rho^-\mu^+\nu_\mu)$ &
$\mathcal B(D^+\to\rho^0e^+\nu_e)$ &
$\mathcal B(D^+\to\rho^0\mu^+\nu_\mu)$ &
$\mathcal B(D^+\to\omega e^+\nu_e)$ &
$\mathcal B(D^+\to\omega\mu^+\nu_\mu)$ \\
\hline
RQM~\cite{Faustov2020}
& $2.21\pm0.25$ & $2.11\pm0.23$ & $2.78\pm0.32$ & $2.66\pm0.29$ & $2.47\pm0.29$ & $2.37\pm0.26$ \\
LCSR/HQEFT~\cite{Wu2006}
& $1.81^{+0.18}_{-0.13}$ & $1.73^{+0.17}_{-0.13}$
& $2.29^{+0.23}_{-0.16}$ & $2.20^{+0.21}_{-0.16}$
& $1.93^{+0.20}_{-0.14}$ & $1.85^{+0.19}_{-0.13}$ \\
CCQM~\cite{Ivanov2019}
& $1.62$ & $1.55$ & $2.09$ & $2.01$ & $1.85$ & $1.78$ \\
CLFQM~\cite{Cheng2017}
&  &  & $2.3\pm0.2$ & $2.2\pm0.2$ & $2.1\pm0.2$ & $2.0\pm0.2$ \\
LCSR~\cite{Fu2020}
& $1.749^{+0.421}_{-0.297}$ &  & $2.217^{+0.534}_{-0.376}$ &  &  &  \\
SCI~\cite{Xing2022}
& $1.27$ & $1.21$ & $1.64$ & $1.56$ &  &  \\
LCSR~\cite{Zhong2024}
& $1.889^{+0.176}_{-0.170}$ & $1.881^{+0.174}_{-0.168}$
& $2.380^{+0.221}_{-0.214}$ & $2.369^{+0.219}_{-0.211}$ &  &  \\
LCSR~\cite{Lin2025}
&  &  & $2.30^{+0.32}_{-0.25}$ & $2.20^{+0.30}_{-0.23}$ &  &  \\
CLFQM~\cite{Yang2026}
& $1.69\pm0.06$ & $1.60\pm0.06$
& $2.17\pm0.09$ & $2.07^{+0.09}_{-0.08}$
& $2.02^{+0.07}_{-0.09}$ & $2.00^{+0.08}_{-0.13}$ \\
\hline
BESIII~\cite{BESIII2019Rho,BESIII2021RhoMu,BESIII2015Omega,BESIII2020OmegaMu}
& $1.445\pm0.070$ & $1.35\pm0.13$
& $1.860\pm0.093$ &
& $1.63\pm0.14$ & $1.77\pm0.21$ \\
CLEO~\cite{CLEO2013RhoOmega}
& $1.77\pm0.16$ &
& $2.17^{+0.17}_{-0.25}$ &
& $1.82\pm0.19$ & \\
PDG 2026~\cite{PDG2026}
& $1.46\pm0.08$ & $1.35\pm0.13$
& $1.87\pm0.06$ & $1.64\pm0.16$
& $1.69\pm0.11$ & $1.77\pm0.21$ \\
\end{tabular}%
}
\end{ruledtabular}
\end{table}

\begin{table}[htbp]
\caption{Branching fractions for $D\to K^*\ell\nu_\ell$ decays, in units of $10^{-2}$.}
\label{tab:BRDKstar}
\scriptsize
\begin{ruledtabular}
\resizebox{\textwidth}{!}{%
\begin{tabular}{ccccc}
Model&
$\mathcal B(D^0\to K^{*-}e^+\nu_e)$ &
$\mathcal B(D^0\to K^{*-}\mu^+\nu_\mu)$ &
$\mathcal B(D^+\to\bar K^{*0}e^+\nu_e)$ &
$\mathcal B(D^+\to\bar K^{*0}\mu^+\nu_\mu)$ \\
\hline
RQM~\cite{Faustov2020}
& $2.03\pm0.22$ & $1.93\pm0.21$ & $5.12\pm0.56$ & $4.85\pm0.53$ \\
LCSR/HQEFT~\cite{Wu2006}
& $2.12^{+0.09}_{-0.09}$ & $2.01^{+0.09}_{-0.08}$
& $5.37^{+0.24}_{-0.23}$ & $5.10^{+0.23}_{-0.21}$ \\
CCQM~\cite{Ivanov2019}
& $2.96$ & $2.80$ & $7.61$ & $7.21$ \\
CLFQM~\cite{Cheng2017}
&  &  & $7.5\pm0.7$ & $7.0\pm0.7$ \\
SCI~\cite{Xing2022}
& $2.10$ & $1.97$ & $5.45$ & $5.13$ \\
LCSR~\cite{Lin2025}
& $1.76^{+0.24}_{-0.19}$ & $1.66^{+0.22}_{-0.18}$
& $4.52^{+0.62}_{-0.50}$ & $4.27^{+0.57}_{-0.45}$ \\
CLFQM~\cite{Yang2026}
& $2.59^{+0.04}_{-0.05}$ & $2.44^{+0.04}_{-0.05}$
& $6.65\pm0.13$ & $6.26^{+0.44}_{-0.24}$ \\
\hline
BESIII~\cite{BESIII2026Kstar}
& $2.043\pm0.022$ & $1.964\pm0.022$ &  &  \\
CLEO~\cite{CLEO2005Kstar,CLEO2010Kstar}
& $2.16\pm0.17$ &
& $5.52\pm0.15$ & $5.27\pm0.16$ \\
PDG 2026~\cite{PDG2026}
& $2.16\pm0.16$ & $2.01\pm0.06$
& $5.40\pm0.10$ & $5.26\pm0.15$ \\
\end{tabular}%
}
\end{ruledtabular}
\end{table}

\begin{table}[htbp]
\caption{Branching fractions for $D_s^+\to\phi,K^{*0}$ semileptonic decays, in units of $10^{-2}$ for $\phi$ and $10^{-3}$ for $K^{*0}$.}
\label{tab:BRDsVector}
\scriptsize
\begin{ruledtabular}
\resizebox{\textwidth}{!}{%
\begin{tabular}{ccccc}
Model&
$\mathcal B(D_s^+\to\phi e^+\nu_e)$ &
$\mathcal B(D_s^+\to\phi\mu^+\nu_\mu)$ &
$\mathcal B(D_s^+\to K^{*0}e^+\nu_e)$ &
$\mathcal B(D_s^+\to K^{*0}\mu^+\nu_\mu)$ \\
\hline
RQM~\cite{Faustov2020}
& $2.76\pm0.29$ & $2.60\pm0.27$ & $2.25\pm0.24$ & $2.15\pm0.22$ \\
LCSR/HQEFT~\cite{Wu2006}
& $2.53^{+0.37}_{-0.40}$ & $2.40^{+0.35}_{-0.37}$
& $2.33^{+0.29}_{-0.30}$ & $2.24^{+0.27}_{-0.29}$ \\
CCQM~\cite{Ivanov2019}
& $3.01$ & $2.85$ & $1.80$ & $1.70$ \\
CLFQM~\cite{Cheng2017}
& $3.1\pm0.3$ & $2.9\pm0.3$ & $1.9\pm0.2$ & $1.9\pm0.2$ \\
SCI~\cite{Xing2022}
& $2.45$ & $2.30$ & $1.76$ & $1.68$ \\
LCSR~\cite{Lin2025}
&  &  & $1.55^{+0.30}_{-0.20}$ & $1.48^{+0.29}_{-0.19}$ \\
QCDSR~\cite{Du2004}
& $1.8\pm0.5$ & $1.8\pm0.5$ &  &  \\
LCSR~\cite{Aliev2004}
& $2.78^{+1.42}_{-1.13}$ & &  &  \\
LQCD (CLQCD)~\cite{Fan2025}
& $2.493\pm0.073$ & $2.351\pm0.067$ &  &  \\
CLFQM~\cite{Yang2026}
& $2.88\pm0.06$ & $2.70\pm0.05$
& $1.85\pm0.08$ & $1.76^{+0.07}_{-0.06}$ \\
\hline
BESIII~\cite{BESIII2018DsPhi,BESIII2023DsPhi,BESIII2026DsKstar}
& $2.26\pm0.46$ & $2.25\pm0.11$
& $2.14\pm0.21$ & $2.07\pm0.24$ \\
CLEO~\cite{CLEO2015Ds}
& $2.14\pm0.19$ &  & $1.80\pm0.41$ &  \\
BaBar~\cite{BaBar2008DsPhi}
& $2.61\pm0.17$ &  &  &  \\
PDG 2026~\cite{PDG2026}
& $2.34\pm0.12$ & $2.24\pm0.11$
& $2.05\pm0.20$ &  \\
\end{tabular}%
}
\end{ruledtabular}
\end{table}

\subsection{$D_{(s)}\to V$}
\label{subsec:comparison_vector}

The vector-meson channels are compared at the level of form-factor values at $q^2=0$, ratios $r_V=V(0)/A_1(0)$ and $r_2=A_2(0)/A_1(0)$, and branching fractions in Tables~\ref{tab:FFD0Kstar}--\ref{tab:FFDsPhi} and \ref{tab:BRDRhoOmega}--\ref{tab:BRDsVector}. For the Cabibbo-favored $D\to K^*$ transitions, the RQM ratios and branching fractions are broadly consistent with the available BESIII, CLEO, and PDG results. The same is true for the integrated $D_s\to\phi$ and $D_s\to K^*$ rates, although the $r_2$ measurement for $D_s\to K^*$ lies close to the $2\sigma$ level below the RQM prediction. The $D_s\to\phi$ form-factor value at $q^2=0$ is also reasonably close to the LQCD (CLQCD) results in Table~\ref{tab:FFDsPhi}.

The largest differences within the vector-meson sample occur in the $D\to\rho$ channels. The RQM prediction for $A_1(0)$ exceeds the CLEO central value with a discrepancy approaching the $3\sigma$ level when the upper CLEO uncertainty and the theoretical uncertainty are combined, despite agreement of the ratios $r_V$ and $r_2$ at the present level of precision (Table~\ref{tab:FFD0Rho}). The RQM branching fractions for $D^0\to\rho^-e^+\nu_e$, $D^0\to\rho^-\mu^+\nu_\mu$, and $D^+\to\rho^0e^+\nu_e$ exceed the BESIII values, with deviations approaching the $3\sigma$ level. For $D^+\to\omega e^+\nu_e$, the discrepancy with the BESIII measurement lies between the $2\sigma$ and $3\sigma$ levels. Therefore, the agreement obtained in the $K^*$ and $\phi$ channels does not extend to all vector final states. Theoretical calculations based on LCSR and CLFQM often give smaller rates for these channels (Table~\ref{tab:BRDRhoOmega}), illustrating the model dependence of these predictions. The comparison is also subject to differences in the treatment of broad vector resonances, notably the $\rho$, which is experimentally reconstructed through its hadronic decay products. A quantitative attribution of the observed deviations to finite-width or nonresonant effects would require a dedicated analysis.

\section{Conclusion}
\label{sec:conclusion}

We have compared the relativistic quark model predictions for semileptonic $D$ and $D_s$ transitions to pseudoscalar and vector mesons with the available experimental determinations of form factors, differential distributions, angular observables, and branching fractions. The $D\to\pi$ and $D\to K$ channels, as well as several $K^*$ and $\phi$ modes, are generally compatible with the available experimental data and with several other theoretical predictions. However, departures beyond two combined standard deviations occur in the $D\to\eta'$ form-factor at $q^2=0$, selected $D_{(s)}\to\eta^{(\prime)}$ and $D_s\to K$ branching fractions, and particularly in $D\to\rho$ rates. These differences limit the extent to which a common level of agreement can be claimed across the full set of decay channels.

The $\eta'$ predictions presented in the tables do not include a gluonic admixture. With a fixed mixing angle of $\phi_G=28.1^\circ$, including this contribution improves the agreement of the $D\to\eta'$ branching fractions with BESIII, although the $D\to\eta'$ form-factor normalization at $q^2=0$ still differs by more than $2\sigma$, approaching the $3\sigma$ level in the fixed-angle comparison. By contrast, the unmixed $D_s\to\eta'$ form factors and rates are already close to experiment, and the same gluonic suppression worsens their agreement, although the deviations remain below $2\sigma$ in the fixed-angle comparison. The present analysis therefore identifies channel-dependent differences rather than a uniform pattern of agreement. Further differential measurements and analyses that retain the relevant correlations could help to clarify their origin.

The detailed comparison of different theoretical results with recent experimental data in Tables \ref{tab:FFDP}-\ref{tab:BRDsVector} shows that the RQM predictions provide a competitive description of many measured semileptonic charm-meson decay observables.

Now we extract the CKM matrix elements $|V_{cd}|$ and $|V_{cs}|$
from the measured branching fractions and from the weighted
average of the values obtained in individual $q^2$ bins
(Mean method).
The results are averaged over all available decay channels
using weights determined solely by experimental uncertainties.
The dominant contributions to the extracted values of $|V_{cd}|$ and $|V_{cs}|$ come from the $D\to\pi$ and $D\to K$ decay channels, respectively, owing to their significantly smaller experimental uncertainties. The 5\% RQM uncertainty of the form-factor values is included separately. The resulting CKM matrix elements are presented in Table~\ref{tab:CKM_average} and compared with the PDG values.

\begin{table}[t]
\centering
\caption{CKM matrix elements extracted from
$D_{(s)}\to P\ell\nu_\ell$ decays.}
\label{tab:CKM_average}
\begin{tabular}{lcc}
\hline\hline
Method & $|V_{cd}|$ & $|V_{cs}|$ \\
\hline
Branching fractions
& $0.2274\pm0.0006\pm0.0114$
& $0.949\pm0.001\pm0.047$ \\

Mean
& $0.2275\pm0.0005\pm0.0114$
& $0.946\pm0.001\pm0.047$ \\

PDG~\cite{PDG2026}
& $0.2247\pm0.0022$
& $0.969\pm0.005$ \\
\hline\hline
\end{tabular}
\end{table}

Summing up, the comparison with recent experimental data demonstrates that the RQM provides a consistent description of a wide range of semileptonic $D$ and $D_s$ decays into pseudoscalar and vector mesons within a unified theoretical framework, without channel-dependent adjustments of the model parameters. Although some discrepancies remain in several channels, the overall agreement supports the applicability of this approach to charm-meson semileptonic transitions.

\begin{acknowledgments}
The authors are grateful to  A. V. Berezhnoy, X.-W. Kang, D. I. Melikhov and M. A. Ivanov for useful discussions.
\end{acknowledgments}
\bibliography{WeakD}

\begin{thebibliography}{59}%
\makeatletter
\providecommand \@ifxundefined [1]{%
 \@ifx{#1\undefined}
}%
\providecommand \@ifnum [1]{%
 \ifnum #1\expandafter \@firstoftwo
 \else \expandafter \@secondoftwo
 \fi
}%
\providecommand \@ifx [1]{%
 \ifx #1\expandafter \@firstoftwo
 \else \expandafter \@secondoftwo
 \fi
}%
\providecommand \natexlab [1]{#1}%
\providecommand \enquote  [1]{``#1''}%
\providecommand \bibnamefont  [1]{#1}%
\providecommand \bibfnamefont [1]{#1}%
\providecommand \citenamefont [1]{#1}%
\providecommand \href@noop [0]{\@secondoftwo}%
\providecommand \href [0]{\begingroup \@sanitize@url \@href}%
\providecommand \@href[1]{\@@startlink{#1}\@@href}%
\providecommand \@@href[1]{\endgroup#1\@@endlink}%
\providecommand \@sanitize@url [0]{\catcode `\\12\catcode `\$12\catcode
  `\&12\catcode `\#12\catcode `\^12\catcode `\_12\catcode `\%12\relax}%
\providecommand \@@startlink[1]{}%
\providecommand \@@endlink[0]{}%
\providecommand \url  [0]{\begingroup\@sanitize@url \@url }%
\providecommand \@url [1]{\endgroup\@href {#1}{\urlprefix }}%
\providecommand \urlprefix  [0]{URL }%
\providecommand \Eprint [0]{\href }%
\providecommand \doibase [0]{https://doi.org/}%
\providecommand \selectlanguage [0]{\@gobble}%
\providecommand \bibinfo  [0]{\@secondoftwo}%
\providecommand \bibfield  [0]{\@secondoftwo}%
\providecommand \translation [1]{[#1]}%
\providecommand \BibitemOpen [0]{}%
\providecommand \bibitemStop [0]{}%
\providecommand \bibitemNoStop [0]{.\EOS\space}%
\providecommand \EOS [0]{\spacefactor3000\relax}%
\providecommand \BibitemShut  [1]{\csname bibitem#1\endcsname}%
\let\auto@bib@innerbib\@empty
%</preamble>
\bibitem [{\citenamefont {Widhalm}\ \emph {et~al.}(2006)\citenamefont {Widhalm}
  \emph {et~al.}}]{Belle2006}%
  \BibitemOpen
  \bibfield  {author} {\bibinfo {author} {\bibfnamefont {L.}~\bibnamefont
  {Widhalm}} \emph {et~al.} (\bibinfo {collaboration} {Belle Collaboration}),\
  }\bibfield  {title} {\bibinfo {title} {Measurement of {$D^0\to\pi\ell\nu$}
  ({$K\ell\nu$}) form factors and absolute branching fractions},\ }\href
  {https://doi.org/10.1103/PhysRevLett.97.061804} {\bibfield  {journal}
  {\bibinfo  {journal} {Phys. Rev. Lett.}\ }\textbf {\bibinfo {volume} {97}},\
  \bibinfo {pages} {061804} (\bibinfo {year} {2006})},\ \Eprint
  {https://arxiv.org/abs/hep-ex/0604049} {arXiv:hep-ex/0604049} \BibitemShut
  {NoStop}%
\bibitem [{\citenamefont {Lees}\ \emph {et~al.}(2015)\citenamefont {Lees} \emph
  {et~al.}}]{BaBar2015Pi}%
  \BibitemOpen
  \bibfield  {author} {\bibinfo {author} {\bibfnamefont {J.~P.}\ \bibnamefont
  {Lees}} \emph {et~al.} (\bibinfo {collaboration} {BABAR Collaboration}),\
  }\bibfield  {title} {\bibinfo {title} {Measurement of the
  {$D^0\to\pi^-e^+\nu_e$} differential decay branching fraction as a function
  of {$q^2$} and study of form factor parameterizations},\ }\href
  {https://doi.org/10.1103/PhysRevD.91.052022} {\bibfield  {journal} {\bibinfo
  {journal} {Phys. Rev. D}\ }\textbf {\bibinfo {volume} {91}},\ \bibinfo
  {pages} {052022} (\bibinfo {year} {2015})},\ \Eprint
  {https://arxiv.org/abs/1412.5502} {arXiv:1412.5502} \BibitemShut {NoStop}%
\bibitem [{\citenamefont {Besson}\ \emph {et~al.}(2009)\citenamefont {Besson}
  \emph {et~al.}}]{CLEO2009PiK}%
  \BibitemOpen
  \bibfield  {author} {\bibinfo {author} {\bibfnamefont {D.}~\bibnamefont
  {Besson}} \emph {et~al.} (\bibinfo {collaboration} {CLEO Collaboration}),\
  }\bibfield  {title} {\bibinfo {title} {Improved measurements of {$D$} meson
  semileptonic decays to {$\pi$} and {$K$} mesons},\ }\href
  {https://doi.org/10.1103/PhysRevD.80.032005} {\bibfield  {journal} {\bibinfo
  {journal} {Phys. Rev. D}\ }\textbf {\bibinfo {volume} {80}},\ \bibinfo
  {pages} {032005} (\bibinfo {year} {2009})},\ \Eprint
  {https://arxiv.org/abs/0906.2983} {arXiv:0906.2983} \BibitemShut {NoStop}%
\bibitem [{\citenamefont {Ablikim}\ \emph
  {et~al.}(2026{\natexlab{a}})\citenamefont {Ablikim} \emph
  {et~al.}}]{BESIII2026Pi}%
  \BibitemOpen
  \bibfield  {author} {\bibinfo {author} {\bibfnamefont {M.}~\bibnamefont
  {Ablikim}} \emph {et~al.} (\bibinfo {collaboration} {BESIII Collaboration}),\
  }\href@noop {} {\bibinfo {title} {Precision measurements of semileptonic
  decays {$D^0\to\pi^-\ell^+\nu_\ell$} and {$D^+\to\pi^0\ell^+\nu_\ell$}
  ({$\ell=e,\mu$})}} (\bibinfo {year} {2026}{\natexlab{a}}),\ \Eprint
  {https://arxiv.org/abs/2607.23945} {arXiv:2607.23945 [hep-ex]} \BibitemShut
  {NoStop}%
\bibitem [{\citenamefont {Ablikim}\ \emph
  {et~al.}(2026{\natexlab{b}})\citenamefont {Ablikim} \emph
  {et~al.}}]{BESIII2026K}%
  \BibitemOpen
  \bibfield  {author} {\bibinfo {author} {\bibfnamefont {M.}~\bibnamefont
  {Ablikim}} \emph {et~al.} (\bibinfo {collaboration} {BESIII Collaboration}),\
  }\bibfield  {title} {\bibinfo {title} {Precise measurements of {$D^0\to
  K^-\ell^+\nu_\ell$} and {$D^+\to\bar K^0\ell^+\nu_\ell$} decays},\ }\href
  {https://doi.org/10.1103/PhysRevD.114.033008} {\bibfield  {journal} {\bibinfo
   {journal} {Phys. Rev. D}\ }\textbf {\bibinfo {volume} {114}},\ \bibinfo
  {pages} {033008} (\bibinfo {year} {2026}{\natexlab{b}})},\ \Eprint
  {https://arxiv.org/abs/2601.21196} {arXiv:2601.21196} \BibitemShut {NoStop}%
\bibitem [{\citenamefont {Ablikim}\ \emph
  {et~al.}(2019{\natexlab{a}})\citenamefont {Ablikim} \emph
  {et~al.}}]{BESIII2019Rho}%
  \BibitemOpen
  \bibfield  {author} {\bibinfo {author} {\bibfnamefont {M.}~\bibnamefont
  {Ablikim}} \emph {et~al.} (\bibinfo {collaboration} {BESIII Collaboration}),\
  }\bibfield  {title} {\bibinfo {title} {Observation of {$D^+\to
  f_0(500)e^+\nu_e$} and improved measurements of {$D\to\rho e^+\nu_e$}},\
  }\href {https://doi.org/10.1103/PhysRevLett.122.062001} {\bibfield  {journal}
  {\bibinfo  {journal} {Phys. Rev. Lett.}\ }\textbf {\bibinfo {volume} {122}},\
  \bibinfo {pages} {062001} (\bibinfo {year} {2019}{\natexlab{a}})},\ \Eprint
  {https://arxiv.org/abs/1809.06496} {arXiv:1809.06496} \BibitemShut {NoStop}%
\bibitem [{\citenamefont {Ablikim}\ \emph
  {et~al.}(2019{\natexlab{b}})\citenamefont {Ablikim} \emph
  {et~al.}}]{BESIII2019DsKstar}%
  \BibitemOpen
  \bibfield  {author} {\bibinfo {author} {\bibfnamefont {M.}~\bibnamefont
  {Ablikim}} \emph {et~al.} (\bibinfo {collaboration} {BESIII Collaboration}),\
  }\bibfield  {title} {\bibinfo {title} {First measurement of the form factors
  in {$D_s^+\to K^0e^+\nu_e$} and {$D_s^+\to K^{*0}e^+\nu_e$} decays},\ }\href
  {https://doi.org/10.1103/PhysRevLett.122.061801} {\bibfield  {journal}
  {\bibinfo  {journal} {Phys. Rev. Lett.}\ }\textbf {\bibinfo {volume} {122}},\
  \bibinfo {pages} {061801} (\bibinfo {year} {2019}{\natexlab{b}})},\ \Eprint
  {https://arxiv.org/abs/1811.02911} {arXiv:1811.02911} \BibitemShut {NoStop}%
\bibitem [{\citenamefont {Ablikim}\ \emph {et~al.}(2020)\citenamefont {Ablikim}
  \emph {et~al.}}]{BESIII2020OmegaMu}%
  \BibitemOpen
  \bibfield  {author} {\bibinfo {author} {\bibfnamefont {M.}~\bibnamefont
  {Ablikim}} \emph {et~al.} (\bibinfo {collaboration} {BESIII Collaboration}),\
  }\bibfield  {title} {\bibinfo {title} {Observation of the semi-muonic decay
  {$D^+\to\omega\mu^+\nu_\mu$}},\ }\href
  {https://doi.org/10.1103/PhysRevD.101.072005} {\bibfield  {journal} {\bibinfo
   {journal} {Phys. Rev. D}\ }\textbf {\bibinfo {volume} {101}},\ \bibinfo
  {pages} {072005} (\bibinfo {year} {2020})},\ \Eprint
  {https://arxiv.org/abs/2002.10578} {arXiv:2002.10578 [hep-ex]} \BibitemShut
  {NoStop}%
\bibitem [{\citenamefont {Ablikim}\ \emph
  {et~al.}(2023{\natexlab{a}})\citenamefont {Ablikim} \emph
  {et~al.}}]{BESIII2023DsEta}%
  \BibitemOpen
  \bibfield  {author} {\bibinfo {author} {\bibfnamefont {M.}~\bibnamefont
  {Ablikim}} \emph {et~al.} (\bibinfo {collaboration} {BESIII Collaboration}),\
  }\bibfield  {title} {\bibinfo {title} {Precision measurements of
  {$D_s^+\to\eta e^+\nu_e$} and {$D_s^+\to\eta^\prime e^+\nu_e$}},\ }\href
  {https://doi.org/10.1103/PhysRevD.108.092003} {\bibfield  {journal} {\bibinfo
   {journal} {Phys. Rev. D}\ }\textbf {\bibinfo {volume} {108}},\ \bibinfo
  {pages} {092003} (\bibinfo {year} {2023}{\natexlab{a}})},\ \Eprint
  {https://arxiv.org/abs/2306.05194} {arXiv:2306.05194} \BibitemShut {NoStop}%
\bibitem [{\citenamefont {Ablikim}\ \emph
  {et~al.}(2023{\natexlab{b}})\citenamefont {Ablikim} \emph
  {et~al.}}]{BESIII2023DsPhi}%
  \BibitemOpen
  \bibfield  {author} {\bibinfo {author} {\bibfnamefont {M.}~\bibnamefont
  {Ablikim}} \emph {et~al.} (\bibinfo {collaboration} {BESIII Collaboration}),\
  }\bibfield  {title} {\bibinfo {title} {Studies of the decay {$D_s^+\to
  K^+K^-\mu^+\nu_\mu$}},\ }\href {https://doi.org/10.1007/JHEP12(2023)072}
  {\bibfield  {journal} {\bibinfo  {journal} {J. High Energy Phys.}\ }\textbf
  {\bibinfo {volume} {12}}\bibfield  {number} {\bibinfo  {number} { (2023)},\
  \bibinfo {pages} {072}},\ }\Eprint {https://arxiv.org/abs/2307.03024}
  {arXiv:2307.03024} \BibitemShut {NoStop}%
\bibitem [{\citenamefont {Ablikim}\ \emph {et~al.}(2018)\citenamefont {Ablikim}
  \emph {et~al.}}]{BESIII2018DsPhi}%
  \BibitemOpen
  \bibfield  {author} {\bibinfo {author} {\bibfnamefont {M.}~\bibnamefont
  {Ablikim}} \emph {et~al.} (\bibinfo {collaboration} {BESIII Collaboration}),\
  }\bibfield  {title} {\bibinfo {title} {Measurements of the branching
  fractions for the semileptonic decays {$D_s^+\to\phi e^+\nu_e$},
  {$\phi\mu^+\nu_\mu$}, {$\eta\mu^+\nu_\mu$} and {$\eta'\mu^+\nu_\mu$}},\
  }\href {https://doi.org/10.1103/PhysRevD.97.012006} {\bibfield  {journal}
  {\bibinfo  {journal} {Phys. Rev. D}\ }\textbf {\bibinfo {volume} {97}},\
  \bibinfo {pages} {012006} (\bibinfo {year} {2018})},\ \Eprint
  {https://arxiv.org/abs/1709.03680} {arXiv:1709.03680} \BibitemShut {NoStop}%
\bibitem [{\citenamefont {Ablikim}\ \emph {et~al.}(2021)\citenamefont {Ablikim}
  \emph {et~al.}}]{BESIII2021RhoMu}%
  \BibitemOpen
  \bibfield  {author} {\bibinfo {author} {\bibfnamefont {M.}~\bibnamefont
  {Ablikim}} \emph {et~al.} (\bibinfo {collaboration} {BESIII Collaboration}),\
  }\bibfield  {title} {\bibinfo {title} {Observation of the decay
  {$D^0\to\rho^-\mu^+\nu_\mu$}},\ }\href
  {https://doi.org/10.1103/PhysRevD.104.L091103} {\bibfield  {journal}
  {\bibinfo  {journal} {Phys. Rev. D}\ }\textbf {\bibinfo {volume} {104}},\
  \bibinfo {pages} {L091103} (\bibinfo {year} {2021})},\ \Eprint
  {https://arxiv.org/abs/2106.02292} {arXiv:2106.02292 [hep-ex]} \BibitemShut
  {NoStop}%
\bibitem [{\citenamefont {Ablikim}\ \emph
  {et~al.}(2024{\natexlab{a}})\citenamefont {Ablikim} \emph
  {et~al.}}]{BESIII2024DsKe}%
  \BibitemOpen
  \bibfield  {author} {\bibinfo {author} {\bibfnamefont {M.}~\bibnamefont
  {Ablikim}} \emph {et~al.} (\bibinfo {collaboration} {BESIII Collaboration}),\
  }\bibfield  {title} {\bibinfo {title} {Improved measurement of the
  semileptonic decay {$D_s^+\to K^0 e^+\nu_e$}},\ }\href
  {https://doi.org/10.1103/PhysRevD.110.052012} {\bibfield  {journal} {\bibinfo
   {journal} {Phys. Rev. D}\ }\textbf {\bibinfo {volume} {110}},\ \bibinfo
  {pages} {052012} (\bibinfo {year} {2024}{\natexlab{a}})},\ \Eprint
  {https://arxiv.org/abs/2406.19190} {arXiv:2406.19190} \BibitemShut {NoStop}%
\bibitem [{\citenamefont {Ablikim}\ \emph
  {et~al.}(2024{\natexlab{b}})\citenamefont {Ablikim} \emph
  {et~al.}}]{BESIII2024DsEtaMu}%
  \BibitemOpen
  \bibfield  {author} {\bibinfo {author} {\bibfnamefont {M.}~\bibnamefont
  {Ablikim}} \emph {et~al.} (\bibinfo {collaboration} {BESIII Collaboration}),\
  }\bibfield  {title} {\bibinfo {title} {Observation of
  {$D_s^+\to\eta^\prime\mu^+\nu_\mu$}, precision test of lepton flavor
  universality with {$D_s^+\to\eta^{(\prime)}\ell^+\nu_\ell$}, and first
  measurements of {$D_s^+\to\eta^{(\prime)}\mu^+\nu_\mu$} decay dynamics},\
  }\href {https://doi.org/10.1103/PhysRevLett.132.091802} {\bibfield  {journal}
  {\bibinfo  {journal} {Phys. Rev. Lett.}\ }\textbf {\bibinfo {volume} {132}},\
  \bibinfo {pages} {091802} (\bibinfo {year} {2024}{\natexlab{b}})},\ \Eprint
  {https://arxiv.org/abs/2307.12852} {arXiv:2307.12852} \BibitemShut {NoStop}%
\bibitem [{\citenamefont {Ablikim}\ \emph
  {et~al.}(2025{\natexlab{a}})\citenamefont {Ablikim} \emph
  {et~al.}}]{BESIII2025Eta}%
  \BibitemOpen
  \bibfield  {author} {\bibinfo {author} {\bibfnamefont {M.}~\bibnamefont
  {Ablikim}} \emph {et~al.} (\bibinfo {collaboration} {BESIII Collaboration}),\
  }\href@noop {} {\bibinfo {title} {Improved measurements of {$D^+\to\eta
  e^+\nu_e$} and {$D^+\to\eta\mu^+\nu_\mu$}}} (\bibinfo {year}
  {2025}{\natexlab{a}}),\ \Eprint {https://arxiv.org/abs/2506.02521}
  {arXiv:2506.02521 [hep-ex]} \BibitemShut {NoStop}%
\bibitem [{\citenamefont {Ablikim}\ \emph
  {et~al.}(2025{\natexlab{b}})\citenamefont {Ablikim} \emph
  {et~al.}}]{BESIII2025EtaPrime}%
  \BibitemOpen
  \bibfield  {author} {\bibinfo {author} {\bibfnamefont {M.}~\bibnamefont
  {Ablikim}} \emph {et~al.} (\bibinfo {collaboration} {BESIII Collaboration}),\
  }\bibfield  {title} {\bibinfo {title} {Combined dynamic analysis of
  {$D^+\to\eta^\prime\ell^+\nu_\ell$}},\ }\href
  {https://doi.org/10.1103/PhysRevLett.134.111801} {\bibfield  {journal}
  {\bibinfo  {journal} {Phys. Rev. Lett.}\ }\textbf {\bibinfo {volume} {134}},\
  \bibinfo {pages} {111801} (\bibinfo {year} {2025}{\natexlab{b}})},\ \Eprint
  {https://arxiv.org/abs/2410.08603} {arXiv:2410.08603} \BibitemShut {NoStop}%
\bibitem [{\citenamefont {Ablikim}\ \emph
  {et~al.}(2026{\natexlab{c}})\citenamefont {Ablikim} \emph
  {et~al.}}]{BESIII2026Kstar}%
  \BibitemOpen
  \bibfield  {author} {\bibinfo {author} {\bibfnamefont {M.}~\bibnamefont
  {Ablikim}} \emph {et~al.} (\bibinfo {collaboration} {BESIII Collaboration}),\
  }\bibfield  {title} {\bibinfo {title} {Precise measurement of the form
  factors in {$D^0\to K^*(892)^-\ell^+\nu_\ell$} and observation of {$D^0\to
  K_2^*(1430)^-\ell^+\nu_\ell$}},\ }\href {https://doi.org/10.1103/9mzg-5jvr}
  {\bibfield  {journal} {\bibinfo  {journal} {Phys. Rev. D}\ }\textbf {\bibinfo
  {volume} {114}},\ \bibinfo {pages} {L071101} (\bibinfo {year}
  {2026}{\natexlab{c}})},\ \Eprint {https://arxiv.org/abs/2603.04136}
  {arXiv:2603.04136} \BibitemShut {NoStop}%
\bibitem [{\citenamefont {Ablikim}\ \emph
  {et~al.}(2026{\natexlab{d}})\citenamefont {Ablikim} \emph
  {et~al.}}]{BESIII2025DsK}%
  \BibitemOpen
  \bibfield  {author} {\bibinfo {author} {\bibfnamefont {M.}~\bibnamefont
  {Ablikim}} \emph {et~al.} (\bibinfo {collaboration} {BESIII Collaboration}),\
  }\bibfield  {title} {\bibinfo {title} {First measurement of the {$D_s^+\to
  K^0\mu^+\nu_\mu$} decay},\ }\href
  {https://doi.org/10.1103/PhysRevLett.136.161802} {\bibfield  {journal}
  {\bibinfo  {journal} {Phys. Rev. Lett.}\ }\textbf {\bibinfo {volume} {136}},\
  \bibinfo {pages} {161802} (\bibinfo {year} {2026}{\natexlab{d}})},\ \Eprint
  {https://arxiv.org/abs/2510.05904} {arXiv:2510.05904} \BibitemShut {NoStop}%
\bibitem [{\citenamefont {Ablikim}\ \emph
  {et~al.}(2026{\natexlab{e}})\citenamefont {Ablikim} \emph
  {et~al.}}]{BESIII2026DsKstar}%
  \BibitemOpen
  \bibfield  {author} {\bibinfo {author} {\bibfnamefont {M.}~\bibnamefont
  {Ablikim}} \emph {et~al.} (\bibinfo {collaboration} {BESIII Collaboration}),\
  }\href@noop {} {\bibinfo {title} {First measurement of the {$D_s^+\to
  K^*(892)^0\mu^+\nu_\mu$} decay, study of dynamics and test of lepton
  universality with {$D_s^+\to K^*(892)^0\ell^+\nu_\ell$} decays}} (\bibinfo
  {year} {2026}{\natexlab{e}}),\ \Eprint {https://arxiv.org/abs/2605.07176}
  {arXiv:2605.07176 [hep-ex]} \BibitemShut {NoStop}%
\bibitem [{\citenamefont {Donald}\ \emph {et~al.}(2014)\citenamefont {Donald},
  \citenamefont {Davies}, \citenamefont {Koponen},\ and\ \citenamefont
  {Lepage}}]{HPQCD2014}%
  \BibitemOpen
  \bibfield  {author} {\bibinfo {author} {\bibfnamefont {G.~C.}\ \bibnamefont
  {Donald}}, \bibinfo {author} {\bibfnamefont {C.~T.~H.}\ \bibnamefont
  {Davies}}, \bibinfo {author} {\bibfnamefont {J.}~\bibnamefont {Koponen}},\
  and\ \bibinfo {author} {\bibfnamefont {G.~P.}\ \bibnamefont {Lepage}}
  (\bibinfo {collaboration} {HPQCD Collaboration}),\ }\bibfield  {title}
  {\bibinfo {title} {{$V_{cs}$} from {$D_s\to\phi\ell\nu$} semileptonic decay
  and full lattice {QCD}},\ }\href {https://doi.org/10.1103/PhysRevD.90.074506}
  {\bibfield  {journal} {\bibinfo  {journal} {Phys. Rev. D}\ }\textbf {\bibinfo
  {volume} {90}},\ \bibinfo {pages} {074506} (\bibinfo {year} {2014})},\
  \Eprint {https://arxiv.org/abs/1311.6669} {arXiv:1311.6669} \BibitemShut
  {NoStop}%
\bibitem [{\citenamefont {Lubicz}\ \emph {et~al.}(2017)\citenamefont {Lubicz},
  \citenamefont {Riggio}, \citenamefont {Salerno}, \citenamefont {Simula},\
  and\ \citenamefont {Tarantino}}]{Lubicz2017}%
  \BibitemOpen
  \bibfield  {author} {\bibinfo {author} {\bibfnamefont {V.}~\bibnamefont
  {Lubicz}}, \bibinfo {author} {\bibfnamefont {L.}~\bibnamefont {Riggio}},
  \bibinfo {author} {\bibfnamefont {G.}~\bibnamefont {Salerno}}, \bibinfo
  {author} {\bibfnamefont {S.}~\bibnamefont {Simula}},\ and\ \bibinfo {author}
  {\bibfnamefont {C.}~\bibnamefont {Tarantino}},\ }\bibfield  {title} {\bibinfo
  {title} {Scalar and vector form factors of {$D\to\pi(K)\ell\nu$} decays with
  {$N_f=2+1+1$} twisted fermions},\ }\href
  {https://doi.org/10.1103/PhysRevD.96.054514} {\bibfield  {journal} {\bibinfo
  {journal} {Phys. Rev. D}\ }\textbf {\bibinfo {volume} {96}},\ \bibinfo
  {pages} {054514} (\bibinfo {year} {2017})},\ \Eprint
  {https://arxiv.org/abs/1706.03017} {arXiv:1706.03017} \BibitemShut {NoStop}%
\bibitem [{\citenamefont {Fan}\ \emph {et~al.}(2025)\citenamefont {Fan},
  \citenamefont {Meng}, \citenamefont {Liu}, \citenamefont {Liu}, \citenamefont
  {Shen}, \citenamefont {Wang}, \citenamefont {Zhang},\ and\ \citenamefont
  {Zhang}}]{Fan2025}%
  \BibitemOpen
  \bibfield  {author} {\bibinfo {author} {\bibfnamefont {G.}~\bibnamefont
  {Fan}}, \bibinfo {author} {\bibfnamefont {Y.}~\bibnamefont {Meng}}, \bibinfo
  {author} {\bibfnamefont {C.}~\bibnamefont {Liu}}, \bibinfo {author}
  {\bibfnamefont {Z.}~\bibnamefont {Liu}}, \bibinfo {author} {\bibfnamefont
  {T.}~\bibnamefont {Shen}}, \bibinfo {author} {\bibfnamefont {T.-X.}\
  \bibnamefont {Wang}}, \bibinfo {author} {\bibfnamefont {K.-L.}\ \bibnamefont
  {Zhang}},\ and\ \bibinfo {author} {\bibfnamefont {L.}~\bibnamefont {Zhang}},\
  }\href@noop {} {\bibinfo {title} {Study of the {$D_s\to\phi\ell\nu_\ell$}
  semileptonic decay with (2+1)-flavor lattice {QCD}}} (\bibinfo {year}
  {2025}),\ \Eprint {https://arxiv.org/abs/2510.14478} {arXiv:2510.14478
  [hep-lat]} \BibitemShut {NoStop}%
\bibitem [{\citenamefont {Du}\ \emph {et~al.}(2004)\citenamefont {Du},
  \citenamefont {Li},\ and\ \citenamefont {Yang}}]{Du2004}%
  \BibitemOpen
  \bibfield  {author} {\bibinfo {author} {\bibfnamefont {D.-S.}\ \bibnamefont
  {Du}}, \bibinfo {author} {\bibfnamefont {J.-W.}\ \bibnamefont {Li}},\ and\
  \bibinfo {author} {\bibfnamefont {M.-Z.}\ \bibnamefont {Yang}},\ }\bibfield
  {title} {\bibinfo {title} {Form factors and semileptonic decay of
  {$D_s^+\to\phi\bar{\ell}\nu$} from {QCD} sum rule},\ }\href
  {https://doi.org/10.1140/epjc/s2004-01979-9} {\bibfield  {journal} {\bibinfo
  {journal} {Eur. Phys. J. C}\ }\textbf {\bibinfo {volume} {37}},\ \bibinfo
  {pages} {173} (\bibinfo {year} {2004})},\ \Eprint
  {https://arxiv.org/abs/hep-ph/0308259} {arXiv:hep-ph/0308259} \BibitemShut
  {NoStop}%
\bibitem [{\citenamefont {Khodjamirian}\ \emph {et~al.}(2009)\citenamefont
  {Khodjamirian}, \citenamefont {Klein}, \citenamefont {Mannel},\ and\
  \citenamefont {Offen}}]{Khodjamirian2009}%
  \BibitemOpen
  \bibfield  {author} {\bibinfo {author} {\bibfnamefont {A.}~\bibnamefont
  {Khodjamirian}}, \bibinfo {author} {\bibfnamefont {C.}~\bibnamefont {Klein}},
  \bibinfo {author} {\bibfnamefont {T.}~\bibnamefont {Mannel}},\ and\ \bibinfo
  {author} {\bibfnamefont {N.}~\bibnamefont {Offen}},\ }\bibfield  {title}
  {\bibinfo {title} {Semileptonic charm decays {$D\to\pi\ell\nu_\ell$} and
  {$D\to K\ell\nu_\ell$} from {QCD} light-cone sum rules},\ }\href
  {https://doi.org/10.1103/PhysRevD.80.114005} {\bibfield  {journal} {\bibinfo
  {journal} {Phys. Rev. D}\ }\textbf {\bibinfo {volume} {80}},\ \bibinfo
  {pages} {114005} (\bibinfo {year} {2009})},\ \Eprint
  {https://arxiv.org/abs/0907.2842} {arXiv:0907.2842} \BibitemShut {NoStop}%
\bibitem [{\citenamefont {Fu}\ \emph {et~al.}(2020)\citenamefont {Fu},
  \citenamefont {Zeng}, \citenamefont {Lue}, \citenamefont {Cheng},\ and\
  \citenamefont {Wu}}]{Fu2020}%
  \BibitemOpen
  \bibfield  {author} {\bibinfo {author} {\bibfnamefont {H.-B.}\ \bibnamefont
  {Fu}}, \bibinfo {author} {\bibfnamefont {L.}~\bibnamefont {Zeng}}, \bibinfo
  {author} {\bibfnamefont {R.}~\bibnamefont {Lue}}, \bibinfo {author}
  {\bibfnamefont {W.}~\bibnamefont {Cheng}},\ and\ \bibinfo {author}
  {\bibfnamefont {X.-G.}\ \bibnamefont {Wu}},\ }\bibfield  {title} {\bibinfo
  {title} {The {$D\to\rho$} semileptonic and radiative decays within the
  light-cone sum rules},\ }\href
  {https://doi.org/10.1140/epjc/s10052-020-7758-4} {\bibfield  {journal}
  {\bibinfo  {journal} {Eur. Phys. J. C}\ }\textbf {\bibinfo {volume} {80}},\
  \bibinfo {pages} {194} (\bibinfo {year} {2020})},\ \Eprint
  {https://arxiv.org/abs/1808.06412} {arXiv:1808.06412} \BibitemShut {NoStop}%
\bibitem [{\citenamefont {Zhong}\ \emph {et~al.}(2024)\citenamefont {Zhong},
  \citenamefont {Dai},\ and\ \citenamefont {Fu}}]{Zhong2024}%
  \BibitemOpen
  \bibfield  {author} {\bibinfo {author} {\bibfnamefont {T.}~\bibnamefont
  {Zhong}}, \bibinfo {author} {\bibfnamefont {Y.-H.}\ \bibnamefont {Dai}},\
  and\ \bibinfo {author} {\bibfnamefont {H.-B.}\ \bibnamefont {Fu}},\
  }\bibfield  {title} {\bibinfo {title} {{$\rho$}-meson longitudinal
  leading-twist distribution amplitude revisited and the {$D\to\rho$}
  semileptonic decay},\ }\href {https://doi.org/10.1088/1674-1137/ad34be}
  {\bibfield  {journal} {\bibinfo  {journal} {Chin. Phys. C}\ }\textbf
  {\bibinfo {volume} {48}},\ \bibinfo {pages} {063108} (\bibinfo {year}
  {2024})},\ \Eprint {https://arxiv.org/abs/2308.14032} {arXiv:2308.14032}
  \BibitemShut {NoStop}%
\bibitem [{\citenamefont {Lin}\ \emph {et~al.}(2025)\citenamefont {Lin},
  \citenamefont {Huang}, \citenamefont {Cheng},\ and\ \citenamefont
  {Yao}}]{Lin2025}%
  \BibitemOpen
  \bibfield  {author} {\bibinfo {author} {\bibfnamefont {W.}~\bibnamefont
  {Lin}}, \bibinfo {author} {\bibfnamefont {X.-E.}\ \bibnamefont {Huang}},
  \bibinfo {author} {\bibfnamefont {S.}~\bibnamefont {Cheng}},\ and\ \bibinfo
  {author} {\bibfnamefont {D.-L.}\ \bibnamefont {Yao}},\ }\bibfield  {title}
  {\bibinfo {title} {Semileptonic decays of {$D\to\rho\ell^+\nu$} and
  {$D_{(s)}\to K^*\ell^+\nu$} from light-cone sum rules},\ }\href
  {https://doi.org/10.1103/jp8y-j9g8} {\bibfield  {journal} {\bibinfo
  {journal} {Phys. Rev. D}\ }\textbf {\bibinfo {volume} {111}},\ \bibinfo
  {pages} {113005} (\bibinfo {year} {2025})},\ \Eprint
  {https://arxiv.org/abs/2505.01329} {arXiv:2505.01329} \BibitemShut {NoStop}%
\bibitem [{\citenamefont {Wu}\ \emph {et~al.}(2006)\citenamefont {Wu},
  \citenamefont {Zhong},\ and\ \citenamefont {Zuo}}]{Wu2006}%
  \BibitemOpen
  \bibfield  {author} {\bibinfo {author} {\bibfnamefont {Y.-L.}\ \bibnamefont
  {Wu}}, \bibinfo {author} {\bibfnamefont {M.}~\bibnamefont {Zhong}},\ and\
  \bibinfo {author} {\bibfnamefont {Y.-B.}\ \bibnamefont {Zuo}},\ }\bibfield
  {title} {\bibinfo {title}
  {{$B_{(s)},D_{(s)}\to\pi,K,\eta,\rho,K^*,\omega,\phi$} transition form
  factors and decay rates with extraction of the {CKM} parameters},\ }\href
  {https://doi.org/10.1142/S0217751X06033209} {\bibfield  {journal} {\bibinfo
  {journal} {Int. J. Mod. Phys. A}\ }\textbf {\bibinfo {volume} {21}},\
  \bibinfo {pages} {6125} (\bibinfo {year} {2006})},\ \Eprint
  {https://arxiv.org/abs/hep-ph/0604007} {arXiv:hep-ph/0604007} \BibitemShut
  {NoStop}%
\bibitem [{\citenamefont {Cheng}\ and\ \citenamefont {Kang}(2017)}]{Cheng2017}%
  \BibitemOpen
  \bibfield  {author} {\bibinfo {author} {\bibfnamefont {H.-Y.}\ \bibnamefont
  {Cheng}}\ and\ \bibinfo {author} {\bibfnamefont {X.-W.}\ \bibnamefont
  {Kang}},\ }\bibfield  {title} {\bibinfo {title} {Branching fractions of
  semileptonic {$D$} and {$D_s$} decays from the covariant light-front quark
  model},\ }\href {https://doi.org/10.1140/epjc/s10052-017-5170-5} {\bibfield
  {journal} {\bibinfo  {journal} {Eur. Phys. J. C}\ }\textbf {\bibinfo {volume}
  {77}},\ \bibinfo {pages} {587} (\bibinfo {year} {2017})},\ \bibinfo {note}
  {erratum: Eur. Phys. J. C \textbf{77}, 863 (2017)},\ \Eprint
  {https://arxiv.org/abs/1707.02851} {arXiv:1707.02851} \BibitemShut {NoStop}%
\bibitem [{\citenamefont {Yang}\ \emph {et~al.}(2026)\citenamefont {Yang},
  \citenamefont {Guo},\ and\ \citenamefont {Zhang}}]{Yang2026}%
  \BibitemOpen
  \bibfield  {author} {\bibinfo {author} {\bibfnamefont {H.}~\bibnamefont
  {Yang}}, \bibinfo {author} {\bibfnamefont {S.-Q.}\ \bibnamefont {Guo}},\ and\
  \bibinfo {author} {\bibfnamefont {Z.-Q.}\ \bibnamefont {Zhang}},\ }\bibfield
  {title} {\bibinfo {title} {Systematic analysis of {$D_{(s)}$} meson
  semileptonic decays in the covariant light-front quark model},\ }\href
  {https://doi.org/10.1140/epjc/s10052-026-15484-8} {\bibfield  {journal}
  {\bibinfo  {journal} {Eur. Phys. J. C}\ }\textbf {\bibinfo {volume} {86}},\
  \bibinfo {pages} {363} (\bibinfo {year} {2026})},\ \Eprint
  {https://arxiv.org/abs/2511.14165} {arXiv:2511.14165} \BibitemShut {NoStop}%
\bibitem [{\citenamefont {Ivanov}\ \emph {et~al.}(2019)\citenamefont {Ivanov},
  \citenamefont {K{\"o}rner}, \citenamefont {Pandya}, \citenamefont
  {Santorelli}, \citenamefont {Soni},\ and\ \citenamefont {Tran}}]{Ivanov2019}%
  \BibitemOpen
  \bibfield  {author} {\bibinfo {author} {\bibfnamefont {M.~A.}\ \bibnamefont
  {Ivanov}}, \bibinfo {author} {\bibfnamefont {J.~G.}\ \bibnamefont
  {K{\"o}rner}}, \bibinfo {author} {\bibfnamefont {J.~N.}\ \bibnamefont
  {Pandya}}, \bibinfo {author} {\bibfnamefont {P.}~\bibnamefont {Santorelli}},
  \bibinfo {author} {\bibfnamefont {N.~R.}\ \bibnamefont {Soni}},\ and\
  \bibinfo {author} {\bibfnamefont {C.-T.}\ \bibnamefont {Tran}},\ }\bibfield
  {title} {\bibinfo {title} {Exclusive semileptonic decays of {$D$} and {$D_s$}
  mesons in the covariant confining quark model},\ }\href
  {https://doi.org/10.1007/s11467-019-0908-1} {\bibfield  {journal} {\bibinfo
  {journal} {Front. Phys. (Beijing)}\ }\textbf {\bibinfo {volume} {14}},\
  \bibinfo {pages} {64401} (\bibinfo {year} {2019})},\ \Eprint
  {https://arxiv.org/abs/1904.07740} {arXiv:1904.07740} \BibitemShut {NoStop}%
\bibitem [{\citenamefont {Yao}\ \emph {et~al.}(2022)\citenamefont {Yao},
  \citenamefont {Binosi}, \citenamefont {Cui},\ and\ \citenamefont
  {Roberts}}]{Yao2022}%
  \BibitemOpen
  \bibfield  {author} {\bibinfo {author} {\bibfnamefont {Z.-Q.}\ \bibnamefont
  {Yao}}, \bibinfo {author} {\bibfnamefont {D.}~\bibnamefont {Binosi}},
  \bibinfo {author} {\bibfnamefont {Z.-F.}\ \bibnamefont {Cui}},\ and\ \bibinfo
  {author} {\bibfnamefont {C.~D.}\ \bibnamefont {Roberts}},\ }\bibfield
  {title} {\bibinfo {title} {Semileptonic transitions: {$B_{(s)}\to\pi(K)$};
  {$D_s\to K$}; {$D\to\pi,K$}; and {$K\to\pi$}},\ }\href
  {https://doi.org/10.1016/j.physletb.2021.136793} {\bibfield  {journal}
  {\bibinfo  {journal} {Phys. Lett. B}\ }\textbf {\bibinfo {volume} {824}},\
  \bibinfo {pages} {136793} (\bibinfo {year} {2022})},\ \Eprint
  {https://arxiv.org/abs/2111.06473} {arXiv:2111.06473} \BibitemShut {NoStop}%
\bibitem [{\citenamefont {Xing}\ \emph {et~al.}(2022)\citenamefont {Xing},
  \citenamefont {Xu}, \citenamefont {Cui}, \citenamefont {Roberts},\ and\
  \citenamefont {Xu}}]{Xing2022}%
  \BibitemOpen
  \bibfield  {author} {\bibinfo {author} {\bibfnamefont {H.-Y.}\ \bibnamefont
  {Xing}}, \bibinfo {author} {\bibfnamefont {Z.-N.}\ \bibnamefont {Xu}},
  \bibinfo {author} {\bibfnamefont {Z.-F.}\ \bibnamefont {Cui}}, \bibinfo
  {author} {\bibfnamefont {C.~D.}\ \bibnamefont {Roberts}},\ and\ \bibinfo
  {author} {\bibfnamefont {C.}~\bibnamefont {Xu}},\ }\bibfield  {title}
  {\bibinfo {title} {Heavy + heavy and heavy + light pseudoscalar to vector
  semileptonic transitions},\ }\href
  {https://doi.org/10.1140/epjc/s10052-022-10844-6} {\bibfield  {journal}
  {\bibinfo  {journal} {Eur. Phys. J. C}\ }\textbf {\bibinfo {volume} {82}},\
  \bibinfo {pages} {889} (\bibinfo {year} {2022})},\ \Eprint
  {https://arxiv.org/abs/2205.13642} {arXiv:2205.13642} \BibitemShut {NoStop}%
\bibitem [{\citenamefont {Faustov}\ \emph {et~al.}(2020)\citenamefont
  {Faustov}, \citenamefont {Galkin},\ and\ \citenamefont {Kang}}]{Faustov2020}%
  \BibitemOpen
  \bibfield  {author} {\bibinfo {author} {\bibfnamefont {R.~N.}\ \bibnamefont
  {Faustov}}, \bibinfo {author} {\bibfnamefont {V.~O.}\ \bibnamefont
  {Galkin}},\ and\ \bibinfo {author} {\bibfnamefont {X.-W.}\ \bibnamefont
  {Kang}},\ }\bibfield  {title} {\bibinfo {title} {Semileptonic decays of {$D$}
  and {$D_s$} mesons in the relativistic quark model},\ }\href
  {https://doi.org/10.1103/PhysRevD.101.013004} {\bibfield  {journal} {\bibinfo
   {journal} {Phys. Rev. D}\ }\textbf {\bibinfo {volume} {101}},\ \bibinfo
  {pages} {013004} (\bibinfo {year} {2020})},\ \Eprint
  {https://arxiv.org/abs/1911.08209} {arXiv:1911.08209} \BibitemShut {NoStop}%
\bibitem [{\citenamefont {Ebert}\ \emph
  {et~al.}(2003{\natexlab{a}})\citenamefont {Ebert}, \citenamefont {Faustov},\
  and\ \citenamefont {Galkin}}]{Ebert2003b}%
  \BibitemOpen
  \bibfield  {author} {\bibinfo {author} {\bibfnamefont {D.}~\bibnamefont
  {Ebert}}, \bibinfo {author} {\bibfnamefont {R.~N.}\ \bibnamefont {Faustov}},\
  and\ \bibinfo {author} {\bibfnamefont {V.~O.}\ \bibnamefont {Galkin}},\
  }\bibfield  {title} {\bibinfo {title} {Weak decays of the {$B_c$} meson to
  charmonium and {$D$} mesons in the relativistic quark model},\ }\href
  {https://doi.org/10.1103/PhysRevD.68.094020} {\bibfield  {journal} {\bibinfo
  {journal} {Phys. Rev. D}\ }\textbf {\bibinfo {volume} {68}},\ \bibinfo
  {pages} {094020} (\bibinfo {year} {2003}{\natexlab{a}})},\ \Eprint
  {https://arxiv.org/abs/hep-ph/0306306} {arXiv:hep-ph/0306306} \BibitemShut
  {NoStop}%
\bibitem [{\citenamefont {Ebert}\ \emph
  {et~al.}(2003{\natexlab{b}})\citenamefont {Ebert}, \citenamefont {Faustov},\
  and\ \citenamefont {Galkin}}]{Ebert2003c}%
  \BibitemOpen
  \bibfield  {author} {\bibinfo {author} {\bibfnamefont {D.}~\bibnamefont
  {Ebert}}, \bibinfo {author} {\bibfnamefont {R.~N.}\ \bibnamefont {Faustov}},\
  and\ \bibinfo {author} {\bibfnamefont {V.~O.}\ \bibnamefont {Galkin}},\
  }\bibfield  {title} {\bibinfo {title} {Weak decays of the {$B_c$} meson to
  {$B_s$} and {$B$} mesons in the relativistic quark model},\ }\href
  {https://doi.org/10.1140/epjc/s2003-01347-5} {\bibfield  {journal} {\bibinfo
  {journal} {Eur. Phys. J. C}\ }\textbf {\bibinfo {volume} {32}},\ \bibinfo
  {pages} {29} (\bibinfo {year} {2003}{\natexlab{b}})},\ \Eprint
  {https://arxiv.org/abs/hep-ph/0308149} {arXiv:hep-ph/0308149} \BibitemShut
  {NoStop}%
\bibitem [{\citenamefont {Ebert}\ \emph
  {et~al.}(2003{\natexlab{c}})\citenamefont {Ebert}, \citenamefont {Faustov},\
  and\ \citenamefont {Galkin}}]{Ebert2003a}%
  \BibitemOpen
  \bibfield  {author} {\bibinfo {author} {\bibfnamefont {D.}~\bibnamefont
  {Ebert}}, \bibinfo {author} {\bibfnamefont {R.~N.}\ \bibnamefont {Faustov}},\
  and\ \bibinfo {author} {\bibfnamefont {V.~O.}\ \bibnamefont {Galkin}},\
  }\bibfield  {title} {\bibinfo {title} {Properties of heavy quarkonia and
  {$B_c$} mesons in the relativistic quark model},\ }\href
  {https://doi.org/10.1103/PhysRevD.67.014027} {\bibfield  {journal} {\bibinfo
  {journal} {Phys. Rev. D}\ }\textbf {\bibinfo {volume} {67}},\ \bibinfo
  {pages} {014027} (\bibinfo {year} {2003}{\natexlab{c}})},\ \Eprint
  {https://arxiv.org/abs/hep-ph/0210381} {arXiv:hep-ph/0210381} \BibitemShut
  {NoStop}%
\bibitem [{\citenamefont {Ebert}\ \emph {et~al.}(2009)\citenamefont {Ebert},
  \citenamefont {Faustov},\ and\ \citenamefont {Galkin}}]{Ebert2009}%
  \BibitemOpen
  \bibfield  {author} {\bibinfo {author} {\bibfnamefont {D.}~\bibnamefont
  {Ebert}}, \bibinfo {author} {\bibfnamefont {R.~N.}\ \bibnamefont {Faustov}},\
  and\ \bibinfo {author} {\bibfnamefont {V.~O.}\ \bibnamefont {Galkin}},\
  }\bibfield  {title} {\bibinfo {title} {Mass spectra and {Regge} trajectories
  of light mesons in the relativistic quark model},\ }\href
  {https://doi.org/10.1103/PhysRevD.79.114029} {\bibfield  {journal} {\bibinfo
  {journal} {Phys. Rev. D}\ }\textbf {\bibinfo {volume} {79}},\ \bibinfo
  {pages} {114029} (\bibinfo {year} {2009})},\ \Eprint
  {https://arxiv.org/abs/0903.5183} {arXiv:0903.5183} \BibitemShut {NoStop}%
\bibitem [{\citenamefont {Ebert}\ \emph {et~al.}(2010)\citenamefont {Ebert},
  \citenamefont {Faustov},\ and\ \citenamefont {Galkin}}]{Ebert2010}%
  \BibitemOpen
  \bibfield  {author} {\bibinfo {author} {\bibfnamefont {D.}~\bibnamefont
  {Ebert}}, \bibinfo {author} {\bibfnamefont {R.~N.}\ \bibnamefont {Faustov}},\
  and\ \bibinfo {author} {\bibfnamefont {V.~O.}\ \bibnamefont {Galkin}},\
  }\bibfield  {title} {\bibinfo {title} {Heavy-light meson spectroscopy and
  {Regge} trajectories in the relativistic quark model},\ }\href
  {https://doi.org/10.1140/epjc/s10052-010-1233-6} {\bibfield  {journal}
  {\bibinfo  {journal} {Eur. Phys. J. C}\ }\textbf {\bibinfo {volume} {66}},\
  \bibinfo {pages} {197} (\bibinfo {year} {2010})},\ \Eprint
  {https://arxiv.org/abs/0910.5612} {arXiv:0910.5612} \BibitemShut {NoStop}%
\bibitem [{\citenamefont {Takahashi}\ \emph {et~al.}(2026)\citenamefont
  {Takahashi} \emph {et~al.}}]{PDG2026}%
  \BibitemOpen
  \bibfield  {author} {\bibinfo {author} {\bibfnamefont {F.}~\bibnamefont
  {Takahashi}} \emph {et~al.} (\bibinfo {collaboration} {Particle Data
  Group}),\ }\bibfield  {title} {\bibinfo {title} {Review of particle
  physics},\ }\href {https://doi.org/10.1142/S0217751X26300115} {\bibfield
  {journal} {\bibinfo  {journal} {Int. J. Mod. Phys. A}\ }\textbf {\bibinfo
  {volume} {41}},\ \bibinfo {pages} {2630011} (\bibinfo {year} {2026})},\
  \bibinfo {note} {current branching-fraction averages taken from PDG Live,
  \url{https://pdglive.lbl.gov/}}\BibitemShut {NoStop}%
\bibitem [{\citenamefont {Aaij}\ \emph {et~al.}(2025)\citenamefont {Aaij} \emph
  {et~al.}}]{LHCb2025EtaMix}%
  \BibitemOpen
  \bibfield  {author} {\bibinfo {author} {\bibfnamefont {R.}~\bibnamefont
  {Aaij}} \emph {et~al.} (\bibinfo {collaboration} {LHCb Collaboration}),\
  }\bibfield  {title} {\bibinfo {title} {Improved measurement of
  {$\eta$}/{$\eta'$} mixing in {$B^0_{(s)} \to J/\psi\eta^{(\prime)}$}
  decays},\ }\href {https://doi.org/10.1007/JHEP10(2025)113} {\bibfield
  {journal} {\bibinfo  {journal} {J. High Energy Phys.}\ }\textbf {\bibinfo
  {volume} {10}}\bibfield  {number} {\bibinfo  {number} { (2025)},\ \bibinfo
  {pages} {113}},\ }\Eprint {https://arxiv.org/abs/2507.13914}
  {arXiv:2507.13914 [hep-ex]} \BibitemShut {NoStop}%
\bibitem [{\citenamefont {Escribano}\ and\ \citenamefont
  {Nadal}(2007)}]{Escribano2007}%
  \BibitemOpen
  \bibfield  {author} {\bibinfo {author} {\bibfnamefont {R.}~\bibnamefont
  {Escribano}}\ and\ \bibinfo {author} {\bibfnamefont {J.}~\bibnamefont
  {Nadal}},\ }\bibfield  {title} {\bibinfo {title} {On the gluon content of the
  {$\eta$} and {$\eta'$} mesons},\ }\href
  {https://doi.org/10.1088/1126-6708/2007/05/006} {\bibfield  {journal}
  {\bibinfo  {journal} {J. High Energy Phys.}\ }\textbf {\bibinfo {volume}
  {05}}\bibfield  {number} {\bibinfo  {number} { (2007)},\ \bibinfo {pages}
  {006}},\ }\Eprint {https://arxiv.org/abs/hep-ph/0703187}
  {arXiv:hep-ph/0703187 [hep-ph]} \BibitemShut {NoStop}%
\bibitem [{\citenamefont {Offen}\ \emph {et~al.}(2013)\citenamefont {Offen},
  \citenamefont {Porkert},\ and\ \citenamefont {Sch{\"a}fer}}]{Offen2013}%
  \BibitemOpen
  \bibfield  {author} {\bibinfo {author} {\bibfnamefont {N.}~\bibnamefont
  {Offen}}, \bibinfo {author} {\bibfnamefont {F.~A.}\ \bibnamefont {Porkert}},\
  and\ \bibinfo {author} {\bibfnamefont {A.}~\bibnamefont {Sch{\"a}fer}},\
  }\bibfield  {title} {\bibinfo {title} {Light-cone sum rules for the
  {$D_{(s)}\to\eta^{(\prime)}\ell\nu_\ell$} form factor},\ }\href
  {https://doi.org/10.1103/PhysRevD.88.034023} {\bibfield  {journal} {\bibinfo
  {journal} {Phys. Rev. D}\ }\textbf {\bibinfo {volume} {88}},\ \bibinfo
  {pages} {034023} (\bibinfo {year} {2013})},\ \Eprint
  {https://arxiv.org/abs/1307.2797} {arXiv:1307.2797} \BibitemShut {NoStop}%
\bibitem [{\citenamefont {Aubert}\ \emph {et~al.}(2007)\citenamefont {Aubert}
  \emph {et~al.}}]{BaBar2007K}%
  \BibitemOpen
  \bibfield  {author} {\bibinfo {author} {\bibfnamefont {B.}~\bibnamefont
  {Aubert}} \emph {et~al.} (\bibinfo {collaboration} {BABAR Collaboration}),\
  }\bibfield  {title} {\bibinfo {title} {Measurement of the {$D^0\to
  K^-e^+\nu_e$} decay form factor},\ }\href
  {https://doi.org/10.1103/PhysRevD.76.052005} {\bibfield  {journal} {\bibinfo
  {journal} {Phys. Rev. D}\ }\textbf {\bibinfo {volume} {76}},\ \bibinfo
  {pages} {052005} (\bibinfo {year} {2007})},\ \Eprint
  {https://arxiv.org/abs/0704.0020} {arXiv:0704.0020} \BibitemShut {NoStop}%
\bibitem [{\citenamefont {Yelton}\ \emph {et~al.}(2011)\citenamefont {Yelton}
  \emph {et~al.}}]{CLEO2011Eta}%
  \BibitemOpen
  \bibfield  {author} {\bibinfo {author} {\bibfnamefont {J.}~\bibnamefont
  {Yelton}} \emph {et~al.} (\bibinfo {collaboration} {CLEO Collaboration}),\
  }\bibfield  {title} {\bibinfo {title} {Studies of {$D^+\to\eta^\prime
  e^+\nu_e$}, {$D^+\to\eta e^+\nu_e$}, and {$D^+\to\phi e^+\nu_e$}},\ }\href
  {https://doi.org/10.1103/PhysRevD.84.032001} {\bibfield  {journal} {\bibinfo
  {journal} {Phys. Rev. D}\ }\textbf {\bibinfo {volume} {84}},\ \bibinfo
  {pages} {032001} (\bibinfo {year} {2011})},\ \Eprint
  {https://arxiv.org/abs/1011.1195} {arXiv:1011.1195} \BibitemShut {NoStop}%
\bibitem [{\citenamefont {Bali}\ \emph {et~al.}(2015)\citenamefont {Bali},
  \citenamefont {Collins}, \citenamefont {D{\"u}rr},\ and\ \citenamefont
  {Kanamori}}]{Bali2015}%
  \BibitemOpen
  \bibfield  {author} {\bibinfo {author} {\bibfnamefont {G.~S.}\ \bibnamefont
  {Bali}}, \bibinfo {author} {\bibfnamefont {S.}~\bibnamefont {Collins}},
  \bibinfo {author} {\bibfnamefont {S.}~\bibnamefont {D{\"u}rr}},\ and\
  \bibinfo {author} {\bibfnamefont {I.}~\bibnamefont {Kanamori}},\ }\bibfield
  {title} {\bibinfo {title} {{$D_s\to\eta,\eta^\prime$} semileptonic decay form
  factors with disconnected quark loop contributions},\ }\href
  {https://doi.org/10.1103/PhysRevD.91.014503} {\bibfield  {journal} {\bibinfo
  {journal} {Phys. Rev. D}\ }\textbf {\bibinfo {volume} {91}},\ \bibinfo
  {pages} {014503} (\bibinfo {year} {2015})},\ \Eprint
  {https://arxiv.org/abs/1406.5449} {arXiv:1406.5449} \BibitemShut {NoStop}%
\bibitem [{\citenamefont {Tian}\ \emph {et~al.}(2024)\citenamefont {Tian},
  \citenamefont {Yang}, \citenamefont {Hu}, \citenamefont {Fu}, \citenamefont
  {Zhong},\ and\ \citenamefont {Wu}}]{Tian2024}%
  \BibitemOpen
  \bibfield  {author} {\bibinfo {author} {\bibfnamefont {H.~J.}\ \bibnamefont
  {Tian}}, \bibinfo {author} {\bibfnamefont {Y.~L.}\ \bibnamefont {Yang}},
  \bibinfo {author} {\bibfnamefont {D.~D.}\ \bibnamefont {Hu}}, \bibinfo
  {author} {\bibfnamefont {H.~B.}\ \bibnamefont {Fu}}, \bibinfo {author}
  {\bibfnamefont {T.}~\bibnamefont {Zhong}},\ and\ \bibinfo {author}
  {\bibfnamefont {X.~G.}\ \bibnamefont {Wu}},\ }\bibfield  {title} {\bibinfo
  {title} {Searching for {$|V_{cd}|$} through the exclusive decay {$D_s^+\to
  K^0 e^+\nu_e$} within {QCD} sum rules},\ }\href
  {https://doi.org/10.1016/j.physletb.2024.138975} {\bibfield  {journal}
  {\bibinfo  {journal} {Phys. Lett. B}\ }\textbf {\bibinfo {volume} {857}},\
  \bibinfo {pages} {138975} (\bibinfo {year} {2024})},\ \Eprint
  {https://arxiv.org/abs/2405.07154} {arXiv:2405.07154} \BibitemShut {NoStop}%
\bibitem [{\citenamefont {Duplan{\v c}i{\'c}}\ and\ \citenamefont
  {Meli{\'c}}(2015)}]{Duplancic2015}%
  \BibitemOpen
  \bibfield  {author} {\bibinfo {author} {\bibfnamefont {G.}~\bibnamefont
  {Duplan{\v c}i{\'c}}}\ and\ \bibinfo {author} {\bibfnamefont
  {B.}~\bibnamefont {Meli{\'c}}},\ }\bibfield  {title} {\bibinfo {title} {Form
  factors of {$B,B_s\to\eta^{(\prime)}$} and {$D,D_s\to\eta^{(\prime)}$}
  transitions from {QCD} light-cone sum rules},\ }\href
  {https://doi.org/10.1007/JHEP11(2015)138} {\bibfield  {journal} {\bibinfo
  {journal} {J. High Energy Phys.}\ }\textbf {\bibinfo {volume} {11}}\bibfield
  {number} {\bibinfo  {number} { (2015)},\ \bibinfo {pages} {138}},\ }\Eprint
  {https://arxiv.org/abs/1508.05287} {arXiv:1508.05287} \BibitemShut {NoStop}%
\bibitem [{\citenamefont {Hu}\ \emph {et~al.}(2022)\citenamefont {Hu},
  \citenamefont {Fu}, \citenamefont {Zhong}, \citenamefont {Zeng},
  \citenamefont {Cheng},\ and\ \citenamefont {Wu}}]{Hu2022}%
  \BibitemOpen
  \bibfield  {author} {\bibinfo {author} {\bibfnamefont {D.-D.}\ \bibnamefont
  {Hu}}, \bibinfo {author} {\bibfnamefont {H.-B.}\ \bibnamefont {Fu}}, \bibinfo
  {author} {\bibfnamefont {T.}~\bibnamefont {Zhong}}, \bibinfo {author}
  {\bibfnamefont {L.}~\bibnamefont {Zeng}}, \bibinfo {author} {\bibfnamefont
  {W.}~\bibnamefont {Cheng}},\ and\ \bibinfo {author} {\bibfnamefont {X.-G.}\
  \bibnamefont {Wu}},\ }\bibfield  {title} {\bibinfo {title}
  {{$\eta^{(\prime)}$}-meson twist-2 distribution amplitude and its application
  to {$D_s^+\to\eta^{(\prime)}\ell^+\nu_\ell$}},\ }\href
  {https://doi.org/10.1140/epjc/s10052-021-09958-0} {\bibfield  {journal}
  {\bibinfo  {journal} {Eur. Phys. J. C}\ }\textbf {\bibinfo {volume} {82}},\
  \bibinfo {pages} {12} (\bibinfo {year} {2022})},\ \Eprint
  {https://arxiv.org/abs/2102.05293} {arXiv:2102.05293} \BibitemShut {NoStop}%
\bibitem [{\citenamefont {Cheng}\ \emph {et~al.}(2004)\citenamefont {Cheng},
  \citenamefont {Chua},\ and\ \citenamefont {Hwang}}]{Cheng2004}%
  \BibitemOpen
  \bibfield  {author} {\bibinfo {author} {\bibfnamefont {H.-Y.}\ \bibnamefont
  {Cheng}}, \bibinfo {author} {\bibfnamefont {C.-K.}\ \bibnamefont {Chua}},\
  and\ \bibinfo {author} {\bibfnamefont {C.-W.}\ \bibnamefont {Hwang}},\
  }\bibfield  {title} {\bibinfo {title} {Covariant light-front approach for
  s-wave and p-wave mesons: Its application to decay constants and form
  factors},\ }\href {https://doi.org/10.1103/PhysRevD.69.074025} {\bibfield
  {journal} {\bibinfo  {journal} {Phys. Rev. D}\ }\textbf {\bibinfo {volume}
  {69}},\ \bibinfo {pages} {074025} (\bibinfo {year} {2004})},\ \Eprint
  {https://arxiv.org/abs/hep-ph/0310359} {arXiv:hep-ph/0310359} \BibitemShut
  {NoStop}%
\bibitem [{\citenamefont {Verma}(2012)}]{Verma2012}%
  \BibitemOpen
  \bibfield  {author} {\bibinfo {author} {\bibfnamefont {R.~C.}\ \bibnamefont
  {Verma}},\ }\bibfield  {title} {\bibinfo {title} {Decay constants and form
  factors of s-wave and p-wave mesons in the covariant light-front quark
  model},\ }\href {https://doi.org/10.1088/0954-3899/39/2/025005} {\bibfield
  {journal} {\bibinfo  {journal} {J. Phys. G}\ }\textbf {\bibinfo {volume}
  {39}},\ \bibinfo {pages} {025005} (\bibinfo {year} {2012})},\ \Eprint
  {https://arxiv.org/abs/1103.2973} {arXiv:1103.2973} \BibitemShut {NoStop}%
\bibitem [{\citenamefont {Dobbs}\ \emph {et~al.}(2013)\citenamefont {Dobbs}
  \emph {et~al.}}]{CLEO2013RhoOmega}%
  \BibitemOpen
  \bibfield  {author} {\bibinfo {author} {\bibfnamefont {S.}~\bibnamefont
  {Dobbs}} \emph {et~al.} (\bibinfo {collaboration} {CLEO Collaboration}),\
  }\bibfield  {title} {\bibinfo {title} {First measurement of the form factors
  in the decays {$D^0\to\rho^-e^+\nu_e$} and {$D^+\to\rho^0e^+\nu_e$}},\ }\href
  {https://doi.org/10.1103/PhysRevLett.110.131802} {\bibfield  {journal}
  {\bibinfo  {journal} {Phys. Rev. Lett.}\ }\textbf {\bibinfo {volume} {110}},\
  \bibinfo {pages} {131802} (\bibinfo {year} {2013})},\ \Eprint
  {https://arxiv.org/abs/1112.2884} {arXiv:1112.2884} \BibitemShut {NoStop}%
\bibitem [{\citenamefont {Aliev}\ \emph {et~al.}(2004)\citenamefont {Aliev},
  \citenamefont {{\"O}zpineci},\ and\ \citenamefont {Savc{\i}}}]{Aliev2004}%
  \BibitemOpen
  \bibfield  {author} {\bibinfo {author} {\bibfnamefont {T.~M.}\ \bibnamefont
  {Aliev}}, \bibinfo {author} {\bibfnamefont {A.}~\bibnamefont
  {{\"O}zpineci}},\ and\ \bibinfo {author} {\bibfnamefont {M.}~\bibnamefont
  {Savc{\i}}},\ }\bibfield  {title} {\bibinfo {title} {Form factors of
  {$D_s^+\to\phi\bar{\ell}\nu$} decay in {QCD} light cone sum rule},\ }\href
  {https://doi.org/10.1140/epjc/s2004-02034-9} {\bibfield  {journal} {\bibinfo
  {journal} {Eur. Phys. J. C}\ }\textbf {\bibinfo {volume} {38}},\ \bibinfo
  {pages} {85} (\bibinfo {year} {2004})},\ \Eprint
  {https://arxiv.org/abs/hep-ph/0401181} {arXiv:hep-ph/0401181} \BibitemShut
  {NoStop}%
\bibitem [{\citenamefont {Aubert}\ \emph {et~al.}(2008)\citenamefont {Aubert}
  \emph {et~al.}}]{BaBar2008DsPhi}%
  \BibitemOpen
  \bibfield  {author} {\bibinfo {author} {\bibfnamefont {B.}~\bibnamefont
  {Aubert}} \emph {et~al.} (\bibinfo {collaboration} {BABAR Collaboration}),\
  }\bibfield  {title} {\bibinfo {title} {Study of the decay {$D_s^+\to
  K^+K^-e^+\nu_e$}},\ }\href {https://doi.org/10.1103/PhysRevD.78.051101}
  {\bibfield  {journal} {\bibinfo  {journal} {Phys. Rev. D}\ }\textbf {\bibinfo
  {volume} {78}},\ \bibinfo {pages} {051101} (\bibinfo {year} {2008})},\
  \Eprint {https://arxiv.org/abs/0807.1599} {arXiv:0807.1599} \BibitemShut
  {NoStop}%
\bibitem [{\citenamefont {Azizi}\ \emph {et~al.}(2011)\citenamefont {Azizi},
  \citenamefont {Khosravi},\ and\ \citenamefont {Falahati}}]{Azizi2011}%
  \BibitemOpen
  \bibfield  {author} {\bibinfo {author} {\bibfnamefont {K.}~\bibnamefont
  {Azizi}}, \bibinfo {author} {\bibfnamefont {R.}~\bibnamefont {Khosravi}},\
  and\ \bibinfo {author} {\bibfnamefont {F.}~\bibnamefont {Falahati}},\
  }\bibfield  {title} {\bibinfo {title} {Exclusive
  {$D_s\to(\eta,\eta')\ell\nu$} decays in light cone {QCD}},\ }\href
  {https://doi.org/10.1088/0954-3899/38/9/095001} {\bibfield  {journal}
  {\bibinfo  {journal} {J. Phys. G}\ }\textbf {\bibinfo {volume} {38}},\
  \bibinfo {pages} {095001} (\bibinfo {year} {2011})},\ \Eprint
  {https://arxiv.org/abs/1011.6046} {arXiv:1011.6046} \BibitemShut {NoStop}%
\bibitem [{\citenamefont {Hietala}\ \emph {et~al.}(2015)\citenamefont
  {Hietala}, \citenamefont {Cronin-Hennessy}, \citenamefont {Pedlar},\ and\
  \citenamefont {Shipsey}}]{CLEO2015Ds}%
  \BibitemOpen
  \bibfield  {author} {\bibinfo {author} {\bibfnamefont {J.}~\bibnamefont
  {Hietala}}, \bibinfo {author} {\bibfnamefont {D.}~\bibnamefont
  {Cronin-Hennessy}}, \bibinfo {author} {\bibfnamefont {T.}~\bibnamefont
  {Pedlar}},\ and\ \bibinfo {author} {\bibfnamefont {I.}~\bibnamefont
  {Shipsey}},\ }\bibfield  {title} {\bibinfo {title} {Exclusive {$D_s$}
  semileptonic branching fraction measurements},\ }\href
  {https://doi.org/10.1103/PhysRevD.92.012009} {\bibfield  {journal} {\bibinfo
  {journal} {Phys. Rev. D}\ }\textbf {\bibinfo {volume} {92}},\ \bibinfo
  {pages} {012009} (\bibinfo {year} {2015})},\ \Eprint
  {https://arxiv.org/abs/1505.04205} {arXiv:1505.04205} \BibitemShut {NoStop}%
\bibitem [{\citenamefont {Ablikim}\ \emph {et~al.}(2015)\citenamefont {Ablikim}
  \emph {et~al.}}]{BESIII2015Omega}%
  \BibitemOpen
  \bibfield  {author} {\bibinfo {author} {\bibfnamefont {M.}~\bibnamefont
  {Ablikim}} \emph {et~al.} (\bibinfo {collaboration} {BESIII Collaboration}),\
  }\bibfield  {title} {\bibinfo {title} {Measurement of the form factors in the
  decay {$D^+\to\omega e^+\nu_e$} and search for {$D^+\to\phi e^+\nu_e$}},\
  }\href {https://doi.org/10.1103/PhysRevD.92.071101} {\bibfield  {journal}
  {\bibinfo  {journal} {Phys. Rev. D}\ }\textbf {\bibinfo {volume} {92}},\
  \bibinfo {pages} {071101} (\bibinfo {year} {2015})},\ \Eprint
  {https://arxiv.org/abs/1508.00151} {arXiv:1508.00151} \BibitemShut {NoStop}%
\bibitem [{\citenamefont {Coan}\ \emph {et~al.}(2005)\citenamefont {Coan} \emph
  {et~al.}}]{CLEO2005Kstar}%
  \BibitemOpen
  \bibfield  {author} {\bibinfo {author} {\bibfnamefont {T.~E.}\ \bibnamefont
  {Coan}} \emph {et~al.} (\bibinfo {collaboration} {CLEO Collaboration}),\
  }\bibfield  {title} {\bibinfo {title} {Absolute branching fraction
  measurements of exclusive {$D^0$} semileptonic decays},\ }\href
  {https://doi.org/10.1103/PhysRevLett.95.181802} {\bibfield  {journal}
  {\bibinfo  {journal} {Phys. Rev. Lett.}\ }\textbf {\bibinfo {volume} {95}},\
  \bibinfo {pages} {181802} (\bibinfo {year} {2005})},\ \Eprint
  {https://arxiv.org/abs/hep-ex/0506052} {arXiv:hep-ex/0506052} \BibitemShut
  {NoStop}%
\bibitem [{\citenamefont {Briere}\ \emph {et~al.}(2010)\citenamefont {Briere}
  \emph {et~al.}}]{CLEO2010Kstar}%
  \BibitemOpen
  \bibfield  {author} {\bibinfo {author} {\bibfnamefont {R.~A.}\ \bibnamefont
  {Briere}} \emph {et~al.} (\bibinfo {collaboration} {CLEO Collaboration}),\
  }\bibfield  {title} {\bibinfo {title} {Analysis of {$D^+\to
  K^-\pi^+e^+\nu_e$} and {$D^+\to K^-\pi^+\mu^+\nu_\mu$} semileptonic decays},\
  }\href {https://doi.org/10.1103/PhysRevD.81.112001} {\bibfield  {journal}
  {\bibinfo  {journal} {Phys. Rev. D}\ }\textbf {\bibinfo {volume} {81}},\
  \bibinfo {pages} {112001} (\bibinfo {year} {2010})},\ \Eprint
  {https://arxiv.org/abs/1004.1954} {arXiv:1004.1954 [hep-ex]} \BibitemShut
  {NoStop}%
\end{thebibliography}%

\end{document}